\documentclass[prd,amsmath,aps,floats,amssymb, floatfix, superscriptaddress,nofootinbib,twocolumn,preprintnumbers]{revtex4-1}

\usepackage{amsmath,amssymb,natbib,latexsym,times,subfigure}

\usepackage[switch,columnwise]{lineno}%DH commented this since didn't compile with it
\usepackage[T1]{fontenc}
\usepackage{aecompl}
\newcommand{\aap}{A\&A}
\newcommand{\apjs}{ApJS}
\newcommand{\aj}{A.J.}
\newcommand{\mnras}{MNRAS}

\newcommand{\jcap}{JCAP}
\newcommand{\apjl}{ApJ}
\newcommand{\pasp}{PASP}
\newcommand{\procspie}{SPIE}
\usepackage{verbatim}
\usepackage{cprotect}
\usepackage{hhline,cuted}

\usepackage{verbatim}
\usepackage[usenames,dvipsnames]{xcolor}
\usepackage{tabulary}
\usepackage{tabularx}
\usepackage{todonotes}
\usepackage{hyperref}
\numberwithin{equation}{section}

\newcommand\be{\begin{equation}}
\newcommand\ee{\end{equation}}
\def\bea{\begin{eqnarray}}
\def\eea{\end{eqnarray}}
\newcommand{\green}[1]{#1}

\newcommand{\jab}[1]{#1}

\newcommand{\blue}[1]{#1}
\newcommand{\bblue}[1]{#1}

\newcommand{\orange}[1]{#1}

\newcommand{\last}[1]{#1}

\allowdisplaybreaks
\begin{document}
%\linenumbers

%\textcolor{blue}{OL: a slightly modified title}
\title[Density split statistics: joint model of counts and lensing in cells]{Density split statistics:\\joint model of counts and lensing in cells}
%\title{Cosmology from Shear and Clustering in DES Year 1 Data}

% Author list file generated with: mkauthlist 1.2.2 
% mkauthlist DES-2017-0268_author_list.csv -a order.csv -j prd 

\author{O.~Friedrich}
\email[Corresponding author: ]{oliverf@usm.uni-muenchen.de}
\affiliation{Universit\"ats-Sternwarte, Fakult\"at f\"ur Physik, Ludwig-Maximilians Universit\"at M\"unchen, Scheinerstr. 1, 81679 M\"unchen, Germany}
\affiliation{Max Planck Institute for Extraterrestrial Physics, Giessenbachstrasse, 85748 Garching, Germany}
\author{D.~Gruen}
\email{Einstein Fellow}
\affiliation{Kavli Institute for Particle Astrophysics \& Cosmology, P. O. Box 2450, Stanford University, Stanford, CA 94305, USA}
\affiliation{SLAC National Accelerator Laboratory, Menlo Park, CA 94025, USA}
\author{J.~DeRose}
\affiliation{Department of Physics, Stanford University, 382 Via Pueblo Mall, Stanford, CA 94305, USA}
\affiliation{Kavli Institute for Particle Astrophysics \& Cosmology, P. O. Box 2450, Stanford University, Stanford, CA 94305, USA}
\author{D.~Kirk}
\affiliation{Department of Physics \& Astronomy, University College London, Gower Street, London, WC1E 6BT, UK}
\author{E.~Krause}
\affiliation{Kavli Institute for Particle Astrophysics \& Cosmology, P. O. Box 2450, Stanford University, Stanford, CA 94305, USA}
\author{T.~McClintock}
\affiliation{Department of Physics, University of Arizona, Tucson, AZ 85721, USA}
\author{E.~S.~Rykoff}
\affiliation{Kavli Institute for Particle Astrophysics \& Cosmology, P. O. Box 2450, Stanford University, Stanford, CA 94305, USA}
\affiliation{SLAC National Accelerator Laboratory, Menlo Park, CA 94025, USA}
\author{S.~Seitz}
\affiliation{Max Planck Institute for Extraterrestrial Physics, Giessenbachstrasse, 85748 Garching, Germany}
\affiliation{Universit\"ats-Sternwarte, Fakult\"at f\"ur Physik, Ludwig-Maximilians Universit\"at M\"unchen, Scheinerstr. 1, 81679 M\"unchen, Germany}
\author{R.~H.~Wechsler}
\affiliation{Department of Physics, Stanford University, 382 Via Pueblo Mall, Stanford, CA 94305, USA}
\affiliation{Kavli Institute for Particle Astrophysics \& Cosmology, P. O. Box 2450, Stanford University, Stanford, CA 94305, USA}
\affiliation{SLAC National Accelerator Laboratory, Menlo Park, CA 94025, USA}
\author{G.~M.~Bernstein}
\affiliation{Department of Physics and Astronomy, University of Pennsylvania, Philadelphia, PA 19104, USA}
\author{J.~Blazek}
\affiliation{Center for Cosmology and Astro-Particle Physics, The Ohio State University, Columbus, OH 43210, USA}
\affiliation{Institute of Physics, Laboratory of Astrophysics, \'Ecole Polytechnique F\'ed\'erale de Lausanne (EPFL), Observatoire de Sauverny, 1290 Versoix, Switzerland}
\author{C.~Chang}
\affiliation{Kavli Institute for Cosmological Physics, University of Chicago, Chicago, IL 60637, USA}
\author{S.~Hilbert}
\affiliation{Faculty of Physics, Ludwig-Maximilians-Universit\"at, Scheinerstr. 1, 81679 Munich, Germany}
\affiliation{Excellence Cluster Universe, Boltzmannstr.\ 2, 85748 Garching, Germany}
\author{B.~Jain}
\affiliation{Department of Physics and Astronomy, University of Pennsylvania, Philadelphia, PA 19104, USA}
\author{A.~Kovacs}
\affiliation{Institut de F\'{\i}sica d'Altes Energies (IFAE), The Barcelona Institute of Science and Technology, Campus UAB, 08193 Bellaterra (Barcelona) Spain}
\author{O.~Lahav}
\affiliation{Department of Physics \& Astronomy, University College London, Gower Street, London, WC1E 6BT, UK}
\author{F.~B.~Abdalla}
\affiliation{Department of Physics \& Astronomy, University College London, Gower Street, London, WC1E 6BT, UK}
\affiliation{Department of Physics and Electronics, Rhodes University, PO Box 94, Grahamstown, 6140, South Africa}
\author{S.~Allam}
\affiliation{Fermi National Accelerator Laboratory, P. O. Box 500, Batavia, IL 60510, USA}
\author{J.~Annis}
\affiliation{Fermi National Accelerator Laboratory, P. O. Box 500, Batavia, IL 60510, USA}
\author{K.~Bechtol}
\affiliation{LSST, 933 North Cherry Avenue, Tucson, AZ 85721, USA}
\author{A.~Benoit-L{\'e}vy}
\affiliation{CNRS, UMR 7095, Institut d'Astrophysique de Paris, F-75014, Paris, France}
\affiliation{Department of Physics \& Astronomy, University College London, Gower Street, London, WC1E 6BT, UK}
\affiliation{Sorbonne Universit\'es, UPMC Univ Paris 06, UMR 7095, Institut d'Astrophysique de Paris, F-75014, Paris, France}
\author{E.~Bertin}
\affiliation{CNRS, UMR 7095, Institut d'Astrophysique de Paris, F-75014, Paris, France}
\affiliation{Sorbonne Universit\'es, UPMC Univ Paris 06, UMR 7095, Institut d'Astrophysique de Paris, F-75014, Paris, France}
\author{D.~Brooks}
\affiliation{Department of Physics \& Astronomy, University College London, Gower Street, London, WC1E 6BT, UK}
\author{A.~Carnero~Rosell}
\affiliation{Laborat\'orio Interinstitucional de e-Astronomia - LIneA, Rua Gal. Jos\'e Cristino 77, Rio de Janeiro, RJ - 20921-400, Brazil}
\affiliation{Observat\'orio Nacional, Rua Gal. Jos\'e Cristino 77, Rio de Janeiro, RJ - 20921-400, Brazil}
\author{M.~Carrasco~Kind}
\affiliation{Department of Astronomy, University of Illinois, 1002 W. Green Street, Urbana, IL 61801, USA}
\affiliation{National Center for Supercomputing Applications, 1205 West Clark St., Urbana, IL 61801, USA}
\author{J.~Carretero}
\affiliation{Institut de F\'{\i}sica d'Altes Energies (IFAE), The Barcelona Institute of Science and Technology, Campus UAB, 08193 Bellaterra (Barcelona) Spain}
\author{C.~E.~Cunha}
\affiliation{Kavli Institute for Particle Astrophysics \& Cosmology, P. O. Box 2450, Stanford University, Stanford, CA 94305, USA}
\author{C.~B.~D'Andrea}
\affiliation{Department of Physics and Astronomy, University of Pennsylvania, Philadelphia, PA 19104, USA}
\author{L.~N.~da Costa}
\affiliation{Laborat\'orio Interinstitucional de e-Astronomia - LIneA, Rua Gal. Jos\'e Cristino 77, Rio de Janeiro, RJ - 20921-400, Brazil}
\affiliation{Observat\'orio Nacional, Rua Gal. Jos\'e Cristino 77, Rio de Janeiro, RJ - 20921-400, Brazil}
\author{C.~Davis}
\affiliation{Kavli Institute for Particle Astrophysics \& Cosmology, P. O. Box 2450, Stanford University, Stanford, CA 94305, USA}
\author{S.~Desai}
\affiliation{Department of Physics, IIT Hyderabad, Kandi, Telangana 502285, India}
\author{H.~T.~Diehl}
\affiliation{Fermi National Accelerator Laboratory, P. O. Box 500, Batavia, IL 60510, USA}
\author{J.~P.~Dietrich}
\affiliation{Faculty of Physics, Ludwig-Maximilians-Universit\"at, Scheinerstr. 1, 81679 Munich, Germany}
\affiliation{Excellence Cluster Universe, Boltzmannstr.\ 2, 85748 Garching, Germany}
\author{A.~Drlica-Wagner}
\affiliation{Fermi National Accelerator Laboratory, P. O. Box 500, Batavia, IL 60510, USA}
\author{T.~F.~Eifler}
\affiliation{Department of Physics, California Institute of Technology, Pasadena, CA 91125, USA}
\affiliation{Jet Propulsion Laboratory, California Institute of Technology, 4800 Oak Grove Dr., Pasadena, CA 91109, USA}
\author{P.~Fosalba}
\affiliation{Institute of Space Sciences, IEEC-CSIC, Campus UAB, Carrer de Can Magrans, s/n,  08193 Barcelona, Spain}
\author{J.~Frieman}
\affiliation{Fermi National Accelerator Laboratory, P. O. Box 500, Batavia, IL 60510, USA}
\affiliation{Kavli Institute for Cosmological Physics, University of Chicago, Chicago, IL 60637, USA}
\author{J.~Garc\'ia-Bellido}
\affiliation{Instituto de Fisica Teorica UAM/CSIC, Universidad Autonoma de Madrid, 28049 Madrid, Spain}
\author{E.~Gaztanaga}
\affiliation{Institute of Space Sciences, IEEC-CSIC, Campus UAB, Carrer de Can Magrans, s/n,  08193 Barcelona, Spain}
\author{D.~W.~Gerdes}
\affiliation{Department of Astronomy, University of Michigan, Ann Arbor, MI 48109, USA}
\affiliation{Department of Physics, University of Michigan, Ann Arbor, MI 48109, USA}
\author{T.~Giannantonio}
\affiliation{Institute of Astronomy, University of Cambridge, Madingley Road, Cambridge CB3 0HA, UK}
\affiliation{Kavli Institute for Cosmology, University of Cambridge, Madingley Road, Cambridge CB3 0HA, UK}
\affiliation{Universit\"ats-Sternwarte, Fakult\"at f\"ur Physik, Ludwig-Maximilians Universit\"at M\"unchen, Scheinerstr. 1, 81679 M\"unchen, Germany}
\author{R.~A.~Gruendl}
\affiliation{Department of Astronomy, University of Illinois, 1002 W. Green Street, Urbana, IL 61801, USA}
\affiliation{National Center for Supercomputing Applications, 1205 West Clark St., Urbana, IL 61801, USA}
\author{J.~Gschwend}
\affiliation{Laborat\'orio Interinstitucional de e-Astronomia - LIneA, Rua Gal. Jos\'e Cristino 77, Rio de Janeiro, RJ - 20921-400, Brazil}
\affiliation{Observat\'orio Nacional, Rua Gal. Jos\'e Cristino 77, Rio de Janeiro, RJ - 20921-400, Brazil}
\author{G.~Gutierrez}
\affiliation{Fermi National Accelerator Laboratory, P. O. Box 500, Batavia, IL 60510, USA}
\author{K.~Honscheid}
\affiliation{Center for Cosmology and Astro-Particle Physics, The Ohio State University, Columbus, OH 43210, USA}
\affiliation{Department of Physics, The Ohio State University, Columbus, OH 43210, USA}
\author{D.~J.~James}
\affiliation{Astronomy Department, University of Washington, Box 351580, Seattle, WA 98195, USA}
\author{M.~Jarvis}
\affiliation{Department of Physics and Astronomy, University of Pennsylvania, Philadelphia, PA 19104, USA}
\author{K.~Kuehn}
\affiliation{Australian Astronomical Observatory, North Ryde, NSW 2113, Australia}
\author{N.~Kuropatkin}
\affiliation{Fermi National Accelerator Laboratory, P. O. Box 500, Batavia, IL 60510, USA}
\author{M.~Lima}
\affiliation{Departamento de F\'isica Matem\'atica, Instituto de F\'isica, Universidade de S\~ao Paulo, CP 66318, S\~ao Paulo, SP, 05314-970, Brazil}
\affiliation{Laborat\'orio Interinstitucional de e-Astronomia - LIneA, Rua Gal. Jos\'e Cristino 77, Rio de Janeiro, RJ - 20921-400, Brazil}
\author{M.~March}
\affiliation{Department of Physics and Astronomy, University of Pennsylvania, Philadelphia, PA 19104, USA}
\author{J.~L.~Marshall}
\affiliation{George P. and Cynthia Woods Mitchell Institute for Fundamental Physics and Astronomy, and Department of Physics and Astronomy, Texas A\&M University, College Station, TX 77843,  USA}
\author{P.~Melchior}
\affiliation{Department of Astrophysical Sciences, Princeton University, Peyton Hall, Princeton, NJ 08544, USA}
\author{F.~Menanteau}
\affiliation{Department of Astronomy, University of Illinois, 1002 W. Green Street, Urbana, IL 61801, USA}
\affiliation{National Center for Supercomputing Applications, 1205 West Clark St., Urbana, IL 61801, USA}
\author{R.~Miquel}
\affiliation{Instituci\'o Catalana de Recerca i Estudis Avan\c{c}ats, E-08010 Barcelona, Spain}
\affiliation{Institut de F\'{\i}sica d'Altes Energies (IFAE), The Barcelona Institute of Science and Technology, Campus UAB, 08193 Bellaterra (Barcelona) Spain}
\author{J.~J.~Mohr}
\affiliation{Excellence Cluster Universe, Boltzmannstr.\ 2, 85748 Garching, Germany}
\affiliation{Faculty of Physics, Ludwig-Maximilians-Universit\"at, Scheinerstr. 1, 81679 Munich, Germany}
\affiliation{Max Planck Institute for Extraterrestrial Physics, Giessenbachstrasse, 85748 Garching, Germany}
\author{B.~Nord}
\affiliation{Fermi National Accelerator Laboratory, P. O. Box 500, Batavia, IL 60510, USA}
\author{A.~A.~Plazas}
\affiliation{Jet Propulsion Laboratory, California Institute of Technology, 4800 Oak Grove Dr., Pasadena, CA 91109, USA}
\author{E.~Sanchez}
\affiliation{Centro de Investigaciones Energ\'eticas, Medioambientales y Tecnol\'ogicas (CIEMAT), Madrid, Spain}
\author{V.~Scarpine}
\affiliation{Fermi National Accelerator Laboratory, P. O. Box 500, Batavia, IL 60510, USA}
\author{R.~Schindler}
\affiliation{SLAC National Accelerator Laboratory, Menlo Park, CA 94025, USA}
\author{M.~Schubnell}
\affiliation{Department of Physics, University of Michigan, Ann Arbor, MI 48109, USA}
\author{I.~Sevilla-Noarbe}
\affiliation{Centro de Investigaciones Energ\'eticas, Medioambientales y Tecnol\'ogicas (CIEMAT), Madrid, Spain}
\author{E.~Sheldon}
\affiliation{Brookhaven National Laboratory, Bldg 510, Upton, NY 11973, USA}
\author{M.~Smith}
\affiliation{School of Physics and Astronomy, University of Southampton,  Southampton, SO17 1BJ, UK}
\author{M.~Soares-Santos}
\affiliation{Fermi National Accelerator Laboratory, P. O. Box 500, Batavia, IL 60510, USA}
\author{F.~Sobreira}
\affiliation{Instituto de F\'isica Gleb Wataghin, Universidade Estadual de Campinas, 13083-859, Campinas, SP, Brazil}
\affiliation{Laborat\'orio Interinstitucional de e-Astronomia - LIneA, Rua Gal. Jos\'e Cristino 77, Rio de Janeiro, RJ - 20921-400, Brazil}
\author{E.~Suchyta}
\affiliation{Computer Science and Mathematics Division, Oak Ridge National Laboratory, Oak Ridge, TN 37831}
\author{M.~E.~C.~Swanson}
\affiliation{National Center for Supercomputing Applications, 1205 West Clark St., Urbana, IL 61801, USA}
\author{G.~Tarle}
\affiliation{Department of Physics, University of Michigan, Ann Arbor, MI 48109, USA}
\author{D.~Thomas}
\affiliation{Institute of Cosmology \& Gravitation, University of Portsmouth, Portsmouth, PO1 3FX, UK}
\author{M.~A.~Troxel}
\affiliation{Center for Cosmology and Astro-Particle Physics, The Ohio State University, Columbus, OH 43210, USA}
\affiliation{Department of Physics, The Ohio State University, Columbus, OH 43210, USA}
\author{V.~Vikram}
\affiliation{Argonne National Laboratory, 9700 South Cass Avenue, Lemont, IL 60439, USA}
\author{J.~Weller}
\affiliation{Excellence Cluster Universe, Boltzmannstr.\ 2, 85748 Garching, Germany}
\affiliation{Max Planck Institute for Extraterrestrial Physics, Giessenbachstrasse, 85748 Garching, Germany}
\affiliation{Universit\"ats-Sternwarte, Fakult\"at f\"ur Physik, Ludwig-Maximilians Universit\"at M\"unchen, Scheinerstr. 1, 81679 M\"unchen, Germany}

\collaboration{DES Collaboration}

% Bibliography and bibfile
% Taken from aa.cls
\def\aj{AJ}%
          % Astronomical Journal
\def\araa{ARA\&A}%
          % Annual Review of Astron and Astrophys
\def\apj{ApJ}%
          % Astrophysical Journal
\def\apjl{ApJ}%
          % Astrophysical Journal, Letters
\def\apjs{ApJS}%
          % Astrophysical Journal, Supplement
\def\ao{Appl.~Opt.}%
          % Applied Optics
\def\apss{Ap\&SS}%
          % Astrophysics and Space Science
\def\aap{A\&A}%
          % Astronomy and Astrophysics
\def\aapr{A\&A~Rev.}%
          % Astronomy and Astrophysics Reviews
\def\aaps{A\&AS}%
          % Astronomy and Astrophysics, Supplement
\def\azh{AZh}%
          % Astronomicheskii Zhurnal
\def\baas{BAAS}%
          % Bulletin of the AAS
\def\jrasc{JRASC}%
          % Journal of the RAS of Canada
\def\memras{MmRAS}%
          % Memoirs of the RAS
\def\mnras{MNRAS}%
          % Monthly Notices of the RAS
\def\pra{Phys.~Rev.~A}%
          % Physical Review A: General Physics
\def\prb{Phys.~Rev.~B}%
          % Physical Review B: Solid State
\def\prc{Phys.~Rev.~C}%
          % Physical Review C
\def\prd{Phys.~Rev.~D}%
          % Physical Review D
\def\pre{Phys.~Rev.~E}%
          % Physical Review E
\def\prl{Phys.~Rev.~Lett.}%
          % Physical Review Letters
\def\pasp{PASP}%
          % Publications of the ASP
\def\pasj{PASJ}%
          % Publications of the ASJ
\def\qjras{QJRAS}%
          % Quarterly Journal of the RAS
\def\skytel{S\&T}%
          % Sky and Telescope
\def\solphys{Sol.~Phys.}%
          % Solar Physics
\def\sovast{Soviet~Ast.}%
          % Soviet Astronomy
\def\ssr{Space~Sci.~Rev.}%
          % Space Science Reviews
\def\zap{ZAp}%
          % Zeitschrift fuer Astrophysik
\def\nat{Nature}%
          % Nature
\def\iaucirc{IAU~Circ.}%
          % IAU Cirulars
\def\aplett{Astrophys.~Lett.}%
          % Astrophysics Letters
\def\apspr{Astrophys.~Space~Phys.~Res.}%
          % Astrophysics Space Physics Research
\def\bain{Bull.~Astron.~Inst.~Netherlands}%
          % Bulletin Astronomical Institute of the Netherlands
\def\fcp{Fund.~Cosmic~Phys.}%
          % Fundamental Cosmic Physics
\def\gca{Geochim.~Cosmochim.~Acta}%
          % Geochimica Cosmochimica Acta
\def\grl{Geophys.~Res.~Lett.}%
          % Geophysics Research Letters
\def\jcap{JCAP}%
          % Journal of Cosmology and Astroparticle Physics
\def\jcp{J.~Chem.~Phys.}%
          % Journal of Chemical Physics
\def\jgr{J.~Geophys.~Res.}%
          % Journal of Geophysics Research
\def\jqsrt{J.~Quant.~Spec.~Radiat.~Transf.}%
          % Journal of Quantitiative Spectroscopy and Radiative Trasfer
\def\memsai{Mem.~Soc.~Astron.~Italiana}%
          % Mem. Societa Astronomica Italiana
\def\nphysa{Nucl.~Phys.~A}%
          % Nuclear Physics A
\def\physrep{Phys.~Rep.}%
          % Physics Reports
\def\physscr{Phys.~Scr}%
          % Physica Scripta
\def\planss{Planet.~Space~Sci.}%
          % Planetary Space Science
\def\procspie{Proc.~SPIE}%
          % Proceedings of the SPIE

%%%\author{Dark Energy Survey Collaboration}

\date{\today}

%\pubyear{2017}

\label{firstpage}
\begin{abstract}
We present density split statistics, a framework that studies lensing and counts-in-cells as a function of foreground galaxy density, thereby providing a large-scale measurement of both 2-point and 3-point statistics. Our method extends our earlier work on trough lensing and is summarized as follows: given a foreground (low redshift) population of galaxies, we divide the sky into subareas of equal size but distinct galaxy density. We then measure lensing around uniformly spaced points separately in each of these subareas, as well as counts-in-cells statistics (CiC). The lensing signals trace the matter density contrast around regions of fixed galaxy density. Through the CiC measurements this can be related to the density profile around regions of fixed matter density. Together, these measurements constitute a powerful probe of cosmology, the skewness of the density field and the connection of galaxies and matter.

In this paper we show how to model both the density split lensing signal and CiC from basic ingredients: a non-linear power spectrum, clustering hierarchy coefficients from perturbation theory and a parametric model for galaxy bias and shot-noise. Using N-body simulations, we demonstrate that this model is sufficiently accurate for a cosmological analysis on year 1 data from the Dark Energy Survey. 
\end{abstract}

\pacs{Valid PACS appear here}

\preprint{DES-2017-0268}
%\preprint{FERMILAB-PUB-17-444-AE}
\preprint{FERMILAB-PUB-17-445-A}

\maketitle

\section{Introduction}
\label{sec:introduction}

The large-scale structure (LSS) observed today is thought to originate from almost perfectly Gaussian density perturbations in the early Universe. This means that there was a complete symmetry in the abundance and amplitude of underdense and overdense regions in very early times. Gravitational attraction then caused initial overdensities to collapse to small but highly overdense structures such as galaxy clusters, while initial underdensities expanded but stayed moderately underdense and e.g. became voids. As a consequence the majority of the volume in the late-time Universe is underdense, compensated by the presence of few highly overdense spots. Or, in other words, a positive skewness in the distribution of density fluctuations emerges due to gravitational collapse. 

A variety of probes have been used to study the statistical properties of the late-time density field and to thereby understand the physics of gravitational collapse as well as the processes responsible for the properties of the initial density fluctuations. So far, the most extensive studies have been carried out on the 2-point statistics of density fluctuations, i.e. on measuring the variance of density fluctuations as a function of scale. This has e.g. been done through measurements of cosmic shear 2-point correlation functions \citep[e.g.][]{Wittman2000,Vanwaerbeke2000,Benjamin2007,Fu2008,Schrabback2010,Kilbinger2013, Becker2015, Jee2016, Hildebrandt2017, Troxel2017_short}, galaxy clustering \citep[e.g.][]{Crocce2016, EPoole2017_short, Alam2017} and galaxy-galaxy lensing \citep[e.g.][]{Brainerd1996,Hudson1998,Wilson2001,vanUitert2011,Brimioulle2013,Clampitt2017_short, Prat2017_short} as well as combined measurements thereof \citep[e.g.][]{Mandelbaum2013, vanUitert2017, DES2017_short}.

While 2-point statistics are only sensitive to the overall amplitude of density fluctuations, higher-order statistics also know about the skewness arising from the different behaviour of underdense and overdense regions. This does not necessarily mean that higher-order statistics are better than 2-point statistics in discriminating between particular choices of cosmological parameters \citep{Barreira2017}. But they scale differently with parameters such as $\Omega_m$, $\sigma_8$, galaxy bias and galaxy stochasticity than their 2-point counterparts. Hence, in a cosmological analysis that varies a large number of parameters, probes that are sensitive to both 2-point and higher order statistics have the power to break degeneracies between these parameters \citep{Bernardeau1997, Takada2002, Pires2012, Uhlemann2017}.

%\red{Can we instead have a three-fold reason why this is interesting? It (1) adds additional independent statistical information, (2) has a different degeneracy with cosmological and nuisance parameters, and (3) is sensitive to novel physics that is different from Newtonian gravitational collapse in a uniformly expanding universe when acting in either underdense or overdense regions (buf cf. Barreira).}

Observations of higher-order statistical features of the density field include measurements of three point correlation functions \citep{Semboloni2011}, shear peak statistics \citep{Lin2015, Liu2015, Kacprzak2016} and the cluster mass function \citep{Mantz2016}. Also, a number of probes have been suggested (and in some cases measured in data) that study the correlation of 2-point statistics and background density. \citet{Chiang2015} have measured this by means of the integrated bispectrum. \citet{Simpson2011, Simpson2013, Simpson2016} have proposed a clipped power spectrum approach, where 2-point statistics are measured on the sky after excluding high density regions. They have shown that these measurements contain information complementary to the corresponding measurements on the full sky. 

A similar direction was investigated by \citet{Gruen2016} who separately measured the lensing power spectrum in underdense and overdense lines of sight. The framework presented in this paper is based on their concept of trough lensing. \last{We will call it \emph{density split lensing} when only lensing measurements are involved and \emph{density split statistics} when it is combined with counts-in-cells measurements.} This method can be summarized as follows: we consider a foreground (low redshift) population of galaxies and smooth their position field with a circular \bblue{top-hat} aperture. This smoothed density field is then used to divide the sky into sub-areas of equal size but distinct galaxy density. In this paper we consider in particular 5 sub-areas and call them quintiles of galaxy density. As a next step, we use a background (high redshift) population of galaxies to measure the tangential shear of these galaxies around a set of uniformly spaced points within the area of each density quintile. The resulting lensing signals trace the matter density contrast around regions of fixed foreground galaxy density. This data vector is then complemented by the histogram of counts-in-cells of the foreground galaxies to pin down their bias and stochasticity. As we show in this paper, a cosmological analysis based on this density split data vector has a number of desirable features:
\begin{trivlist}
\item[$\bullet$] it allows for an accurate analytic modeling with the help of cosmological perturbation theory and a non-linear power spectrum, 
\item[$\bullet$] it yields high signal-to-noise measurement,
\item[$\bullet$] it avoids systematics common to cosmic shear such as additive shear biases or intrinsic alignments (as long as tracer sample and source sample do not overlap in redshift),
\item[$\bullet$] it has a very intuitive interpretation.
%\item[$\bullet$] \red{can we add a point on complementarity?}
\end{trivlist}
This paper is a companion paper of \citet{Gruen2017}, where we present our actual data analysis, including tests for systematic effects as well as a description of how we estimate the covariance of our signal. This paper is presenting the modeling framework used in that analysis. \orange{Our section \ref{sec:cosmology} gives a general overview of density split statistics: we describe our data vector, explain how it can be modelled and also present forecasts on cosmological parameter constraints, both for a $\Lambda$CDM model and an extended model that allows gravitational collapse to behave differently than within general relativity.} In section \ref{sec:simulations} we describe the simulated data used in this work. \orange{Section \ref{sec:model} explains details of the model presented in section \ref{sec:cosmology}. There we also compare individual \last{components} of this model directly to measurements in N-body simulations.} In section \ref{sec:log-normal} we show that our model for a data vector combining density split lensing and counts-in-cells statistics is accurate enough to recover the cosmology underlying our N-body simulations. Any possible deviation between our model and the simulations is shown to be well within statistical uncertainties of year 1 data of the Dark Energy Survey (DES Y1).

\blue{In appendix \ref{app:differential_equations}, we review a number of differential equations that govern gravitational collapse. In appendix \ref{app:PT_for_skewness}, we review the leading order perturbative calculation of the 3-point statistics of the cosmic density field for a general $\Lambda$CDM model. Appendix \ref{app:appendix_C} qualitatively compares our model of the cosmic density PDF to a second set of N-body simulations as a complement to the comparison carried out in the main text. Appendix \ref{app:stochasticity} derives properties of joint log-normal random fields and appendix \ref{app:validate_alternative_model} repeats the validation of our model for an alternative shot-noise parametrization.}

\section{Density split statistics: data vector, modeling and forecasts}
\label{sec:cosmology}

This section provides an introduction to the program of density \last{split} statistics. In section \ref{sec:measuring_DT} we describe how we obtain the density split lensing signal and how this signal can be further complemented with information on galaxy bias and stochasticity from counts-in-cells. \orange{In section \ref{sec:modeling_DT} we outline our modeling of this signal (but postponing technical details of this model to section \ref{sec:model}).} In section \ref{sec:using_DT} we provide forecasts on the cosmological information that can be obtained with a measurement of density split statistics in year-1 data of the Dark Energy Survey (DES Y1).

\begin{figure*}
  \includegraphics[width=\textwidth]{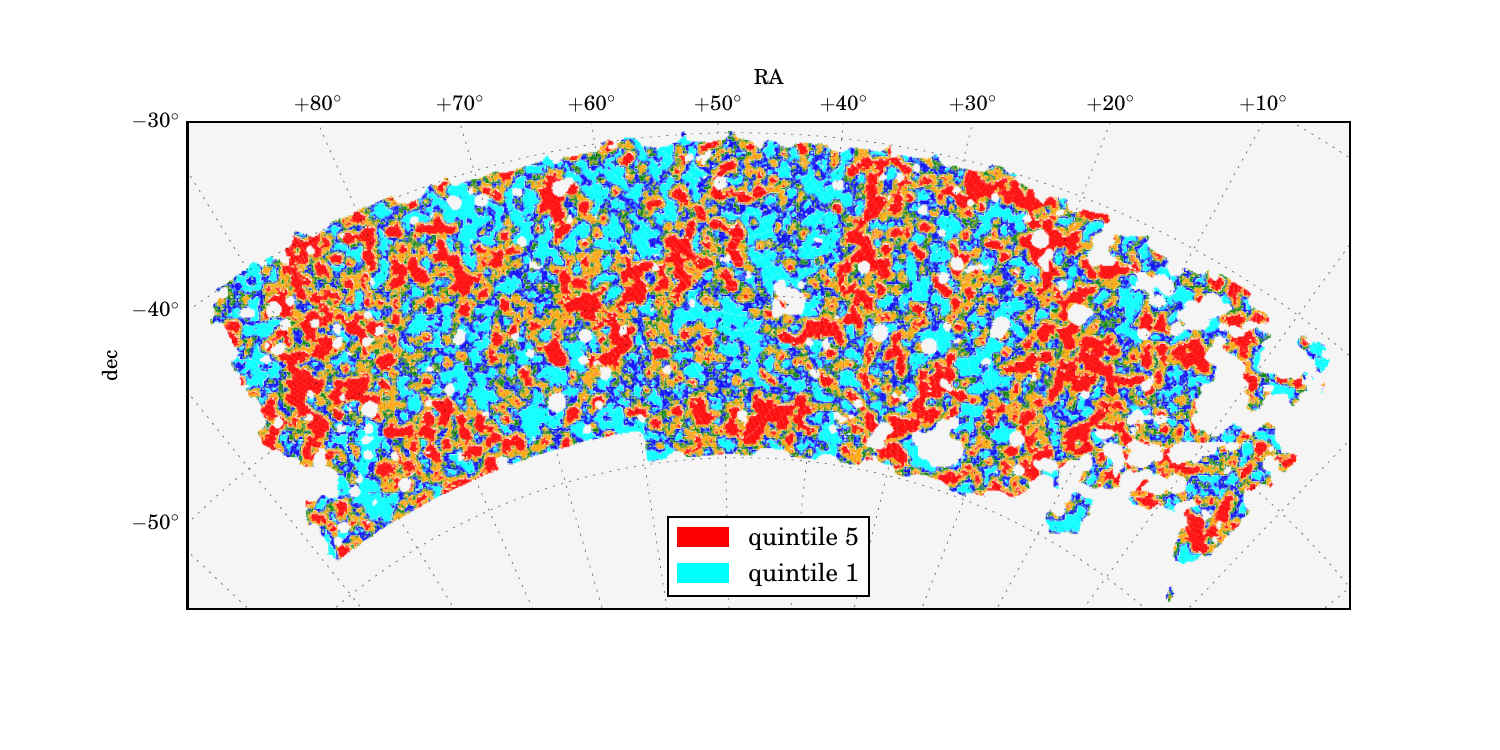}
  
  \includegraphics[width=0.444\textwidth]{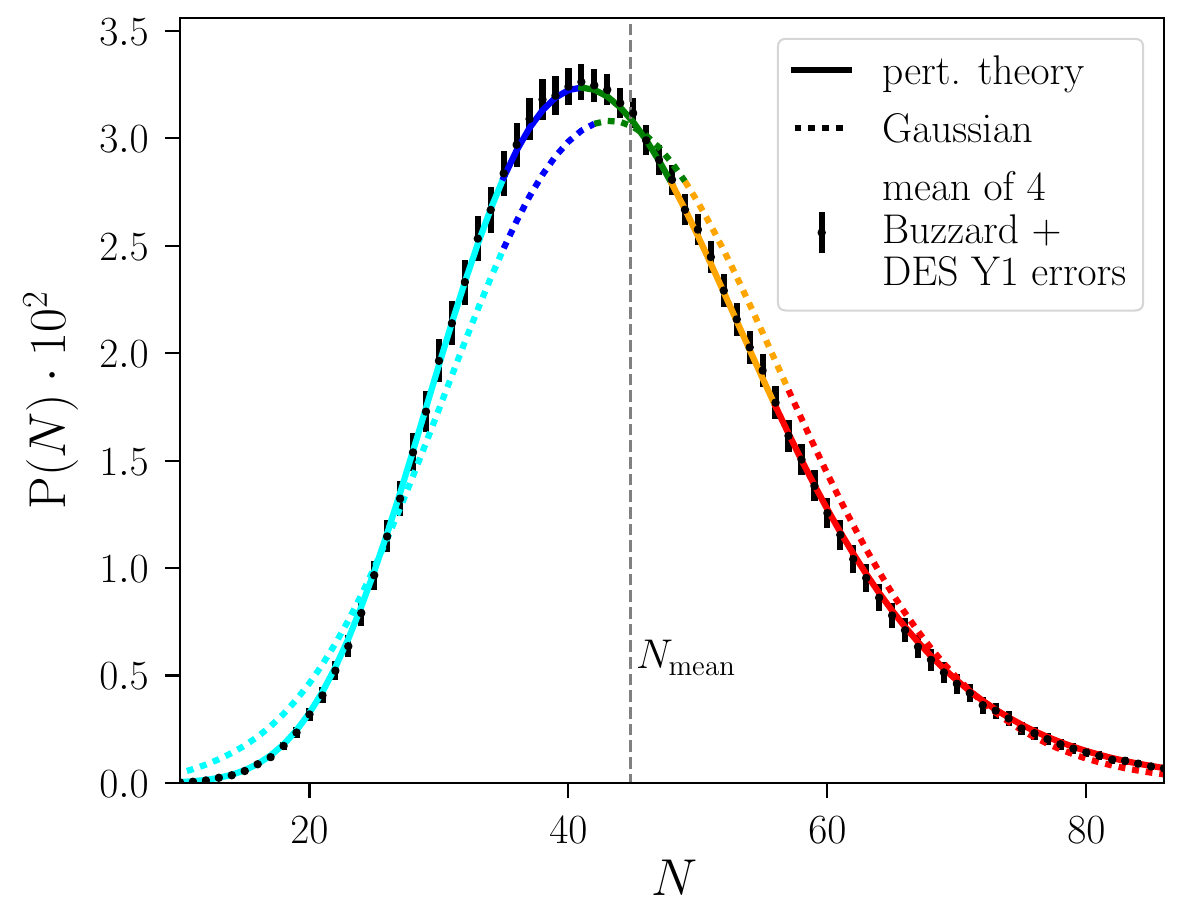}\includegraphics[width=0.555\textwidth]{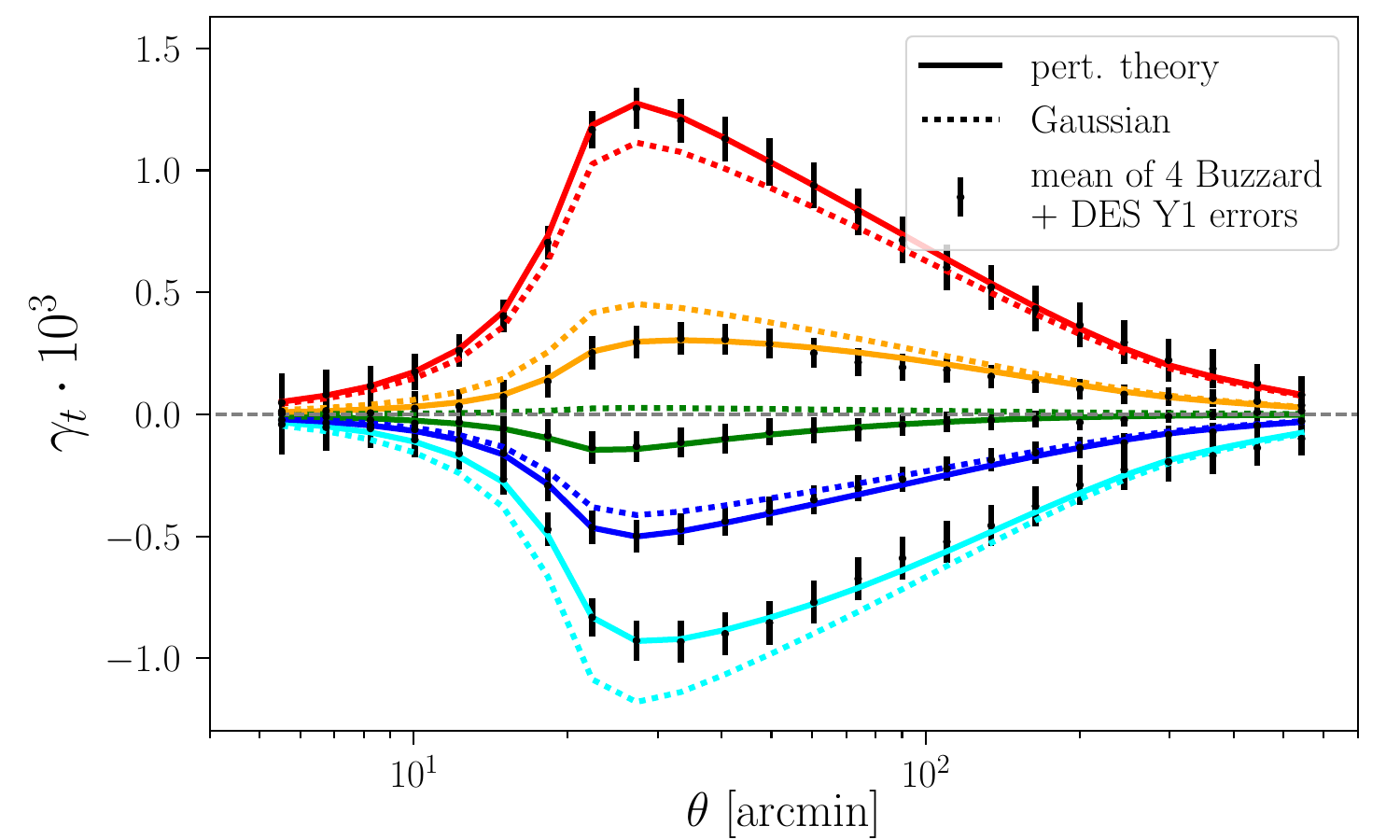}
  %\cprotect
  \caption{\underline{Top panel:} splitting the lines of sight in one DES-Y1 like Buzzard simulation into 5 quantiles of galaxy density (color coding from cyan, most underdense, to red, most overdense). The map uses a 20 arcmin \bblue{top-hat} radius and \textsc{redMaGiC} galaxies with a redshift range of $0.2\lesssim z \lesssim 0.45$. \underline{Bottom left:} histogram of \textsc{redMaGiC} galaxy counts in 20 arcmin radii (counts-in-cells). We show the mean histogram from 4 Buzzard realisations of DES-Y1 (black points), our model based on perturbation theory and cylindrical collapse (solid line) and a model that assumes the projected density contrast to be a Gaussian random field (dotted line). The color coding corresponds exactly to the density quantiles in the top panel. \underline{Bottom right:} Lensing signals around random points split by the density quantile in which these points are located. We show the mean measurement from 4 Buzzard realisations (black points), our perturbation theory model (solid line) and a model that assumes projected density contrast and lensing convergence to be joint Gaussian random variables (dotted line). Color coding is the same as in the other panels. The asymmetry between the lensing signals around the most underdense and most overdense lines-of-sight indicates the skewness of the cosmic density PDF.%\DG{ADD VERTICAL LINE TO BOTTOM RIGHT AT 20'}
  }
  
  \label{fi:quantile_illustration}
\end{figure*}

\begin{figure}  
  \includegraphics[width=0.5\textwidth]{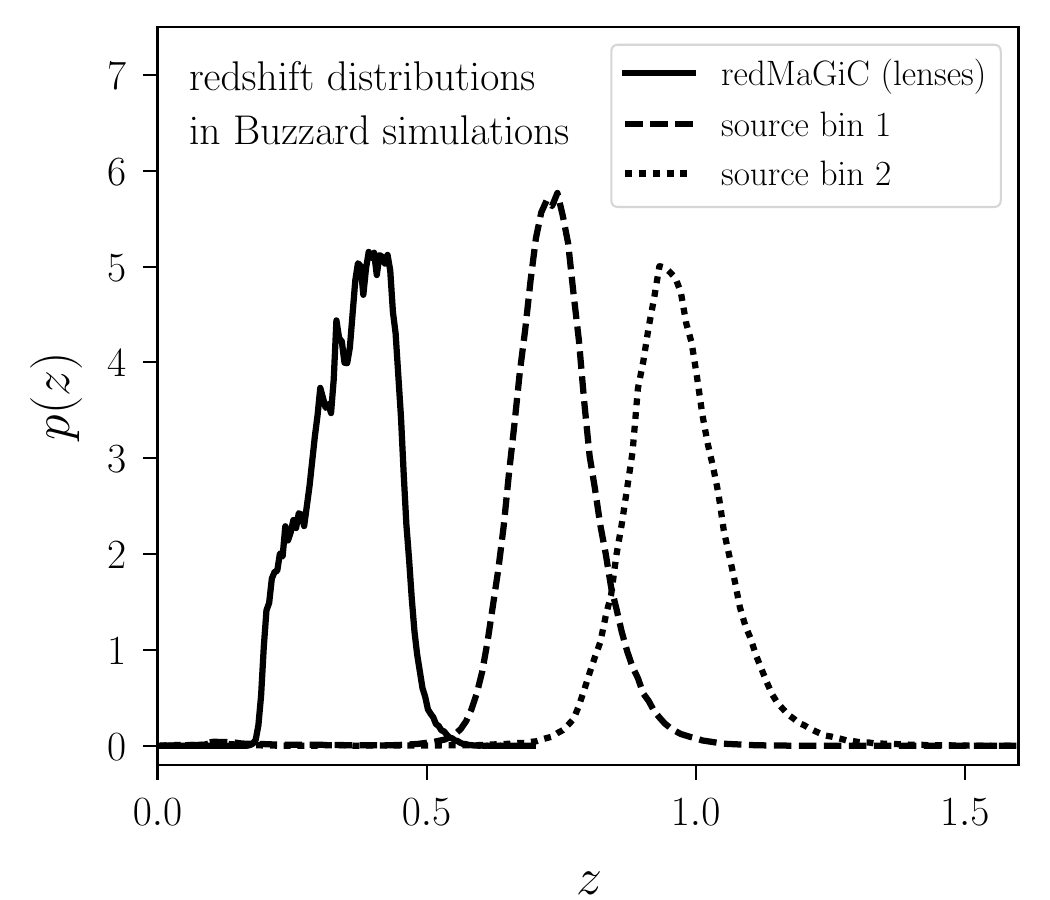}
  %\cprotect
  \caption{\bblue{Redshift distributions of} the tracer galaxy sample and the source samples of our N-body realizations of DES-Y1.}
  \label{fi:pofz}
\end{figure}

\subsection{Measuring density split statistics}
\label{sec:measuring_DT}

Density split lensing is a generalization of trough lensing \citep{Gruen2016} and can be described in three steps:
\\

\noindent \emph{1.) Splitting the sky into quantiles of different foreground galaxy density}
\\

\noindent Consider a sample of low-redshift galaxies that are tracing the line-of-sight density of matter with some redshift distribution $n_{l}(z)$. We will call these galaxies the foreground sample. For an angular radius $\theta_T$, which we will call the \emph{top-hat aperture radius}, we define $N_T(\mathbf{\hat n})$ to be the number of galaxies found within a radius $\theta_T$ around the point on the sky specified by the vector $\mathbf{\hat n}$ on the unit sphere. The field $N_T(\mathbf{\hat n})$ can be used to divide the sky into regions of different galaxy density. \citet{Gruen2016} have done this by discretizing the sky with a \textsc{healpix}\footnote{See \citet{Gorski2005} for details on \textsc{healpix}.} grid and sorting the pixels according to their value of $N_T(\mathbf{\hat n})$. Then they considered the $20\%$ of the pixels with the lowest values of $N_T$, calling them troughs. In the limit of a fine pixelization these pixels can be considered the most underdense quintile of the sky area. This can be generalized to the second most underdense quintile, the third most underdense quintile etc. or even to finer splits using more then just 5 quantiles. 

We stick to a division into 5 quantiles (quintiles) throughout this paper. The upper panel of figure \ref{fi:quantile_illustration} illustrates such a subdivision on a patch of a simulated sky (from the \emph{Buzzard flock}, see section \ref{sec:simulations} and especially \citet{DeRose2017} for details). There we use a top-hat aperture radius $\theta_T = 20'$ and the tracer galaxies have the redshift distribution that is displayed by the solid line in Figure \ref{fi:pofz}. Figure \ref{fi:quantile_illustration} shows the most underdense quintile of the simulated patch in cyan, the most overdense quintile in red and the three intermediate quintiles in blue, green and orange.

\blue{Note that the sum of the 5 lensing signals will vanish on average (we subtracted $1/5$ times the shear around random points from each signal, but due to boundary effects their sum will not vanish exactly). This means that roughly 4 of the 5 signals contain independent information. We have not investigated, whether our choice of 5 quantiles is in any way optimal. Choosing 3 quantiles would leave us with 2 independent signals and would hence suffice to be sensitive to both the variance and skewness of the density field. 5 quantiles enable a sentitivity beyond the 3rd moment of the density field. And they also allow us to explicitly show, that the median universe is underdense (which we could not do with 4 quantiles). In section \ref{sec:model} we investigate different radii of our top-hat aperture and find that $\theta_T = 20'$ is the smallest radius at which our model is reliable (given the redshift distribution we use in \citet{Gruen2017}).}
\\

\noindent \emph{2.) Tracing the mean dark matter density in each sky quintile with gravitational lensing}
\\

\noindent Now consider a second sample of galaxies at higher redshifts than the foreground sample (the \emph{source sample}, see e.g. the dashed and dotted redshift distributions in Figure \ref{fi:pofz}). As the light of these galaxies passes the large-scale structure of the foreground density distribution it undergoes gravitational lensing effects such as gravitational shear \citep[see e.g.][]{BartelmannSchneider2001}. The density split lensing signal around each quintile of the sky is obtained by measuring the stacked radial profile of tangential shear around random points located within that quintile. These points are constrained to lie within the part of the sky covered by a certain quintile of galaxy density but are otherwise random in their location. 
%Usually, when stacking the shear signal around random points across the sky, one expects the signal to vanish. But when 
Because these random points are split according to the density quintile they are located in, their stacked shear signals trace the average profile of density contrast around each quintile.

In the lower right panel of Figure \ref{fi:quantile_illustration} we show the signals measured for each density quintile in our mock data. The points show the average measurement from 4 Buzzard realisations of DES year-1 data (using the highest redshift source population shown in Figure \ref{fi:pofz}) and the solid lines show predictions by the model presented in this paper. The error bars are derived from a set of log-normal realisations (using the \verb|FLASK| tool by \citet{Xavier2016}; in \citet{Gruen2017} we describe in detail how we configured \verb|FLASK| to generate our mock catalogs). Two main features of the density split lensing signals are apparent: first, the amplitude of the radial shear around the $20\%$ most underdense pixels is lower than the amplitude of the tangential shear around the $20\%$ most overdense pixels. This is reflecting the skewness of the cosmic density PDF. Secondly, the signal around points in the third quintile is still significantly negative, which reflects the fact that the median universe is underdense. A more subtle feature is the fact that the underdense signals fall off less rapidly with increasing scale than the overdense signals. This is because on large scales the density field becomes Gaussian and hence recovers its initial symmetry between overdensities and underdensities.
\\

\noindent \emph{3.) Measuring the average counts-in-cells in each density quintile to obtain additional information on galaxy bias and stochasticity}
\\

\noindent If galaxy counts and the matter density field were perfectly correlated, then a split of the sky by galaxy density would be identical to a split by matter density. Hence, in this limit the density split lensing signals would be independent of the bias of the tracer galaxies. In a realistic scenario however, shot-noise of the galaxies smears out our attempts to divide the sky into areas of different matter density. Hence the density split lensing signals obtain a dependence on galaxy bias, but also on galaxy stochasticity. Increasing the linear bias of galaxy clustering will sharpen the tracers' ability to distinguish between overdensities and underdensities. Thus, increasing this bias will increase the amplitudes of the signals. This means that linear bias is to some degree degenerate with the amplitude of density fluctuations, $\sigma_8$. But $\sigma_8$ and bias influence the third moments of the density field in different ways and their degeneracy is not complete. As a consequence, it is possible to obtain constraints on cosmological parameters from the lensing signals alone (cf. section \ref{sec:using_DT} and the blue contour in the left panel of figure \ref{fi:forecasts}).

But additional information on bias and stochasticity nevertheless helps to tighten these constraints. In this paper we decided to add that information in the form of normalized quantiles of the counts-in-cells (CiC) histogram of the tracer galaxies: we measure the histogram of tracer counts within the same aperture that was used to identify our density quintiles. Then we identify the parts of this histogram that correspond to these quintiles (cf. lower left panel of Figure \ref{fi:quantile_illustration}). For each quintile $q$ we then determine the mean galaxy count in that quintile, $N_q$, and normalize it by the overall mean galaxy count in our aperture, $\bar N$. i.e. for each quintile we add $N_q/\bar N$ to our data vector. This indeed helps to tighten constraints on cosmological parameters (cf. section \ref{sec:using_DT} and the green contour in the left panel of figure \ref{fi:forecasts}).

\subsection{Modelling density split statistics}
\label{sec:modeling_DT}

We now outline a general framework for modeling the data vector described above, leaving details of this framework to section \ref{sec:model}. Unless stated differently, we will assume a flat $\Lambda$CDM universe throughout this paper.

Let us start by introducing the quantities whose relations need to be modelled. First, we denote with $\delta_{m,\mathrm{2D}}$ the line-of-sight projection of the 3D density contrast according to the redshift distribution $n_l(z)$ of our foreground galaxy sample, i.e.
\begin{equation}
\delta_{m,\mathrm{2D}}(\hat{\mathbf{n}}) = \int \mathrm{d}w\ q_l(w)\ \delta_{m,\mathrm{3D}}(w\hat{\mathbf{n}}, w)
\end{equation}
where $\hat{\mathbf{n}}$ denotes a unit vector on the sky, $w$ is co-moving distance and the projection kernel $q_l(w)$ is given in terms of $n_l(z)$ as
\begin{equation}
\label{eq:definition_of_q_l}
q_l(w) = n_l(z[w]) \frac{\mathrm d z[w]}{\mathrm d w}\ .
\end{equation}
We furthermore define $\delta_{m, T}$ to be the average of $\delta_{m,\mathrm{2D}}$ over \bblue{top-hat filters with aperture radius} $\theta_T$, i.e.
\begin{equation}
\delta_{m, T}(\hat{\mathbf{n}}) = \underset{|\hat{\mathbf{n}} , \hat{\mathbf{n}}'| < \theta_T}{\int} \mathrm{d}\Omega'\ \frac{\delta_{m,\mathrm{2D}}(\hat{\mathbf{n}}')}{2\pi(1 - \cos \theta_T)}\ .
\end{equation}
Here $|\cdot , \cdot |$ denotes the angular distance between two points on the sky.

We identify regions of different density by means of our foreground galaxy sample. When smoothed \bblue{with a top-hat filter of} radius $\theta_T$, these galaxies are biased and possibly stochastic tracers of $\delta_{m, T}(\hat{\mathbf{n}})$. Hence our model also needs to include a description of how $N_T(\hat{\mathbf{n}})$, the number of tracer galaxies found within an angular radius $\theta_T$ around the line-of-sight $\hat{\mathbf{n}}$, relates to $\delta_{m, T}(\hat{\mathbf{n}})$.

Finally, in order to describe the density split lensing signal, we need to consider the lensing convergence field for our population of source galaxies. Given the source redshift distribution $n_s(z)$, the convergence $\kappa$ is given by the line-of-sight projection
\begin{equation}
\kappa(\hat{\mathbf{n}}) = \int \mathrm{d}w\ W_{s}(w)\ \delta_{m,\mathrm{3D}}(w\hat{\mathbf{n}}, w)\ ,
\end{equation}
where $W_{s}$ is the lensing efficiency, which is defined by
\begin{equation}
\label{eq:lensing_efficiency}
W_{s}(w) = \frac{3 \Omega_m H_0^2}{2c^2} \int_w^\infty \mathrm d w'\ \frac{w (w'-w)}{w'\ a(w)}\ q_s(w')\ ,
\end{equation}
and $\smash{q_s(w) = n_s(z[w]) \frac{\mathrm d z[w]}{\mathrm d w}}$ is the line-of-sight density of the sources. Smoothing the convergence field with a circular aperture of radius $\theta$ results in a field which we will denote by $\kappa_{<\theta}(\hat{\mathbf{n}})$.

Because of the isotropy of the universe, we will now omit the the dependence of the above quantities on $\hat{\mathbf{n}}$. To model the density split lensing signal one needs to answer the following questions:
\begin{trivlist}
\item[$\bullet$] Given the number of galaxies $N_T$ found around a line-of-sight $\hat{\mathbf{n}}$, what distribution can be inferred for the matter density contrast $\delta_{m, T}$ in that line-of-sight? i.e. what is the expectation value $\langle \delta_{m, T} | N_T\rangle$?
\\

\item[$\bullet$] Given the matter density contrast $\delta_{m,T}$ in the line-of-sight $\hat{\mathbf{n}}$, what lensing convergence $\kappa_{<\theta}$ is expected inside an angular distance $\theta$ from that line-of-sight? i.e. what is the expectation value $\langle \kappa_{<\theta} | \delta_{m, T}\rangle$? The tangential shear profile around that line-of-sight can then be inferred from the convergence profile as
\begin{align}
\label{eq:gamma_in_terms_of_kappa}
\langle \gamma_t(\theta)| \delta_{m, T}\rangle &= \langle \kappa_{<\theta} | \delta_{m, T}\rangle - \langle \kappa_{\theta} | \delta_{m, T}\rangle\nonumber \\
&= \frac{\cos\theta - 1}{\sin \theta} \frac{\mathrm d}{\mathrm d \theta} \langle \kappa_{<\theta} | \delta_{m, T}\rangle\ ,
\end{align}
where $\kappa_{\theta}$ is the average convergence at the radius $\theta$.
\end{trivlist}

\noindent The first of the above questions can be answered in the form of a conditional PDF of $\delta_{m,T}$ given a certain value of $N_T$, i.e. $\smash{p(\delta_{m,T}|N_T)}$. Using Bayes' theorem this can be written as
\begin{equation}
\label{eq:Bayes}
p(\delta_{m,T}|N_T) = \frac{P(N_T| \delta_{m,T})\ p(\delta_{m,T})}{P(N_T)}\ ,
\end{equation}
where $P(N_T| \delta_{m,T})$ is the probability of finding a number of galaxies $N_T$ given that the density contrast is $\delta_{m,T}$ and where $p(\delta_{m,T})$ and $P(N_T)$ are the total PDF of $\delta_{m,T}$ and the total probability of finding $N_T$ tracer galaxies. The average convergence profile around a circle with $N_T$ galaxies is then given by
\begin{align}
\label{eq:trough_radius_approx}
\langle\kappa_{<\theta} | N_T\rangle & = \int \mathrm d \delta_{m,T}\  \langle\kappa_{<\theta} | \delta_{m,T}, N_T\rangle\ p(\delta_{m,T}|N_T)\nonumber \\
& \approx \int \mathrm d \delta_{m,T}\  \langle\kappa_{<\theta} | \delta_{m,T}\rangle\ p(\delta_{m,T}|N_T)\ ,
\end{align}
where in the second step we have assumed that the expected convergence within $\theta$ only depends on the total matter density contrast within $\theta_T$.

We now divide the sky into different quintiles of tracer galaxy density. Let us denote with $Q[0.0, 0.2]$ the 20\% of the lines-of-sight on the sky that have the lowest value of $N_T$. There will be a maximal value $N_T \leq N_{\max}$ in this quintile and the stacked convergence profile around these lines-of-sight is given by
\begin{align}
\langle\kappa_{<\theta} | Q[0.0, 0.2]\rangle = \ \ \ \ \ \ \ \ \ \ \ \ \ \ \ \ \ \ \ \ \ \ \ \ \nonumber \\
\frac{1}{0.2} \left(\sum_{N < N_{\max}} P(N) \langle\kappa_{<\theta} | N_T = N \rangle + \alpha\ \langle\kappa_{<\theta} | N_T = N_{\max}\rangle\right)\ .
\end{align}
Here the factor $\alpha$ in the second term accounts for the fact that the lines-of-sight with exactly $N_T = N_{\max}$ might have to be split between the quintile $Q[0.0, 0.2]$ and the quintile $Q[0.2, 0.4]$. It is given by
\begin{equation}
\alpha =  0.2\ - \sum_{N < N_{\max}} P(N)\ .
\end{equation}
This can be easily generalized to the other quintiles $Q[q_{\min}, q_{\max}]$ and also to the case of dividing the sky into more than 5 density regimes. Finally, we also add the average of the counts-in-cells in each quintile normalized by the mean galaxy count $\bar N$ to our data vector. For the quintile $Q[0.0, 0.2]$ this is given by
\begin{align}
\frac{\langle N_T | Q[0.0, 0.2]\rangle}{\bar N} = \frac{1}{0.2 \bar N} \left(\sum_{N < N_{\max}} P(N)\ N + \alpha\ N_{\max}\right)\ ,
\end{align}
which is also straightforward to generalize to other quantiles $Q[q_{\min}, q_{\max}]$.

The probabilities $P(N)$ can be computed from the normalization of equation \ref{eq:Bayes}. Hence, our model for the density split lensing signal needs the following three ingredients:
\begin{trivlist}
\item[(i)] The PDF of matter density contrast, smoothed \bblue{with a top-hat filter} of radius $\theta_T$,
\begin{equation}
p(\delta_{m,T})\ .
\end{equation}
\item[(ii)] The expectation value of convergence inside a distance $\theta$ given the density contrast inside $\theta_T$,
\begin{equation}
\langle\kappa_{<\theta} | \delta_{m,T}\rangle\ .
\end{equation}
\item[(iii)] The distribution of galaxy counts inside the \bblue{top-hat} radius $\theta_T$ given the density contrast within that radius, 
\begin{equation}
P(N_T | \delta_{m,T})\ .
\end{equation}
\end{trivlist}
\citet{Gruen2016} assumed $\delta_{m,T}$ and $\kappa_{<\theta}$ to have a joint Gaussian distribution. This allowed them to compute (i) and (ii) solely from the dark matter clustering power spectrum. To compute (iii) they assumed a linear galaxy bias and Poissonian shot-noise of the tracer galaxies. These assumptions allowed a sufficient model for their measurements made on DES Science Verification data. But as can be seen from the dotted lines in the lower panels of Figure \ref{fi:quantile_illustration}, \bblue{a} Gaussian model for the density PDF is not sufficient within \bblue{the much smaller uncertainty} of DES-Y1. Also, in section \ref{sec:model} we demonstrate that the shot-noise of the tracer galaxies can not necessarily be assumed \bblue{to be} Poissonian. \orange{In this work we hence want to revise their model.}

\orange{For each of the model components (i) and (ii) we investigate two different modeling approaches - a baseline approach and an approach with increased complexity used to check the validity of the baseline model. In the following we briefly outline each approach. The reader interested in details of each modeling ansatz is referred to section \ref{sec:model}. Readers who are not interested in this technical part of the paper should feel free to skip section \ref{sec:model}.}

\begin{trivlist}
\item[(i)]\orange{\underline{Baseline model for $p(\delta_{m,T})$:}}

\noindent In our fiducial model we assume $\delta_{m,T}$ to be a log-normal random field \citep{Hilbert2011}. The PDF of such a variable can e.g. be fixed by specifying the variance $\langle \delta_{m, T}^2 \rangle$ and skewness $\langle \delta_{m, T}^3 \rangle$. We predict the variance of $\delta_{m, T}$ from the the non-linear matter power spectrum \citep[cf.][]{Gruen2016}. The latter is computed using \emph{halofit} \citep{Smith2003,Takahashi2012} and an analytic transfer function \citep{Eisenstein1998}. We then use leading order perturbation theory to compute a scaling relation between the bispectrum and the power spectrum of the density field. Together with our power spectrum this fixes the skewness of $\delta_{m, T}$.
\\

\noindent \underline{Alternative model for $p(\delta_{m,T})$:}

\noindent As an alternative we compute the PDF $p(\delta_{m,T})$ from its cumulant generating function \blue{(see section \ref{sec:model} for a definition and further details)}. To model this function we use a cylindrical collapse approach based on the work of \citet{Bernardeau1994}, \citet{Bernardeau2000} and \citet{Valageas2002}.
\\

\item[(ii)]\underline{Baseline model for $\langle\kappa_{<\theta} | \delta_{m,T}\rangle$:}

\noindent In our fiducial model we assume that $\kappa_{<\theta}$ can be expressed as the sum of two random variables,
\begin{equation}
\kappa_{<\theta} = \kappa_{<\theta, \mathrm{corr.}} + \kappa_{<\theta, \mathrm{uncorr.}}\ ,
\end{equation}
where $\kappa_{<\theta, \mathrm{uncorr.}}$ is assumed to be completely uncorrelated to $\delta_{m,T}$ and hence doesn't contribute to the density split lensing signal. As a result we have
\begin{equation}
\langle\kappa_{<\theta} | \delta_{m,T}\rangle \equiv \langle\kappa_{<\theta, \mathrm{corr.}} | \delta_{m,T}\rangle\ .
\end{equation}
We assume a joint log-normal PDF for the two random variables $\delta_{m,T}$ and $\kappa_{<\theta, \mathrm{corr.}}$. The expectation value $\langle\kappa_{<\theta, \mathrm{corr.}} | \delta_{m,T}\rangle$ is then fixed by specifying the moments 
\begin{equation}
\langle \delta_{m, T}^2 \rangle\ ,\ \langle \delta_{m, T}^3 \rangle
\end{equation}
as well as
\begin{equation}
\langle \kappa_{<\theta} \delta_{m, T} \rangle \equiv \langle \kappa_{<\theta, \mathrm{corr.}} \delta_{m, T} \rangle
\end{equation}
and
\begin{equation}
\langle \kappa_{<\theta} \delta_{m, T}^2 \rangle \equiv \langle \kappa_{<\theta, \mathrm{corr.}} \delta_{m, T}^2 \rangle\ .
\end{equation}
Second order moments are again computed from our non-linear power spectrum while third order moments are inferred from perturbation theory. (The introduction of $\kappa_{<\theta, \mathrm{uncorr.}}$ is only necessary for consistency reasons: a joint log-normal PDF of $\delta_{m,T}$ and $\kappa_{<\theta}$ characterized by the above moments would be inconsistent with the variance $\langle \kappa_{<\theta}^2 \rangle$ derived from our power spectrum.)
\\

\noindent \underline{Alternative model for $\langle\kappa_{<\theta} | \delta_{m,T}\rangle$:}

\noindent As an alternative we compute $\langle\kappa_{<\theta} | \delta_{m,T}\rangle$ from the joint cumulant generating function of the variables $\delta_{m,T}$ and $\kappa_{<\theta}$. This function can also be modelled in a cylindrical collapse approach.
\end{trivlist}

\noindent For model component (iii) we also investigate two different modeling approaches - one ansatz introducing 2 free parameters and one ansatz introducing 3 free parameters. We find that our simulated tracer catalogs are well described by the 2-parametric model. But - anticipating real galaxies to behave more complicated than simulated ones - we do not consider either of these models as our baseline model and instead apply both approaches to DES data in \citet{Gruen2017}. We summarize both ansatzes here, but the interested reader is again referred to section \ref{sec:model} for details of each ansatz.

\begin{trivlist}
\item[(iii)]\orange{\underline{Model 1 for $P(N_T | \delta_{m,T})$:}}

\noindent \orange{In our fiducial model we introduce an auxilliary field $\delta_{g,T}$ such that our foreground galaxies are Poissonian tracers of that field, i.e.
\begin{equation}
P(N_T | \delta_{g,T}) = \frac{\left[\bar N (1+\delta_{g,T})\right]^{N_T}}{N_T!}\ e^{-\bar N (1+\delta_{g,T})}\ .
\end{equation}
$\delta_{g,T}$ can be thought of as a smooth (shot noise free) galaxy density contrast. We then assume that $\delta_{g,T}$ and $\delta_{m,T}$ are joint log-normal random variables with
\begin{equation}
\label{eq:definition_b}
\langle \delta_{g, T}^2 \rangle = b^2 \langle \delta_{m, T}^2 \rangle\ ,\ \langle \delta_{g, T}^3 \rangle = b^3 \langle \delta_{m, T}^3 \rangle
\end{equation}
and
\begin{equation}
\label{eq:definition_r}
\langle \delta_{g, T} \delta_{m, T}  \rangle = b r \langle \delta_{m, T}^2 \rangle\ .
\end{equation}
The parameters $b$ and $r$ will be called \emph{galaxy bias} and \emph{galaxy stochasticity} and are free parameters of the model.
}
\\

\noindent \orange{\underline{Model 2 for $P(N_T | \delta_{m,T})$:}}

\noindent \orange{As an alternative we assume $P(N_T | \delta_{m,T})$ to be a generalization of the Poisson distribution, that allows for
\begin{equation}
\langle N_T^2 | \delta_{m,T}  \rangle \neq \langle N_T | \delta_{m,T}  \rangle + \langle N_T | \delta_{m,T}  \rangle^2\ ,
\end{equation}
i.e. for a shot-noise that is either enhanced or suppressed wrt. the Poisson case. The enhancement of shot-noise is also allowed to be a function of $\delta_{m,T}$ of (approximately) the form
\begin{equation}
\frac{\langle N_T^2 | \delta_{m,T}  \rangle - \langle N_T | \delta_{m,T}  \rangle^2}{\langle N_T | \delta_{m,T}\rangle} \approx \alpha_0 + \alpha_1 \delta_{m,T}\ .
\end{equation}
This model introduces an alternative bias parameter $\tilde b$ such that
\begin{equation}
\langle N_T | \delta_{m,T}\rangle = \bar N [1 + \tilde b \delta_{m,T}]\ .
\end{equation}
}
\\
\end{trivlist}
\orange{For the model components (i) and (ii) and on the scales considered in this paper, the baseline and alternative modeling approaches yield almost indistinguishable predictions (cf. section \ref{sec:model}). For component (iii) the modeling ansatzes 1 and 2 are not necessarily identical, because they introduce a different number of degrees of freedom. Figure \ref{fi:quantile_illustration} as well as all parameter contours shown in this paper are using the baseline model for components (i) and (ii) and ansatz 1 for component (iii). The predictions derived from different modelling approaches are however almost indistinguishable (cf.$\ $figure \ref{fi:p_of_delta}).}

\begin{figure*}  
  \includegraphics[width=0.5\textwidth]{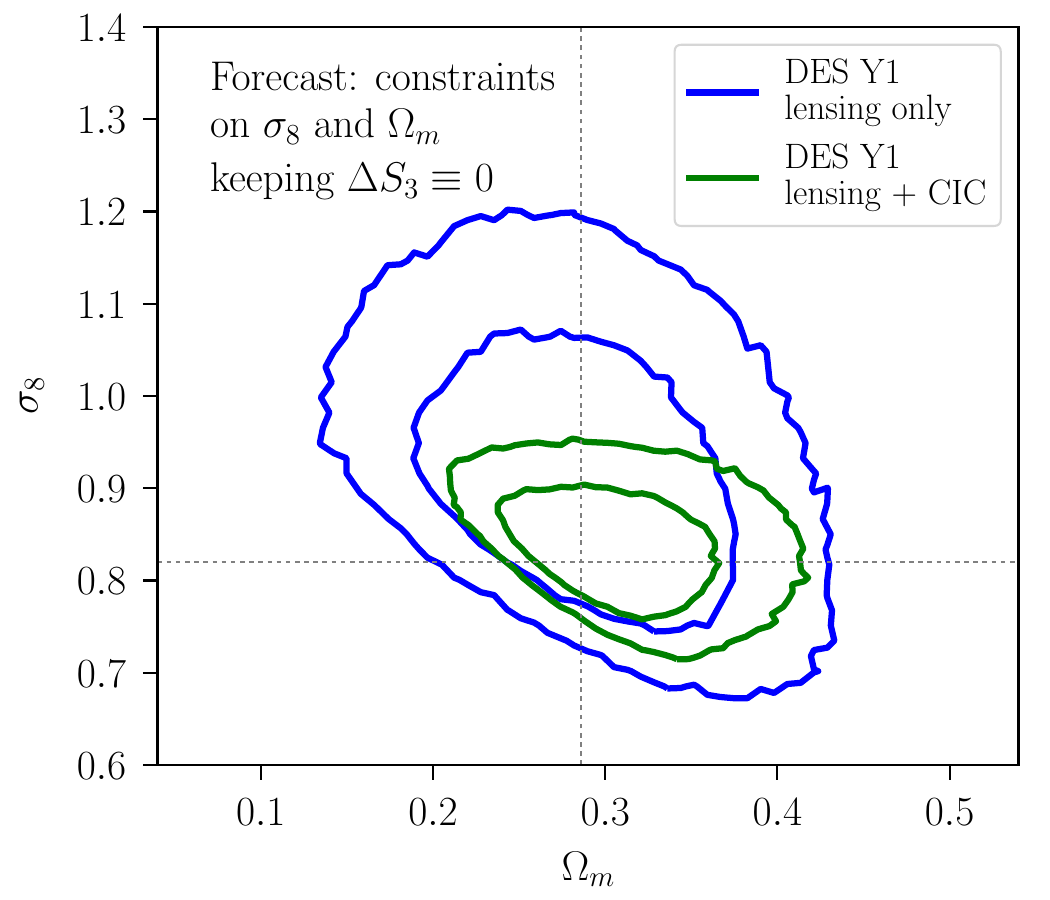}\includegraphics[width=0.5\textwidth]{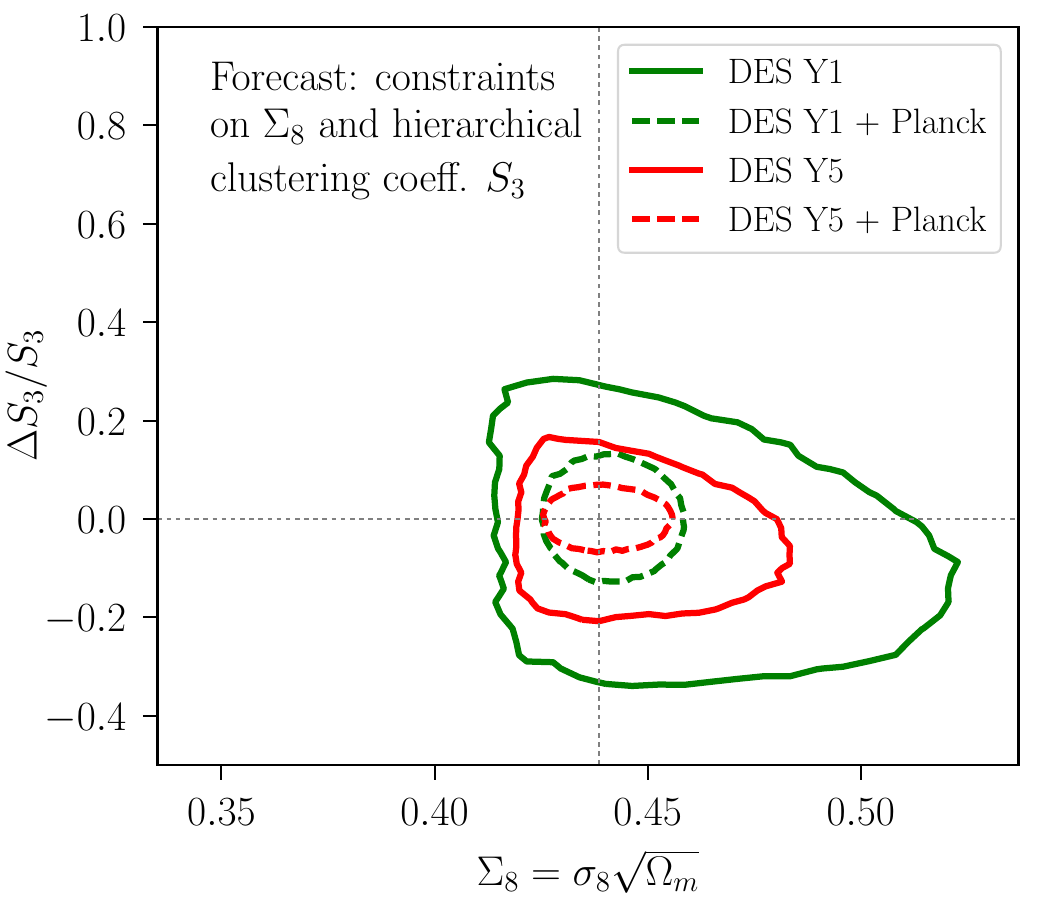}
  %\cprotect
  \caption{\underline{Left panel:} Forecast of $1\sigma$ and $2\sigma$ constraints on $\Omega_m$ and $\sigma_8$ achievable with density split lensing and counts-in-cells in DES Y1 data. The constraints are marginalized over $\Omega_b$, $n_{s}$, $h_{100}$, \textsc{redMaGiC} galaxy bias $b$ and galaxy-matter correlation coefficient $r$. For the parameters $\Omega_b$, $n_{s}$, $h_{100}$ we have assumed the same flat priors as used in the DES Y1 combined probes analysis presented in \citet{DES2017_short}. \underline{Right panel:} \blue{$\Delta S_3 / S_3$ measures relative deviations from our fiducial perturbation theory prediction of the scaling coefficient \smash{$S_3\equiv\frac{\langle\delta^3\rangle}{\langle\delta^2\rangle^2}$} (cf$.$ equation \ref{eq:definition_S3} and section \ref{sec:model} for details). It can hence be thought of as the Bispectrum amplitude. We show $1\sigma$ constraints on this parameter} achievable with density split lensing and counts-in-cells in DES data alone (solid lines) and using additional information on cosmology from Planck (dashed lines, no lensing). The sharp \bblue{cut-off} of the contours for low values of $\Sigma_8$ is caused by the requirement that matter density and a \bblue{shot-noise free galaxy density field} must have a correlation coefficient $r\leq 1$. All likelihoods are centered around our fiducial cosmology, i.e. the parameters describing the Buzzard simulations.}
  
  \label{fi:forecasts}
\end{figure*}

\begin{figure*}  
  \includegraphics[width=0.5\textwidth]{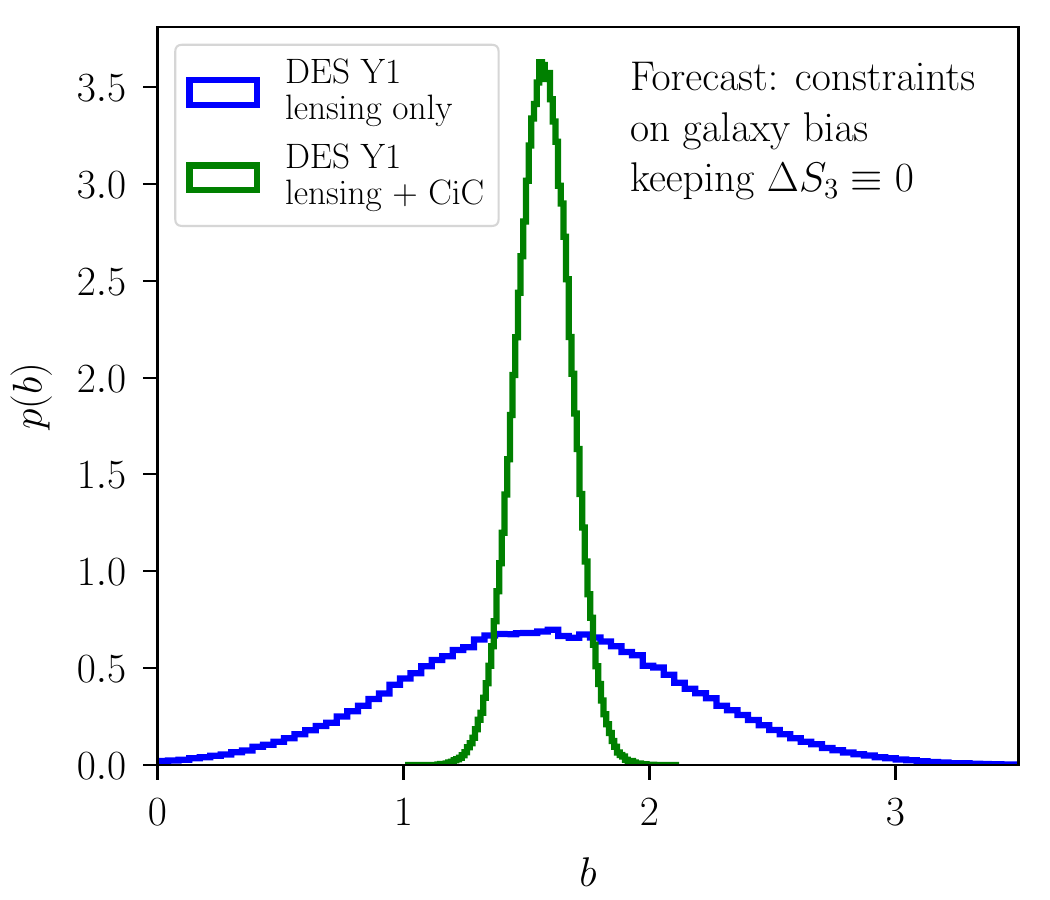}\includegraphics[width=0.5\textwidth]{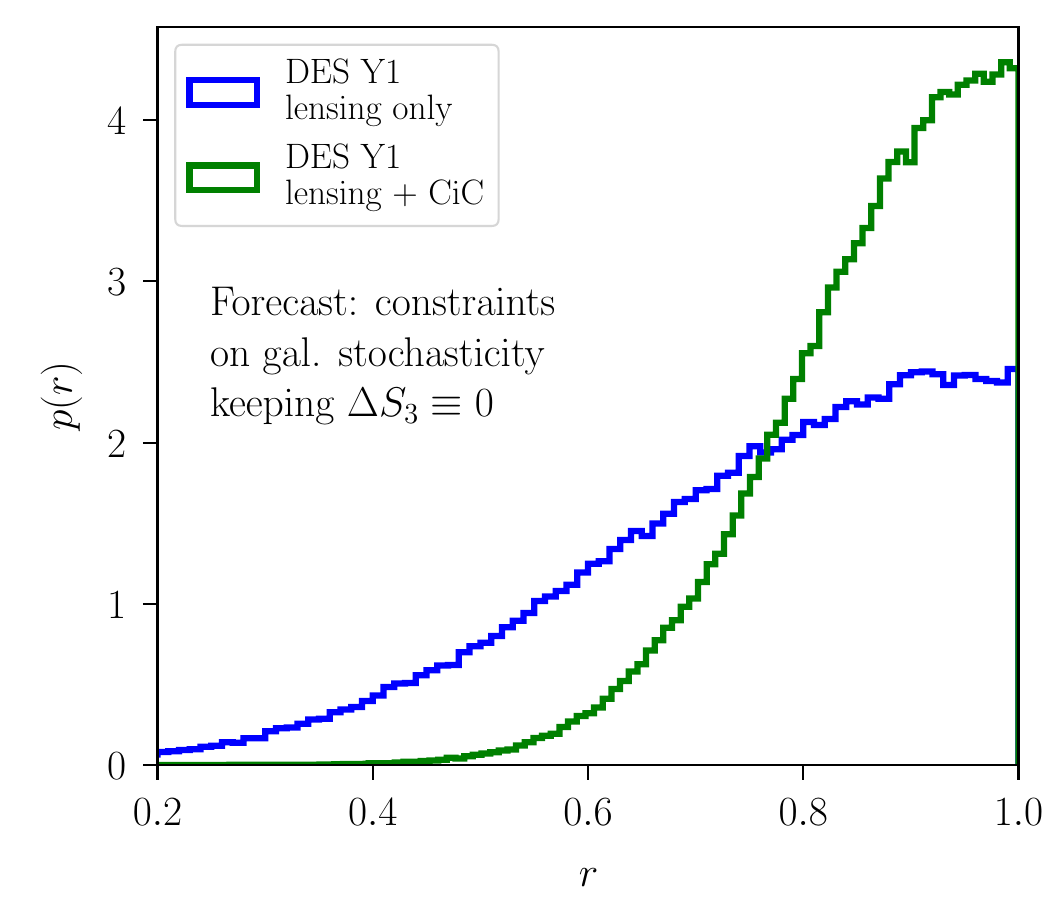}
  %\cprotect
  \caption{Forecast of posterior constraints on galaxy bias $b$ and galaxy-matter correlation coefficient $r$ achievable with density split lensing and counts-in-cells in DES Y1 data. The constraints are marginalized over $\Omega_m$, $\sigma_8$, $\Omega_b$, $n_{s}$ and $h_{100}$, where for the last three parameters we have assumed the same flat priors as used in the DES Y1 combined probes analysis presented in \citet{DES2017_short}.}
  
  \label{fi:forecasts}
\end{figure*}

\subsection{Data vector and forecasts on parameter constraints}
\label{sec:using_DT}

\subsubsection{Binning and scales}

Throughout this paper, we assume that one sample of tracer galaxies is used to identify density quintiles and that the lensing profiles around these quintiles are measured with two source redshift bins (cf. redshift distributions in Figure~\ref{fi:pofz}). To identify the different density quintiles, we use a top-hat filter with fiducial smoothing radius of $\theta_T = 20'$.

We measure the density split lensing signal in 24 log-spaced angular bins with
\begin{equation}
5' < \theta < 600'\ .
\end{equation}
But in the parameter forecasts and likelihood contours shown in the following, we exclude bins with scales $\leq\theta_T$. This reduces all lensing signals to 17 log-spaced angular bins with
\begin{equation}
20' \lesssim \theta < 600'\ .
\end{equation}
The scales with $\theta < \theta_T$ are excluded from our analysis for a number of reasons: first, the signal-to-noise ratio of our lensing profiles drops quickly below $\theta_T$ (cf. Figure \ref{fi:quantile_illustration}). Second, we chose our fiducial value of $\theta_T = 20'$ because we still trust the modeling of the density PDF described in~\autoref{sec:model} on this scale, and we do not want smaller angular scales to contribute to our fiducial data vector. Third, the approximation made in equation \ref{eq:trough_radius_approx} might fail at scales smaller than our aperture.

The sum of all 5 lensing signals will be very close to zero (though not exactly zero because of masking effects), so that they are not an independent set of observations. Hence, we only include the 2 most overdense and the 2 most underdense quintiles in our analysis, i.e. the cyan, blue, orange and red lines in Figure \ref{fi:quantile_illustration}. For the same reason, we also use the normalized mean galaxy count only in four of the five quintiles of the counts-in-cells histogram to complement the lensing measurement. The total number of data points in our data vector is thus
\begin{align}
N_{\mathrm{data}} &= N_{\mathrm{lens}} + N_{\mathrm{CiC}} \nonumber\\
&= [N_{\mathrm{quant.}}-1]\cdot N_{\mathrm{source}}\cdot N_{\mathrm{ang}} + [N_{\mathrm{quant.}}-1] \nonumber\\
&= 4\cdot 2\cdot 17 + 4 \nonumber\\
&= 140 \ .
\end{align}

\subsubsection{Model parameters and forecast on constraining power}
\label{sec:forecast_section}

\begin{table*}
\centering
\begin{tabular}{l|lllll}
\hline\hline
parameter  &  fiducial value & prior in & DES-Y1 constraints & DES-Y1 constraints & DES-Y1 constraints \\
 &   (in Buzzard) & likelihood analysis & without $\Delta S_3 / S_3$ & with $\Delta S_3 / S_3$ & without $\Delta S_3 / S_3$ \\
 &    &  &  &  & (lensing only) \\
\hline\hline
 $\sigma_8$ & $0.82$ & $[0.2,1.6]$ & $\pm0.05$ & $\pm0.10$ & $\pm0.11$\\
 $\Omega_m$  & $0.286$ & $[0.1,0.9]$ & $\pm0.04$ & $\pm0.06$ & $\pm0.06$\\
 $\Omega_b$  & $0.047$ & $[0.01,0.07]$ &  - & - & - \\
 $n_s$ & $0.96$ & $[0.7,1.3]$ & - & - \\
 $h_{100}$ & $0.7$ & $[0.55,0.91]$ & - & - & - \\
\hline
 $b$ & $1.618$ & $[0.8,2.5]$ & $\pm0.11$ & $\pm0.27$& $\pm0.57$ \\
  & & (\emph{lensing only:}$[0.0,4.5]$) \hspace{0.1cm} &  & \\
 $r$ & $0.956$ & $[0.0,1.0]$ & $\pm0.10$ & $\pm0.11$& $\pm0.18$ \\
\hline
 $\Delta S_3 / S_3$ & $0.0$ & $[-1.0,2.0]$ & - & $\pm 0.20$ & - \\
\hline\hline
\end{tabular}
\caption{Model parameters of the forecast described in section \ref{sec:forecast_section}. The second column shows our fiducial values (the values describing the Buzzard simulations). The third column shows the parameter priors used to cut our prediction for the posterior distribution of best-fit parameters. The priors on $\Omega_b$, $n_s$ and $h_{100}$ are informative and chosen to be the same as used by the \citet{DES2017_short}. The prior on $r$ is needed for mathematical consistency. And the other priors only have to be introduced formally since we are approximating our analytic posterior by an MCMC. The 4th column shows the standard deviation of each parameter (as computed from the MCMC) after marginalization over all other parameters. The 5th column shows the same standard deviations but for the case where also $\Delta S_3 / S_3$ is introduced as a free parameter of our model. In column 6, we again fix $\Delta S_3 / S_3$ but assume that only the lensing part of the data vector is used. These forecasts can be compared to the actual errors we find in \citet[][their tables 2 and 3]{Gruen2017} which are close despite marginalizing over systematic uncertainties.}
\label{table:params}
\end{table*}

We now investigate what constraints on model parameters can be expected from the above data vector when measured in DES-Y1 data. For this, we assume the optimistic case that component (iii) of our model is sufficiently described by two parameters, i.e. by modeling approach 1 in the previous section.\footnote{\citet{Gruen2017} will also consider modeling approach 2.} This means that our model is determined by the following 7 parameters:
\begin{trivlist}
\item[1.)] $\Omega_m$: present total matter density in units the critical density of the universe,
\item[2.)] $\sigma_8$: amplitude of present day density fluctuations in spheres of $8$Mpc$/h$ radius as predicted by the linear power spectrum,
\item[3.)] $\Omega_b$: present density of baryonic matter in units of the critical density of the universe,
\item[4.)] $n_s$: the spectral index of the primordial power spectrum,
\item[5.)] $h_{100}$: the present day Hubble parameter in units of $100$Mpc/(km/s),
\item[6.)] $b$: \bblue{linear} bias of the tracers w.r.t. matter density (cf. equation \ref{eq:definition_b}),
\item[7.)] $r$: correlation coefficient between $\delta_{m,T}$ and $\delta_{g,T}$ (cf. equation \ref{eq:definition_r}).
\end{trivlist}
We summarize these parameters and our fiducial values for them in \autoref{table:params}. Throughout this paper, we assume the universe to be flat.

If $\pi_\alpha$ and $\pi_\beta$ are any two of the above parameters then let $\hat \pi_{\alpha, \mathrm{ML}}$ and $\hat\pi_{\beta, \mathrm{ML}}$ be maximum likelihood estimates of these parameters based on a measurement of density split statistics. The covariance of $\hat \pi_{\alpha, \mathrm{ML}}$ and $\hat\pi_{\beta, \mathrm{ML}}$ can be estimated by
\begin{equation}
\label{eq:param_cov}
\mathrm{Cov}[\hat\pi_{\alpha, \mathrm{ML}}, \hat\pi_{\beta, \mathrm{ML}}]^{-1} \approx \frac{\partial \mathbf{d}_{\mathrm{th}}^T}{\partial \pi_\alpha}\cdot \mathbf{C}_d^{-1}\cdot \frac{\partial \mathbf{d}_{\mathrm{th}}}{\partial \pi_\beta}\ ,
\end{equation}
where $\mathbf{d}_{\mathrm{th}}$ is our model of the density split data vector and $\mathbf{C}_d$ is the covariance matrix of a measurement of this signal in DES-Y1. We will in the following use an estimate of $\mathbf{C}_d$ from log-normal mocks and real DES Y1 shape noise \blue{(see section \ref{sec:FLASK} for a brief summary of our covariance estimation and \citealt{Gruen2017} for details)}. The parameter covariance computed with equation \ref{eq:param_cov} can then be used to approximate the expected distribution of our best-fit parameter estimates with a multivariate Gaussian distribution.

Since the three parameters $\Omega_b$, $n_s$ and $h_{100}$ are only poorly constrained by our data vector we are forced to assume prior knowledge on them. To do so, we cut the Gaussian posterior predicted from the parameter covariance with flat informative priors. These priors are chosen to be the same used by the \citet{DES2017_short}. For reasons of mathematical consistency we furthermore have to demand that $r \in [0,1]$. These hard cuts of our originally Gaussian approximation to the posterior distribution of best-fit parameters make it difficult to marginalize over individual parameters. We hence numerically evaluate our analytic posterior with a Monte-Carlo Markov-Chain (MCMC). This chain is used in the following visualizations. Since we are using an MCMC to trace our analytic posterior, we have to formally define priors for all other model parameters. These are chosen to be flat and to extend well beyond the single-parameter standard deviations of the posterior. In the third column of table \ref{table:params}, we summarize the priors chosen for each model parameter.

The left panel of Figure \ref{fi:forecasts} shows the 1$\sigma$ and 2$\sigma$ constraints achievable in the $\Omega_m$-$\sigma_8$ plane. These contours are marginalized over the other model parameters, using the priors mentioned above. The blue contours assume that only the density split lensing signal has been used while the green contours allow for complementary information from the tracer counts-in-cells histogram. In the fourth column of table \ref{table:params} we show the standard deviation of each parameter as found in our approximation to the posterior (and assuming the full data vector, including lensing and counts-in-cell).

Density split statistics is complementary to an analysis based on 2-point statistics \green{not just because it has a different dependence on the connection of galaxies and matter, but also} since it \green{is sensitive to} higher order moments of the density field. We demonstrate this by introducing an additional degree of freedom in our model, described by an additional parameter:
\begin{trivlist}
\item[8.)] $\Delta S_3 / S_3$: \noindent a factor multiplying all third order statistics in our predictions. The notation for this parameter is based on the ratio
\begin{equation}
\label{eq:definition_S3}
S_3\equiv\frac{\langle\delta^3\rangle}{\langle\delta^2\rangle^2}
\end{equation} 
which connects third and second moments of the density contrast and hence characterizes the amplitude of the density bispectrum (see section \ref{sec:Tree_level_for_PDF} for details). In our fiducial setup we compute it from leading order perturbation theory and $\Delta S_3 / S_3$ hence describes a relative deviation from that result.
\end{trivlist}

\noindent Within the $\Lambda$CDM model and at leading order in perturbation theory, the scaling between 2-point and 3-point statistics of the density field is almost independent of the cosmological parameters $\Omega_m$ and $\sigma_8$ \citep{Bernardeau}. Hence, a value of $\Delta S_3 / S_3\neq0$ would allow for deviations from the leading order result that cannot be compensated by changing $\Omega_m$ or $\sigma_8$. Such deviations could be caused non-standard physics of dark matter and dark energy that affect overdensities and underdensities differently (see e.g. \citet{Multamaki2003,Lue2004}; though $f(R)$ modified gravity theories have been shown to largely preserve the $\Lambda$CDM scaling, cf. \citet{Jain2008, Borisov2009}). Alternatively, they could indicate a break down the the perturbative scaling relations due to highly non-linear evolution of the density field or any small scale Baryonic physics that do not follow the scaling relations of perturbation theory \citep[cf.][]{Bernardeau, Uhlemann2017, Jeong2009}.

In the right panel of Figure \ref{fi:forecasts} we show how density split statistics including lensing and counts-in-cells can simultaneously constrain $\Delta S_3 / S_3$ and the parameter
\begin{equation}
\Sigma_8 = \sigma_8 \sqrt{\Omega_m}\ ,
\end{equation}
even after marginalization over the other model parameters. We also project how these constraints will improve when moving to year 5 (Y5) data of DES or when adding cosmological information from the cosmic microwave background. For the latter we estimated the parameter covariance in a Planck chain\cprotect\footnote{\verb|plikHM_TT_lowTEB| in \verb|COM_CosmoParams_base-plikHM_R2.00.tar.gz| from the Planck legacy archive \url{https://pla.esac.esa.int/pla/}, no lensing, cf. \citet{Planck2015}} and added this covariance as an additional Gaussian prior around our fiducial cosmology. 

With DES Y1 alone, we will be able to constrain the aplitude of third order statistics of the density field to about $20\%$ accuracy (cf. last column of table \ref{table:params}). Combining DES Y5 and Planck, this improves to about $5\%$. And this is even underestimating the power of DES-Y5: to project our constraints onto year-5 we simply divided our covariance by a factor of 4 in order to account for the larger area of the final DES survey. But this does not take into account the fact that DES Y5 will also be deeper than DES-Y1, which reduces shape noise and opens up the possibility of analyzing a larger number of redshift bins.

\section{Simulated data and Covariance matrix}
\label{sec:simulations}

In this work we use two sets of simulated data: the Buzzard galaxy catalogs which are constructed from high-resolution N-body simulations \citep{DeRose2017, Wechsler2017, MacCrann2017} and simulated random fields on the sky generated by the \verb|FLASK| tool \citep{Xavier2016}. We briefly describe these data sets in the following sections.

\subsection{Buzzard mock galaxy catalogs}
\label{sec:buzzard_description}

Here we describe the key aspects of the Buzzard galaxy catalogs for the purposes of this work and refer the reader to more detailed descriptions elsewhere \citep{DeRose2017, Wechsler2017, MacCrann2017}. We use four independent realizations of a DES Y1-like survey in version Buzzard-v1.1. These catalogs were constructed from N-body simulations run using \textsc{L-Gadget2} \citep{Springel2005}, a version of \textsc{Gadget2} modified for memory efficiency. Second-order Lagrangian perturbation theory initial conditions \citep{Crocce2006} were employed using \textsc{2LPTIC} \citep{Crocce2006}, and lightcones were output on the fly. Each galaxy catalog is built from a set of three nested lightcones using progressively larger volume and lower resolution at higher redshifts. The force resolution in each box is $20,\, 35\, \textrm{and }\, 53\, \textrm{kpc/h}$ with the boundaries between the lightcones falling at redshifts $z=0.34$ and $z=0.9$. 

The galaxy catalogs are constructed from the lightcones using the \textsc{ADDGALS} algorithm \citep{Wechsler2017} which assigns galaxy luminosities and positions based on the relation between redshift, r-band absolute magnitude, and large-scale density, $p(\delta|M_{r},z)$, found in a subhalo abundance matching (SHAM) model \citep[e.g.]{Conroy2006, Reddick2013}, in a high resolution N-body simulation. Spectral Energy Distributions (SEDs) are given to each simulated galaxy by finding a SDSS DR7 galaxy \citep{Cooper2011} that has a close match in $M_{r}$ and distance to its fifth nearest neighbor galaxy and assigning the SDSS galaxy's SED to the simulated galaxy. Galaxy sizes and ellipticities were assigned by drawing from distributions fit to high resolution SuprimeCam $i^{`}$-band data \citep{spcam}. Observed magnitudes in $griz$ were generated by redshifting the SEDs to the observer frame and integrating over the DES passbands. Photometric errors were added using the DES Y1 Multi Object Fitting (MOF) depth maps. 

The effects of weak gravitational lensing are calculated using the multiple-plane raytracing algorithm \textsc{CALCLENS} \citep{Becker2013}. The raytracing is done on a $n_{side}=4096$ \textsc{HEALPIX} \citep{Gorski2005} grid yielding an effective angular resolution of $0.85^{'}$. At each lens plane, the inverse magnification matrix of the ray closest to every galaxy is interpolated to the galaxy's 3D position and used to shear and magnify the galaxy. 

With galaxy catalogs with magnitudes, sizes and lensing effects in hand, \textsc{\textsc{redMaGiC}} and \textsc{metacalibration} \citep{Zuntz2017, Sheldon2017, Huff2017} samples of galaxies are selected from the simulations in an effort to approximate the selections done in the data. In the case of \textsc{\textsc{redMaGiC}}, the same algorithm which is used for selection in the data is applied to the simulations, \bblue{resulting in a tracer galaxy catalog of equivalent volume density and at least comparable bias}. For the \textsc{metacalibration} sample, as we do not run full image simulations, we must make approximate cuts on signal to noise of the galaxies to create a facsimile of the source sample in the data. For in depth comparisons of these simulated samples to their data counterparts see \citet{DeRose2017}.

For the density split analysis in this work, we use \textsc{\textsc{redMaGiC}} high density tracer galaxies ($L>0.5L_{\rm star}, \rho=10^{-3} \mathrm{Mpc}^{-3}$ comoving density) selected at a \textsc{redMaGiC} photometric redshift estimate of $0.2<z<0.45$. For the source redshift split lensing signals, we bin source galaxies by the expectation value of their $p(z)$ as estimated with BPZ \citep{Benitez2000,Hoyle2017} from the Buzzard mock photometry. Bin limits are chosen such that the true mean redshifts of the bins match the mean redshifts of the two highest redshift source samples defined in \citet{Hoyle2017}.

\subsection{Simulated density and convergence fields from \texttt{FLASK} and Covariance matrix}
\label{sec:FLASK}

For testing the numerical implementation of the model described in the following section and for estimating a covariance matrix of the density split lensing and counts-in-cells signals in the Buzzard simulations and the DES data, we use log-normal realizations of matter density and convergence fields. We summarize their properties here, with details given in appendix A of \citet[]{Gruen2017}.

We generate these fields as all-sky \textsc{healpix} maps of matter density and convergence using the \texttt{FLASK} software \citep{Xavier2016}. For the matter field, we choose the true redshift distribution of \textsc{\textsc{redMaGiC}} galaxies in the Buzzard simulations, selected as described in \autoref{sec:buzzard_description}. The matter field is sampled by a tracer population with \textsc{\textsc{redMaGiC}} density, bias of $b=1.54$, and Poissonian noise, from which lines of sight of different density are identified by the same algorithm as in Buzzard or in the data. For the convergence fields, we choose the estimated redshift distributions of DES source galaxies in the two highest redshift bins of \citet{Hoyle2017}. Log-normal parameters of the density and convergence fields are set by the perturbation theory formalism described in the following section.

960 of these realizations are used to estimate large-scale structure and shot noise contributions to the covariance matrix of the signals modeled herein. The contribution of shape noise is estimated by measuring the lensing signals in actual DES Y1 data with 960 realizations of the \textsc{metacalibration} shape catalog \citep{Zuntz2017} in which each galaxy ellipticity was rotated by a random angle.

\section{Modelling details and comparison to Simulations}
\label{sec:model}

In this section we present the approximations used to compute the ingredients (i), (ii) and (iii) of our model that are listed at the end of section \ref{sec:modeling_DT}. We also test the model ingredients (i) and (iii) directly against our N-body simulations.

Section \ref{sec:PDF_computation} describes our model for the PDF of projected density contrast within the top-hat smoothing radius $\theta_T$, $p(\delta_{m,T})$. Section \ref{sec:convergence_profile} describes how we model the convergence profile $\langle\kappa_{<\theta} | \delta_{m,T}\rangle$ around apertures of fixed density contrast $\delta_{m,T}$. And in section \ref{sec:CiC_computation} we describe our modeling of the probability $P(N_T | \delta_{m,T})$ of finding $N_T$ tracer galaxies in an aperture with density contrast $\delta_{m,T}$.

Section \ref{sec:model_summary} summarizes our fiducial model and the approximations used therein.

\subsection{Projected density PDF}
\label{sec:PDF_computation}

The computation of the PDF of the density field when smoothed by top-hat filters has e.g.$\ $been adressed by \citet{Bernardeau1994}, \citet{Bernardeau2000} and \citet{Valageas2002} (see also more recent developments in \cite{Bernardeau2014, Bernardeau2015, Codis2016a, Codis2016b, Uhlemann2017} which however do not yet enter our formalism). \citet{Bernardeau2000} demonstrated how to extend methods for the computation of the 3D density PDF to the weak lensing aperture mass which is a projected quantity. In the following we show how to modify their formalism in order to compute the line-of-sight density PDF in angular circles of radius $\theta_T$. To do so, we have to consider the \emph{cumulant generating function} (CGF) of the field $\delta_{m, T}(\hat{\mathbf{n}})$. The \emph{moment generating function} (MGF) of $\delta_{m, T}(\hat{\mathbf{n}})$ is defined as
\begin{equation}
\psi(y) = \sum_k \frac{\langle \delta_{m, T}(\hat{\mathbf{n}})^k\rangle }{k!}\ y^k\ .
\end{equation}
Due to the isotropy of the universe it does not depend on $\hat{\mathbf{n}}$. The CGF $\varphi(y)$ is given in terms of the MGF $\psi(y)$ as
\begin{eqnarray}
\label{eq:cgf}
\varphi(y) &=& \ln \psi(y) \nonumber \\
&\equiv& \sum_k \frac{\langle \delta_{m, T}(\hat{\mathbf{n}})^k\rangle_c }{k!}\ y^k\ ,
\end{eqnarray}
where in the last line we have defined the \emph{connected moments} or \emph{cumulants} $\langle \delta_{m, T}(\hat{\mathbf{n}})^k\rangle_c$ of $\delta_{m, T}(\hat{\mathbf{n}})$. The CGF of $\delta_{m, T}(\hat{\mathbf{n}})$ is related to its PDF via
\begin{equation}
\label{eq:CDF_definition}
e^{\varphi(y)} = \int \mathrm d \delta_{m, T}\ e^{y\delta_{m, T}}\ p(\delta_{m, T}) \ ,
\end{equation}
which is the Laplace transform of the PDF. Hence, if $\varphi(y)$ is a known analytic function, then $p(\delta_{m, T})$ can be computed by the inverse Laplace transform
\begin{equation}
\label{eq:inverse_Laplace}
p(\delta_{m, T}) = \int_{-\infty}^{\infty} \frac{\mathrm d y}{2\pi}\ e^{-iy\delta_{m, T} + \varphi(iy)}\ .
\end{equation}
This integral in the complex plain is most efficiently evaluated along the path of steepest descent (cf.$\ $ \cite{Bernardeau2014}, especially their appendix B). The cumulants of $\delta_{m, T}(\hat{\mathbf{n}})$ can be approximated as (cf. \citet{Bernardeau2000})
\begin{eqnarray}
\label{eq:Limber_projection_connected_moment}
\langle \delta_{m,T}(\hat{\mathbf{n}})^k \rangle_c &\approx &
\int \mathrm{d}w\ \frac{\left\langle\left\lbrace\delta_{\mathrm{cy.}, \theta w, L}(w\hat{\mathbf{n}}, w) q_l(w) L\right\rbrace^k\right\rangle_c}{L}\ .\nonumber \\
\end{eqnarray}
Here $q_l$ is the line-of-sight density of tracer galaxies defined in equation \ref{eq:definition_of_q_l} and $\delta_{\mathrm{cy.}, R, L}$ is the average of $\delta_{m,\mathrm{3D}}$ over a cylinder of length $L$ and radius $R$, where $L$ has to be chosen such that correlations of $\delta_{m,\mathrm{3D}}$ over a distance $L$ vanish for all practical purposes. Equation \ref{eq:Limber_projection_connected_moment} employs a small angle approximation and a Limber-like approximation \citep{Limber1953}. This means it assumes that any n-point correlation function between density constrast at different redshifts $z_i$, $i=1...n$, varies much more quickly with the redshift differences $\Delta z_{ij}$ than the projection kernel $q_l$. Comparing \ref{eq:cgf} and \ref{eq:Limber_projection_connected_moment} we see that the CGF of $\delta_{m,T}$ can be computed in terms of the CGF of $\delta_{\mathrm{cy.}, R, L}$ as
\begin{eqnarray}
\label{eq:Limber_projection_generating_function}
\varphi(y) &\approx &
\int \mathrm{d}w\ \frac{\varphi_{\mathrm{cy.}, \theta w, L}(q_l(w) L y, w)}{L}\ .
\end{eqnarray}
Hence, we have reduced the task of computing $p(\delta_{m, T})$ to the task of computing the connected moments of matter contrast in a long 3D cylinder.

To proceed we consider two different ansatzes. The first is to approximate $p(\delta_{m, T})$ by a log-normal distribution which is fixed by the first three connected moments of $\delta_{m, T}(\hat{\mathbf{n}})$. The second ansatz is to compute $\varphi_{\mathrm{cy.}, \theta w, L}(y)$ as a whole in a way similar to the one of \citet{Bernardeau1994} and \citet{Valageas2002} for the matter contrast in a 3-dimensional sphere. Using \ref{eq:Limber_projection_generating_function} we can then attempt to solve the integral in \ref{eq:inverse_Laplace} directly. 
We present details of both approaches in the following subsections.

\subsubsection{Log-normal approximation for the PDF}
\label{sec:lognormal_for_PDF}

Instead of computing the complete cumulant generating function of $\delta_{m, T}$ via equation \ref{eq:Limber_projection_generating_function} we start with an approach that only requires knowledge of the second and third cumulant, i.e. the moments $\langle \delta_{m, T}^2 \rangle_c $ and $\langle \delta_{m, T}^3 \rangle_c $ (implicitly we also assume $\langle \delta_{m, T} \rangle_c \equiv 0$ for the first cumulant). To do so, we approximate $\delta_{m, T}$ as a shifted log-normal random variable \citep{Hilbert2011, Xavier2016}. In this case the PDF of $\delta_{m, T}$ is given by
\begin{equation}
\label{eq:lognormal_PDF}
p(\delta_{m, T}) = \frac{\exp\left(-\frac{[\ln(\delta_{m, T}/\delta_0 - 1) + \sigma^2/2]^2}{2\sigma^2} \right)}{\sqrt{2\pi}\sigma (\delta_{m, T} - \delta_0)} 
\end{equation}
if $\delta_{m, T} > \delta_0$ and $p(\delta_{m, T}) = 0$ otherwise. The expectation value of this PDF is zero, as is appropriate. The variance and skewness of $\delta_{m, T}$ are given in terms of the parameters parameters $\delta_0$ and $\sigma$ by \citep{Hilbert2011}
\begin{align}
\label{eq:variance_for_lognormal_PDF}
\langle \delta_{m, T}^2 \rangle_c &= \delta_0^2 \left( e^{\sigma^2} - 1 \right)\nonumber \\
\langle \delta_{m, T}^3 \rangle_c &= \frac{3}{\delta_0}\ \langle \delta_{m, T}^2 \rangle_c^2 + \frac{1}{\delta_0^3}\ \langle \delta_{m, T}^2 \rangle_c^3\ .
\end{align}
The ansatz of a log-normal PDF has been shown to be consistent with early DES data by \citet{Clerkin2017}. \blue{For the 3-dimensional density contrast $\delta_{m, 3D}$ this can be reasonably motivated from theory by observing that at leading order in perturbation theory the skewness of $\delta_{m, 3D}$ scales as
\begin{equation}
\langle \hat \delta_{m,\mathrm{3D}}^3 \rangle_c \sim \langle \hat \delta_{m,\mathrm{3D}}^2 \rangle_c^2
\end{equation}
with corrections of the order $\langle \hat \delta_{m,\mathrm{3D}}^2 \rangle_c^3$. This is exactly the scaling obeyed by the log-normal distribution and choosing the parameter $\delta_0$ appropriately allows one to exactly reproduce the scaling coefficients predicted by perturbation theory.}

At least for a power law power spectrum, the same kind of scaling is observed also for 2-dimensional, projected versions of the density field. Hence, one of the ansatzes used in this paper to compute the PDF of $\delta_{m, T}$ is to derive its variance and skewness as described in appendix \ref{app:PT_for_skewness} and fix $\delta_0$ and $\sigma$ by demanding that the PDF in \ref{eq:lognormal_PDF} has the same second and third moments.

\subsubsection{Tree level computation of $\varphi_{\mathrm{cy.}, \theta w, L}(y, w)$ in the cylindrical collapse model}
\label{sec:Tree_level_for_PDF}

Let us first consider the cumulant generating function $\varphi_{3D}(y, \tau)$ of the 3-dimensional density contrast $\delta_{3D}(\mathbf{x}, \tau)$. To compute $\varphi_{3D}$ at tree-level in perturbation theory it is sufficient to assume spherical symmetry around a particular point $\mathbf{x}$ \citep[see e.g.][]{Valageas2002}. Doing so, we can directly compute $\delta_{3D}(\mathbf{x}, \tau)$ as a function of the linear density contrast $\delta_{3D, \mathrm{lin.}}(\mathbf{x}, \tau)$ by means of the spherical collapse model \citep{Fosalba1998, Valageas2002}, i.e.
\begin{equation}
\delta_{3D}(\mathbf{x}, \tau) = F[\delta_{3D, \mathrm{lin.}}(\mathbf{x}, \tau), \tau]
\end{equation}
where $F$ is determined by one of the differential equations presented in appendix A. Hence, using the assumption that the linear density contrast has a Gaussian distribution with variance $\sigma_{3D, \mathrm{lin.}}^2(\tau)$ we can express the cumulant generating function as (cf. equation \ref{eq:CDF_definition})
\begin{eqnarray}
\label{eq:phi_in_SC_1}
&& \exp\left\lbrace\varphi_{3D}(y, \tau)\right\rbrace \nonumber \\
&=& \int \mathrm d \delta_{3D}\ e^{y\delta_{3D}}\ p(\delta_{3D}, \tau) \nonumber \\
&=& \int \frac{\mathrm d \delta_{3D, \mathrm{lin.}}}{\sqrt{2\pi\sigma_{3D, \mathrm{lin.}}^2(\tau)}}\ \exp\left(yF[\delta_{3D, \mathrm{lin.}}, \tau] - \frac{\delta_{3D, \mathrm{lin.}}^2}{2\sigma_{3D, \mathrm{lin.}}^2(\tau)}\right)\ ,\nonumber \\
\end{eqnarray}
where in the last step we simply performed a change of variables from $\delta_{3D}$ to $\delta_{3D, \mathrm{lin.}}$. We now employ Laplace's method, which states that a function $f(x)$ which strongly peaks around $x_0$ fulfils
\begin{equation}
\int \mathrm d x\ e^{f(x)} \approx \sqrt{\frac{2\pi}{|f''(x_0)|}} e^{f(x_0)}\ .
\end{equation}
This way we can approximate the last line of \ref{eq:phi_in_SC_1} as
\begin{eqnarray}
\label{eq:phi_in_SC_2}
e^{\varphi_{3D}(y, \tau)} &\approx & \sqrt{\frac{1}{|1 - y F''[\delta^*, \tau]\ \sigma_{3D, \mathrm{lin.}}^2(\tau) |}}\ \times \nonumber \\
&& \times\ \exp\left(y F[\delta^*, \tau] - \frac{{\delta^*}^2}{2\sigma_{3D, \mathrm{lin.}}^2(\tau)}\right)\ ,
\end{eqnarray}
where $\delta^*$ is the linear density contrast that maximizes the exponent in \ref{eq:phi_in_SC_1} and $'$ denotes derivation wrt. $\delta\ $. In the quasi-linear limit of $\sigma_{3D, \mathrm{lin.}}^2 \rightarrow 0$ this gives
\begin{equation}
\label{eq:phi_in_SC_final}
\varphi_{3D}(y, \tau) \approx y F[\delta^*, \tau] - \frac{{\delta^*}^2}{2\sigma_{3D, \mathrm{lin.}}^2(\tau)}\ ,
\end{equation}
where $\delta^*$ has to be determined by the implicit equation
\begin{equation}
\label{eq:speepest_descent_condition}
\frac{\delta^*}{\sigma_{3D, \mathrm{lin.}}^2(\tau)} = y F'[\delta^*, \tau]\ .
\end{equation}
It should be noted that equations \ref{eq:phi_in_SC_final} and \ref{eq:speepest_descent_condition} reproduce \emph{exactly} the tree-level results for the cumulant generating function \citep[cf.][]{Bernardeau1994, Valageas2002, Bernardeau, Bernardeau2015} ! As described in \citet{Bernardeau} the coefficients
\begin{equation}
\label{eq:Bernardeau_rescaling}
S_n = \frac{\langle \delta^n \rangle_c}{\langle \delta^2 \rangle_c^{n-1}}
\end{equation}
are given quite accurately by the lowest order of perturbation theory. Hence, using \emph{halofit} \citep{Smith2003,Takahashi2012} and an analytic transfer function \citep{Eisenstein1998} to compute the non-linear matter power spectrum, we can compute the non-linear variance $\langle \delta^2 \rangle_{c, \mathrm{non}.\mathrm{lin}}$ and then rescale the leading order CGF to its non-linear version via
\begin{equation}
\label{eq:phi_rescaling}
\varphi_{\mathrm{non}.\mathrm{lin}}(y, \tau)\ =\ \frac{\langle \delta^2 \rangle_{c, \mathrm{lin}}}{\langle \delta^2 \rangle_{c, \mathrm{non}.\mathrm{lin}}}\ \varphi_{\mathrm{lead}}\left(y\frac{\langle \delta^2 \rangle_{c, \mathrm{non}.\mathrm{lin}}}{\langle \delta^2 \rangle_{c, \mathrm{lin}}}, \tau\right)\ .
\end{equation}

\begin{figure*}
\center{
  \includegraphics[width=0.48\textwidth]{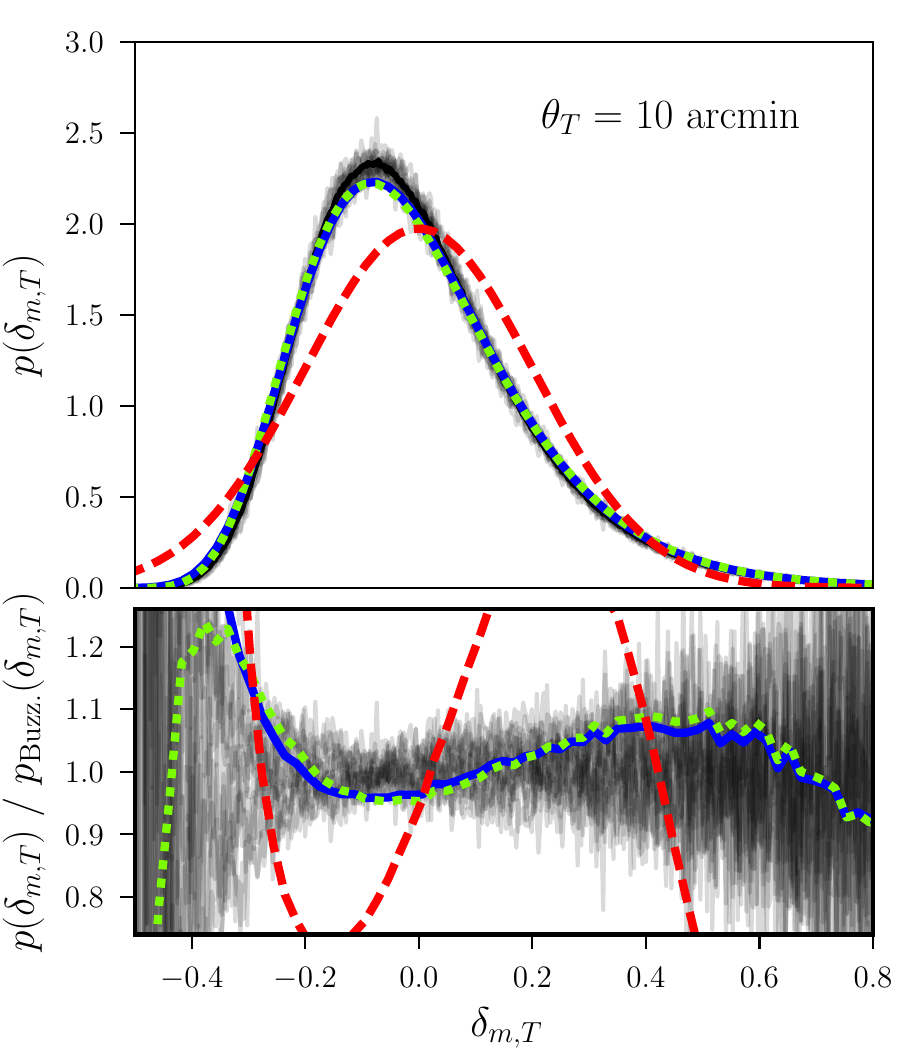}\hspace{0.04\textwidth}\includegraphics[width=0.48\textwidth]{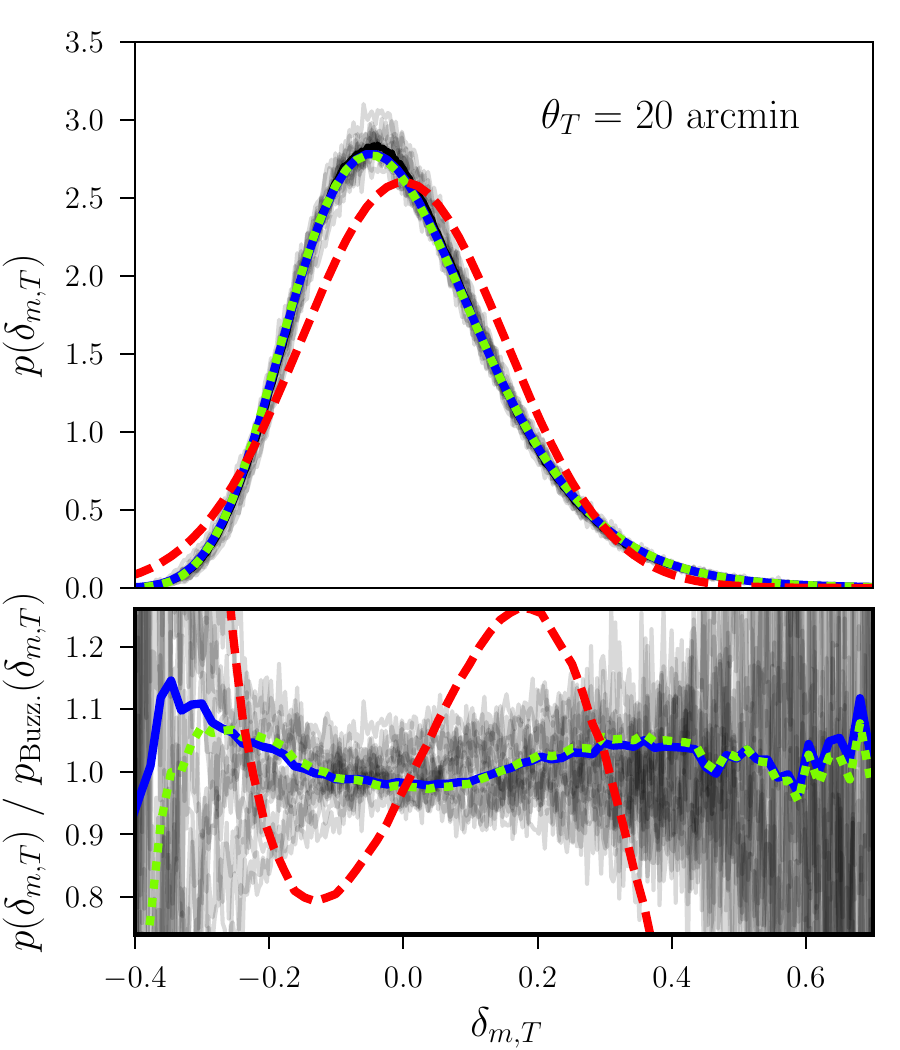}
  \includegraphics[width=0.48\textwidth]{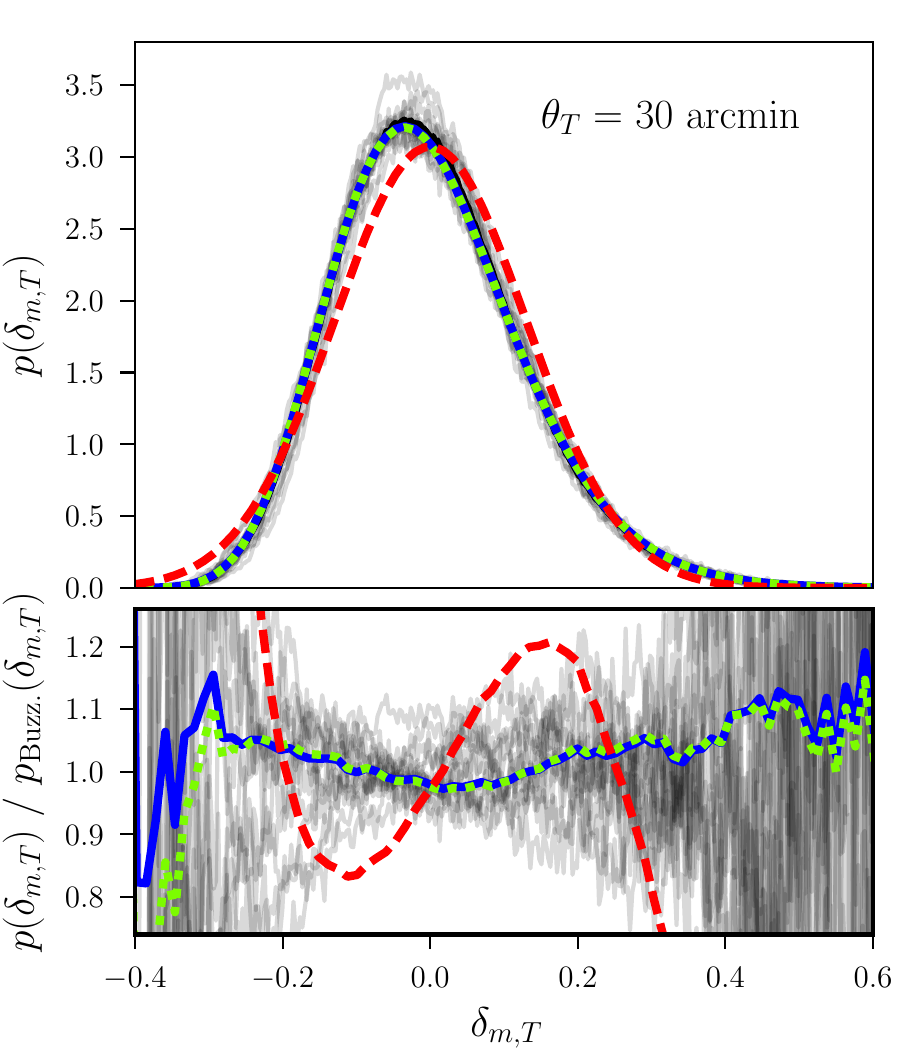}\hspace{0.04\textwidth}\includegraphics[width=0.48\textwidth]{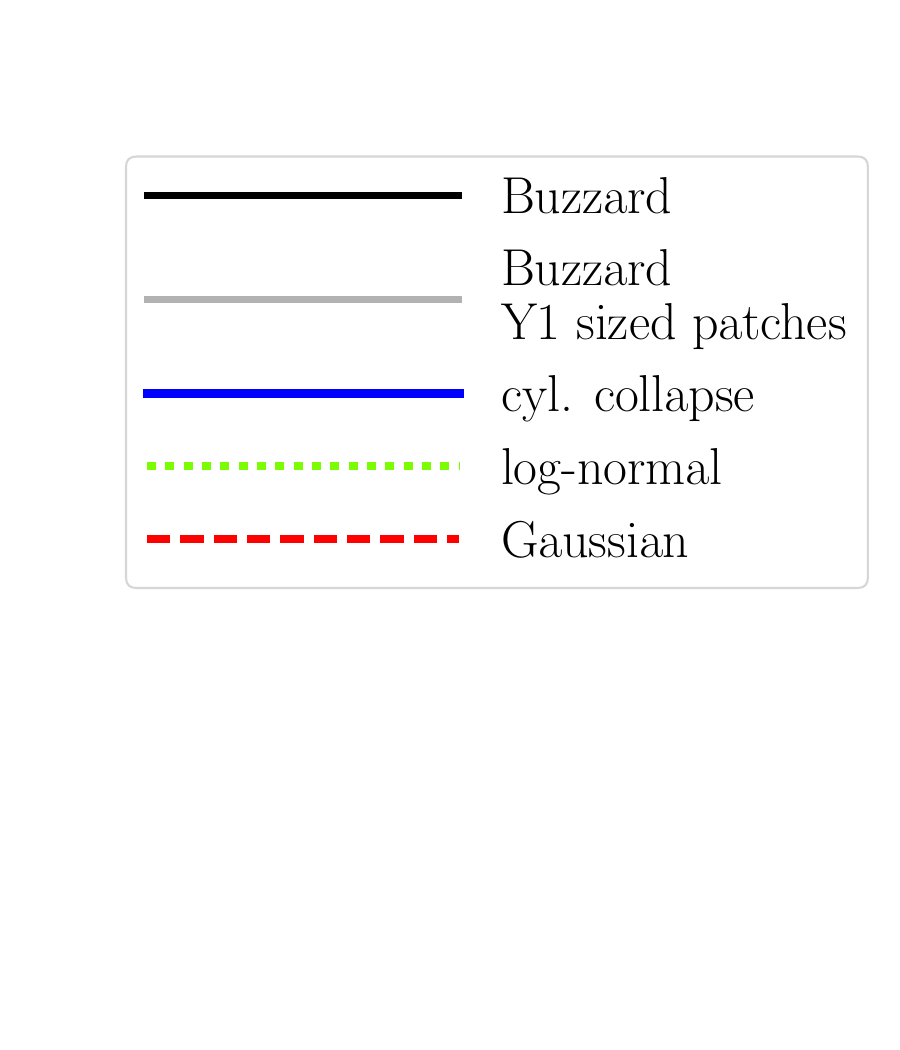}
  }
   \caption{The PDF of projected density contrast in Buzzard compared to several models for various smoothin scales \blue{($\delta_{m,T}$ is the projected density contrast $\delta_{m}$ smoothed by our top-hat radius $\theta_T$)}. In the upper panel of each plot the black line shows a histogram of $\delta_{m,T}$ measured in an all-sky map from Buzzard. The grey lines show histograms measured in $14$ DES year1 shaped patches of that all-sky map. The blue lines show the PDF predicted by our tree-level computation of the cumulant generating function, the green lines show the PT-motivated log-normal model and the red lines show a Gaussian PDF with the same variance as the other two models. The bottom panels of each plot are showing the ratio of each PDF to the one measured in the Buzzard all-sky. For all \last{aperture} radii our halofit power spectrum is predicting a standard deviation of $\delta_{m,T}$ that is $\gtrsim 2\%$ higher than what we find in Buzzard (cf. figure \ref{fi:delta_moments}). For $\theta_T = 20'$ and $30'$ this is the dominant source of \last{mismatch}.}
  \label{fi:p_of_delta}
\end{figure*}

\begin{figure}
\center{
  \includegraphics[width=8.0cm]{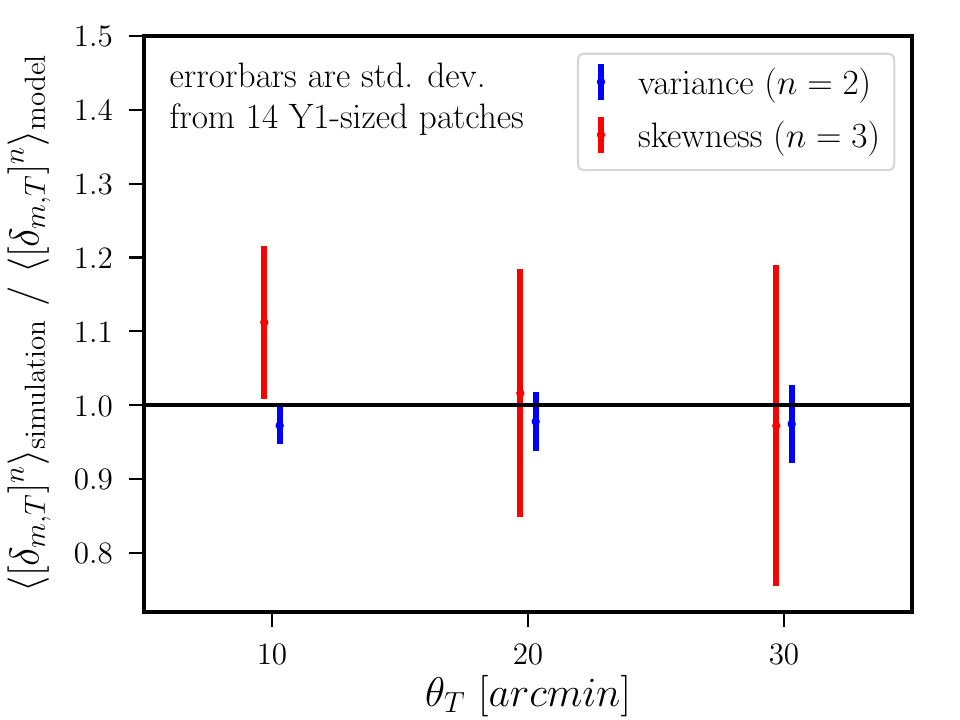}
  }
   \caption{This figure shows the ratio of moments of $\delta_{m,T}$ measured in a Buzzard all-sky density map to the moments predicted by our model as a function of $\theta_T$. The errorbars represent the standard deviations of the moments found in a set of 14 DES Y1-sized patches in the Buzzard map. At our fiducial radius $\theta_T = 20'$ we find a $\sim 2.4\%$ deviation between the variance of $\delta_{m,T}$ measured in Buzzard and that predicted by our model. This would correspond to a $\sim 1.2\%$ deviation in $\sigma_8$.}
  \label{fi:delta_moments}
\end{figure}

To perform the projection integral in equation \ref{eq:Limber_projection_generating_function} we need to know the cumulant generating function of the density contrast in a long cylinder of radius R and length $L >> R$, $\delta_{\mathrm{cy.}, R, L}$. \citeauthor{Bernardeau1994} \citep[1994; see also][and the other references above]{Valageas2002} have generalized equation \ref{eq:phi_in_SC_final} to the case of matter density in a spherical aperture. Their results can easily be transferred to cylindrical apertures, yielding
\begin{equation}
\label{eq:phi_cyl_in_SC_final}
\varphi_{\mathrm{cy.}, R, L}(y, \tau) \approx y F_{\mathrm{cyl.}}[\delta^*, \tau] - \frac{{\delta^*}^2}{2\sigma_{R\sqrt{1+F_{\mathrm{cyl.}}[\delta^*]}, L, \mathrm{lin.}}^2(\tau)}\ ,
\end{equation}
where $\sigma_{R, L, \mathrm{lin.}}^2$ is the variance of linear density contrast in a cylinder, $F_{\mathrm{cyl.}}$ is now determined by cylindrical collapse and $\delta^*$ has to be determined by the implicit equation
\begin{equation}
\frac{1}{2} \frac{\mathrm d}{\mathrm d \delta^*} \frac{{\delta^*}^2}{\sigma_{R\sqrt{1+F_{\mathrm{cyl.}}[\delta^*]}, L, \mathrm{lin.}}^2(\tau)} = y F'[\delta^*, \tau]\ .
\end{equation}
Using equation \ref{eq:phi_rescaling} we can again rescale this leading order result for the generating function to its non-linear counterpart.

The validity of equation \ref{eq:phi_rescaling} is ultimately limiting the accuracy of our model for the distribution of the density contrast inside the aperture radius $\theta_T$, $p(\delta_{m,T})$. In figure \ref{fi:p_of_delta} we compare the PDF measured in the Buzzard simulations to both the log-normal model and the full CGF computation of the PDF for aperture radii $\theta_T = 10', 20', 30'$. For both $\theta_T = 20'$ and $30'$ our model PDF's and the Buzzard simulations agree within DES-Y1 cosmic variance as can be seen in the residuals shown in the lower panels of each plot.  Also, the difference between log-normal approximation and full CGF computation is completely negligible. To investigate the agreement of Buzzard and our models more quantitatively, we also compare the variance and skewness of each PDF. In figure \ref{fi:delta_moments} we show the ratio of these moments as found in Buzzard to our predictions. For $\theta_T = 10'$, the density field in Buzzard seems to have a significantly higher skewness than predicted by our model. We attribute this indeed to the failing of the Quasi-linear rescaling, eq. \ref{eq:phi_rescaling}. For the other radii the skewness values agree to within $2-3\%$. A similar relative agreement is achieved for the variance of the distributions. At our fiducial radius $\theta_T=20'$ the variances of Buzzard and our model differ by about $2.4\%$. This corresponds to a disagreement in the fluctuation amplitude $\sigma_8$ of about $1.2\%$.

For comparison, we also show a Gaussian model for the density PDF in figure \ref{fi:p_of_delta}, using the same variance as for the other PDF models. It can clearly be seen, that $p(\delta_{m,T})$ cannot be well described by a Gaussian distribution for any of the considered smoothing radii.

\subsection{Convergence profile around lines-of-sight of fix density contrast $\delta_{m,T}$}
\label{sec:convergence_profile}

We now want to know the average lensing convergence $\kappa_{<\theta}$ inside a radius $\theta$ around a line-of-sight with a given value of $\delta_{m, T}$. By means of equation \ref{eq:gamma_in_terms_of_kappa} this can be turned into a prediction of the density split lensing signal. We start by looking at the joint moment generating function of $\kappa_{<\theta}$ and $\delta_{m,T}$,
\begin{equation}
\psi_\theta(y, z) = \sum_{k, l} \frac{\langle \delta_{m, T}^k\ \kappa_{<\theta}^l\rangle }{k!\ l!}\ y^k\ z^l\ ,
\end{equation}
and their joint cumulant generating function defined by
\begin{eqnarray}
\label{eq:cgf_joint}
\varphi_\theta(y, z) &=& \ln \psi_\theta(y, z) \nonumber \\
&\equiv& \sum_{k, l} \frac{\langle \delta_{m, T}^k\ \kappa_{<\theta}^l\rangle_c }{k!\ l!}\ y^k\ z^l\ .
\end{eqnarray}
Using a Limber-like approximation similar to the one employed in eq. \ref{eq:Limber_projection_generating_function}, one can write $\varphi_\theta(y, z)$ as a line-of-sight projection of the CGF of matter density contrasts that are smoothed over concentrical cylindrical apertures with length $L$ and radii $R_1, R_2$ , $\varphi_{\mathrm{cyl}., R_1, R_2, L}(y, z, w)$:
\begin{eqnarray}
\label{eq:Limber_projection_generating_function_2_variables}
\varphi_\theta(y, z) &\approx &
\int \mathrm{d}w\ \frac{\varphi_{\mathrm{cyl.}, \theta_T w, \theta w, L}(q_l(w) L y, W_s(w) L z, w)}{L}\ .\nonumber \\
\end{eqnarray}
Here, $q_l(w)$ is again the line-of-sight density of lens galaxies and $W_{s}(w)$ is the lensing efficiency defined in eq. \ref{eq:lensing_efficiency}.

The joint PDF of $\kappa_{<\theta}$ and $\delta_{m,T}$ is related to $\varphi_\theta$ via
\begin{equation}
\label{eq:inverse_Laplace_two_variables}
p(\delta_{m,T}=s,\kappa_{<\theta}=r) = \int_{-\infty}^{\infty} \frac{\mathrm d y\ \mathrm d z}{(2\pi)^2}\ e^{-iys -izr + \varphi_\theta(iy, iz)}\ .
\end{equation}
The expectation value of $\kappa_{<\theta}$ given a certain value of $\delta_{m,T}$ is then given by
\begin{align}
\label{eq:kappa_given_delta_in_terms_of_G}
&\ \langle\kappa_{<\theta} | \delta_{m,T} = s\rangle \nonumber \\
=&\ \frac{1}{p(\delta_{m,T} = s)}\int \mathrm d r\ r\ p(\delta_{m,T}=s,\kappa_{<\theta}=r) \nonumber \\
=&\ \frac{1}{p(\delta_{m,T} = s)}\int_{-\infty}^{\infty} \frac{\mathrm d y\ \mathrm d z}{(2\pi)^2}\ e^{-iys + \varphi_\theta(iy, iz)}\int \mathrm d r\ r\ e^{-izr} \nonumber \\
=&\ \frac{1}{p(\delta_{m,T} = s)}\int_{-\infty}^{\infty} \frac{\mathrm d y\ \mathrm d z}{2\pi}\ e^{-iys + \varphi_\theta(iy, iz)}\ i\frac{\mathrm d}{\mathrm dz} \delta_D(z) \nonumber \\
=&\ \frac{1}{p(\delta_{m,T} = s)}\int_{-\infty}^{\infty} \frac{\mathrm d y}{2\pi}\ e^{-iys + \varphi_{m,T}(iy)}\ G_\theta(iy) \ ,\nonumber \\
\end{align}
with
\begin{eqnarray}
\label{eq:definition_of_G}
G_\theta(y) &=& \left.\frac{\mathrm d}{\mathrm dz} \varphi_\theta(y, z)\right|_{z=0}  \nonumber \\
&=& \sum_{k} \frac{\langle \delta_{m, T}^k\ \kappa_{<\theta}\rangle_c }{k!}\ y^k\ .
\end{eqnarray}
Using equation \ref{eq:Limber_projection_generating_function_2_variables} we can express $G_\theta(y)$ by the corresponding cylindrical, 3-dimensional quantity:
\begin{eqnarray}
\label{eq:Limber_projection_G}
G_\theta(y) &\approx &
\int \mathrm{d}w\ W_s(w)\ G_{\mathrm{cyl.}, \theta_T w, \theta w, L}(q_l(w) L y, w)\ .\nonumber \\
\end{eqnarray}
We again pursue two ansatzes for computing $G_\theta(y)$: one involving a log-normal model for the joint cumulants of $\kappa_{<\theta}$ and $\delta_{m,T}$ and one involving the model of cylindrical collapse to compute a leading order perturbation theory prediction for $G_{\mathrm{cyl.}, \theta_T w, \theta w, L}(y, w)$. We are detailing these ansatzes in the following subsections.

\subsubsection{Log-normal model for the joint cumulants of $\kappa_{<\theta}$ and $\delta_{m,T}$}
\label{sec:lognormal_for_kappa}

From equation \ref{eq:definition_of_G} one can see that only joint cumulants of the form
\begin{equation}
\langle \delta_{m, T}^k\ \kappa_{<\theta}\rangle_c
\end{equation}
enter the convergence profile around a given density contrast. Hence, in a spirit similar to section \ref{sec:lognormal_for_PDF} we only compute the leading order cumulants $\langle \delta_{m, T}\ \kappa_{<\theta}\rangle_c$ and $\langle \delta_{m, T}^2\ \kappa_{<\theta}\rangle_c$ as described in appendix \ref{app:PT_for_skewness} and use these moments to fix a joint log-normal PDF for $\kappa_{<\theta}$ and $\delta_{m,T}$. Note that this is indeed \emph{not} assuming, that $\kappa_{<\theta}$ is a log-normal random variable. It rather assumes that
\begin{equation}
\kappa_{<\theta} = \kappa_{\mathrm{log}-\mathrm{normal}} + \kappa_{\mathrm{uncorr}.}\ ,
\end{equation}
where $\kappa_{\mathrm{log}-\mathrm{normal}}$ is log-normal and $\kappa_{\mathrm{uncorr}.}$ is an unspecified random variable that is uncorrelated with $\delta_{m,T}$. Only $\kappa_{\mathrm{log}-\mathrm{normal}}$ will actually contribute to the density split lensing signal and we can model the expectation value $\langle\kappa_{<\theta} | \delta_{m,T} = s\rangle$ by the following relation holding for two joint log-normal variables:
\begin{equation}
\frac{\langle\kappa_{<\theta} | \delta_{m,T} = s\rangle}{\kappa_0} =  \exp\left(\frac{C (2\log(1+s/\delta_0)+V-C)}{2V}\right)-1
\end{equation}
with
\begin{eqnarray}
V &=& \log\left(1.0 + \frac{\langle \delta_{m, T}^2\rangle_c}{\delta_0^2}\right)\nonumber \\
C &=& \log\left(1.0 + \frac{\langle \delta_{m, T}\ \kappa_{<\theta}\rangle_c}{\delta_0\ \kappa_0}\right)\nonumber \\
\kappa_0 &=& \frac{\langle \delta_{m, T}\ \kappa_{<\theta}\rangle_c^2\ e^V}{\langle \delta_{m, T}^2\ \kappa_{<\theta}\rangle_c - 2\langle \delta_{m, T}\ \kappa_{<\theta}\rangle_c \langle \delta_{m, T}^2\rangle_c/\delta_0}
\end{eqnarray}
and $\delta_0$ determined as described in section \ref{sec:lognormal_for_PDF}.

It should be noted that the log-normal parameter $\kappa_0$ which we use to approximate the contribution of $\kappa_{<\theta}$ to the lensing signal is dependent on the the smoothing scale $\theta$. This indicates even further, that we do indeed not approximate $\kappa$ as a log-normal field.

\subsubsection{Tree level computation of $G_{\mathrm{cyl.}, R_1, R_2, L}(y, w)$ in the cylindrical collapse model}
\label{sec:cyl_collapse_for_kappa}

For convenience we will shorten the notation of section \ref{sec:Tree_level_for_PDF} by defining
\begin{eqnarray}
G(y) &\equiv & G_{\mathrm{cyl.}, R_1, R_2, L}(y, w)\nonumber \\
\varphi(y, z) &\equiv & \varphi_{\mathrm{cyl.}, R_1, R_2, L}(y, z, w)\nonumber \\
F[\delta] &\equiv & F_{\mathrm{cyl.}}[\delta, \tau]\nonumber \\
\sigma_R &\equiv & \sigma_{R, L, \mathrm{lin.}}\nonumber \\
R_{i, \mathrm{L}}(\delta) &\equiv & R_i\sqrt{1+F[\delta]}\ ,\ i = 1,2 \ .
\end{eqnarray}
The joint cumulant generating function of density contrast in concentric cylinders is then \citep[in complete analogy to equation \ref{eq:phi_cyl_in_SC_final}; see also][who present very similar calculations]{Bernardeau2000} given by
\begin{equation}
\label{eq:phi_cyl_in_SC_final_2_variables}
\varphi(y, z) \approx y F[\delta_1^*] +  z F[\delta_2^*] - \frac{1}{2} \sum_{i,j} \delta_i^* \delta_j^* \left(\mathcal{C}^{-1}\right)_{ij}\ ,
\end{equation}
where the elements of the matrix $\mathcal{C}$ are given by $\mathcal{C}_{11} = \sigma_{R_{1, \mathrm{L}}(\delta_1^*)}^2$, $\mathcal{C} = \sigma_{R_{2, \mathrm{L}}(\delta_2^*)}^2$ and $\mathcal{C}_{12} = \mathcal{C}_{21}$ is the linear covariance of density contrasts in concentric cylinders of radii $R_{1, \mathrm{L}}(\delta_1^*)$ and $R_{2, \mathrm{L}}(\delta_2^*)$. This time the critical linear density contrasts $\delta_2^*$ and $\delta_2^*$ are given by the implicit equations
\begin{eqnarray}
\label{eq:saddle_point_2_variables}
\frac{1}{2} \frac{\partial}{\partial \delta_1^*} \sum_{i,j} \delta_i^* \delta_j^* \left(\mathcal{C}^{-1}\right)_{ij} &=& y F'[\delta_1^*] \\
\label{eq:saddle_point_2_variables_b}
\frac{1}{2} \frac{\partial}{\partial \delta_2^*} \sum_{i,j} \delta_i^* \delta_j^* \left(\mathcal{C}^{-1}\right)_{ij} &=& z F'[\delta_2^*]\ .
\end{eqnarray}
Note that these conditions force each $\delta_i^*$ to be a function of both $y$ and $z$, i.e. $\delta_i^* = \delta_i^*(y,z)$. 

To predict the convergence profile around apertures of a given density contrast $\delta_{m,T}$ by means of equations \ref{eq:kappa_given_delta_in_terms_of_G} and \ref{eq:Limber_projection_G} we are interested in computing the function
\begin{equation}
G(y) = \left.\frac{\partial}{\partial z}\varphi(y, z)\right|_{z=0}\ .
\end{equation}
Using the conditions \ref{eq:saddle_point_2_variables} and \ref{eq:saddle_point_2_variables_b} one can see right away that
\begin{equation}
G(y) = F[\delta_2^*(y, 0)]\ .
\end{equation}
Furthermore, for $z=0$ equations \ref{eq:saddle_point_2_variables} and \ref{eq:saddle_point_2_variables_b} can be simplified to
\begin{eqnarray}
\label{eq:saddle_point_2_variables_c}
\frac{1}{2} \frac{\mathrm d}{\mathrm d \delta_1^*} \frac{{\delta_1^*}^2}{\mathcal{C}_{11}} &=& y F'[\delta_1^*] \\
\label{eq:saddle_point_2_variables_d}
\delta_2^* &=& \frac{\mathcal{C}_{12}}{\mathcal{C}_{11}} \delta_1^*\ .
\end{eqnarray}
This way we obtain a solution for $G(y)$ at leading order in perturbation theory. In appendix \ref{app:PT_for_skewness} we argue that for $R_2 \geq R_1$ the cumulants $\langle \delta_{R_1}^k\ \delta_{R_2}\rangle_c$ approximately follow the scaling relation
\begin{equation}
\label{eq:rescaling_mixed_moments}
\langle \delta_{R_1}^k\ \delta_{R_2}\rangle_c\ \sim\ \langle \delta_{R_1}\ \delta_{R_2}\rangle_c\ \langle \delta_{R_1}^2\rangle_c^{k-1}\ .
\end{equation}
This can be used to correct the tree-level approximation of $G(y)$ for the non-linear evolution of the power spectrum. To do so, we first determine the proportionality factors of the relation \ref{eq:rescaling_mixed_moments} at leading order by fitting a polynomial in $y$ to the function $G(y)$ and extracting the cumulants $\langle \delta_{R_1}^k\ \delta_{R_2}\rangle_c$ from the polynomial coefficients. In practice, we do this with a polynomial of degree $10$, but already a polynomial of degree $5$ gives almost identical results\footnote{The coefficients of linear and quadratic order in $y$ are always obtained from the exact perturbation theory computation of appendix \ref{app:PT_for_skewness}.}. Then, we use the revised halofit of \citet{Takahashi2012} to compute a late-time version of the right-hand-side of \ref{eq:rescaling_mixed_moments}. This, together with the tree-level proportionality factors determined before, yields a non-linear approximation of the polynomial coefficients representing $G(y)$. We use those to re-compute $G(y)$ and then carry out the projection integral in equation \ref{eq:Limber_projection_G}.

Our rescaling of the coefficient corresponding to the cumulant $\langle \delta_{R_1}^2\ \delta_{R_2}\rangle_c$ is in fact more complicated than described here, cf. appendix \ref{app:third_moment_rescaling}. But we find our prediction of the density split lensing signal to be insensitive to the details of the rescaling procedure.

\subsection{Shot-noise, stochasticity and Counts-in-Cells}
\label{sec:CiC_computation}

We now want to model the conditional probability $P(N_T| \delta_{m,T})$ of finding $N_T$ galaxies in an angular radius of $\theta_T$, when the projected density contrast in that radius is $\delta_{m,T}$. This is the third ingredient of the framework described in section \ref{sec:cosmology} and completes our modeling of the density split lensing signal as well as the counts-in-cells histogram.

To analyze the relation of $N_T$ and $\delta_{m,T}$ in a systematic way, let us introduce the auxiliary field $\delta_{g,T}$. We assume that $\delta_{g,T}(\hat{\mathbf{n}})$ is a smooth field in the sky and that $N_T$ is a Poissonian tracer of this field. This means we will assume that
\begin{equation}
P(N_T = N | \delta_{g,T}) = \frac{\left[ \bar N (1 + \delta_{g,T}) \right]^N}{N!}\ e^{-\bar N (1 + \delta_{g,T})}\ ,
\end{equation}
where $\bar N \equiv \langle N_T \rangle$. A consequence of this assumption is that the expectation value of $N_T$ for fixed $\delta_{g,T}$ is given by
\begin{equation}
\langle N_T | \delta_{g,T} \rangle = \bar N (1 + \delta_{g,T})
\end{equation}
and that the variance of $N_T$ for fixed $\delta_{g,T}$ fulfils
\begin{equation}
\label{eq:Poisson_ratio}
\frac{\mathrm{Var}\left[ N_T | \delta_{g,T} \right]}{\langle N_T | \delta_{g,T} \rangle} = 1\ .
\end{equation}
To connect the galaxy field to the lensing convergence we however need to know the relation between $N_T$ and $\delta_{m,T}$. Assuming a generic joint PDF of $\delta_{m,T}$ and $\delta_{g,T}$ we can write the expectation values of $N_T$ for fixed $\delta_{m,T}$ as
\begin{align}
\langle N_T | \delta_{m,T} \rangle & = \int \mathrm{d} \delta_{g,T}\ p(\delta_{g,T} | \delta_{m,T}) \langle N_T | \delta_{g,T} \rangle \nonumber \\
& = \bar N (1 + \langle\delta_{g,T}| \delta_{m,T} \rangle)\ . 
\end{align}
Also, it can be shown that the variance of $N_T$ for a fixed value of $\delta_{m,T}$ is given by
\begin{equation}
\label{eq:not_quite_Poisson}
\mathrm{Var}\left[ N_T | \delta_{m,T} \right] = \langle N_T | \delta_{m,T} \rangle + \bar N^2\mathrm{Var}\left[ \delta_{g,T} | \delta_{m,T} \right] \ .
\end{equation}
From equation \ref{eq:not_quite_Poisson} we can see that the distribution of $N_T$ given $\delta_{m,T}$ can only be a Poisson distribution if $\mathrm{Var}\left[ \delta_{g,T} | \delta_{m,T} \right] \equiv 0$. \green{This is the simplest model for the connection of $N_T$ and $\delta_{m,T}$ that we test in this work and in \citet{Gruen2017}.} If $\mathrm{Var}\left[ \delta_{g,T} | \delta_{m,T} \right] \neq 0$ we will say that there is a stochasticity between the galaxy field and the matter density field, and we cannot assume a Poisson distribution for $P(N_T| \delta_{m,T})$. \jab{We note that ``stochasticity'' in this context could arise from a nonlinear biasing relationship between $\delta_{g,T}$ and $\delta_{m,T}$, including e.g.\ a dependence on higher powers of $\delta_{m,T}$ or effects from halo exclusion \citep{Baldauf2013}, or from physical stochasticity in galaxy formation.}

We explore two ways to account for a possible stochasticity (see also \citet{Dekel1999}, who have discussed similar concepts). In our first approach we introduce a free parameter to our model - a Pearson correlation coefficient $r\neq1$ between the random fields $\delta_{g,T}$ and $\delta_{m,T}$. Within our log-normal framework we show that this automatically leads to a $\delta_{m,T}$-dependence of the ratio in eq. \ref{eq:Var_N_ratio}. We explain the details of this in section \ref{sec:stochasticity}.

In our second approach we employ a generalized Poisson distribution for $P(N_T| \delta_{m,T})$ that allows for
\begin{equation}
\frac{\mathrm{Var}\left[ N_T | \delta_{m,T} \right]}{\langle N_T | \delta_{m,T} \rangle} \neq 1\ .
\end{equation}
In this approach we introduce 2 parameters, $\alpha_0$ and $\alpha_1$, to our model such that
\begin{equation}
\label{eq:Var_N_ratio}
\frac{\mathrm{Var}\left[ N_T | \delta_{m,T} \right]}{\langle N_T | \delta_{m,T} \rangle} \approx \alpha_0 + \alpha_1\ \delta_{m,T} \ .
\end{equation}
The details of this are explained in section \ref{sec:parametric_shot_noise}.

Both of our approaches match our simulated data equally well (cf. Figure \ref{fi:subpoisson}). \bblue{This means that, for the galaxies in these realizations, the model based on the correlation coefficient $r$ is a sufficient description. It will thus be the fiducial model in this paper, used in all figures unless otherwise noted. In \citet{Gruen2017} we will nevertheless apply both this and the two-parametric model to account for the possibility that the shot-noise of real galaxies behaves in a more complicated way than that of our simulated galaxies.}

\subsubsection{Shot-noise model 1: correlation $r\neq 1$ between galaxy density and matter density}
\label{sec:stochasticity}

In our fiducial model of $P(N_T| \delta_{m,T})$ we approximate the joint distribution of both $\delta_{m,T}$ and $\delta_{g,T}$ with a joint log-normal distribution (cf. eq. \ref{eq:lognormal_PDF} and \citet{Hilbert2011} for properties of joint log-normal distributions). The joint PDF of two log-normal random variables is characterized by 5 parameters, e.g. by the variance and skewness of each variable and the covariance between the two variables.

In our case, we compute the variance and skewness of $\delta_{m,T}$ as described in section \ref{sec:lognormal_for_PDF} and set the variance and skewness of $\delta_{g,T}$ to
\begin{align}
\label{eq:variance_and_skewness_relation}
\langle \delta_{g,T}^2 \rangle_c & = b^2 \langle \delta_{m,T}^2 \rangle_c \nonumber \\
\langle \delta_{g,T}^3 \rangle_c & = b^3 \langle \delta_{m,T}^3 \rangle_c 
\end{align}
where the \emph{galaxy bias} $b$ is a free parameter.
%\jab{[Does this serve to define $b$ for this model? If so, the assumption is that the same $b$ gives both the variance and the skewness, right?]}
The covariance of $\delta_{m,T}$ and $\delta_{g,T}$ is parametrized by their correlation coefficient
\begin{align}
r & = \frac{\langle \delta_{g,T}\ \delta_{m,T} \rangle_c}{\sqrt{\langle \delta_{g,T}^2 \rangle_c\ \langle \delta_{m,T}^2 \rangle_c }} \nonumber \\
& = \frac{\langle \delta_{g,T}\ \delta_{m,T} \rangle_c}{b\ \langle \delta_{m,T}^2 \rangle_c} \ ,
\end{align}
i.e.
\begin{equation}
\langle \delta_{g,T}\ \delta_{m,T} \rangle_c = r b\ \langle \delta_{m,T}^2 \rangle_c\ .
\end{equation}
The log-normal model for the joint PDF of $\delta_{m,T}$ and $\delta_{g,T}$ now allows us to compute the variance of galaxy counts as a function of $\delta_{m,T}$ and more generally to compute $P(N_T| \delta_{m,T})$. We present the necessary derivations in detail in appendix \ref{app:stochasticity}.

In our data analysis we consider $b$ and $r$ as free parameters. The only restrictions we impose on them are 
\begin{equation}
0 < b\ , \ 0 \leq r \leq 1\ .
\end{equation}
%\jab{[This upper limit on $r$ is justified in the case of counts-in-cells. When defined in terms of correlation functions, $r$ can be greater than 1. It might be useful to add a sentence about this.]}
To test how \bblue{accurately} this model describes the behaviour of our mock \textsc{redMaGiC} catalogs based on the Buzzard N-body simulations we nevertheless want to determine what values of $b$ and $r$ are underlying our simulations. To this end, we generate \textsc{healpix} maps of $\delta_{m,T}$ with different \bblue{top-hat aperture} radii $\theta_T=10,20,30$~arcmin, based on particle count maps at resolution $N_{\rm side}=8192$ in slices of comoving 50$h^{-1}$Mpc thickness. We co-add these maps to reproduce a redshift range close to that of our fiducial analysis, $z=0.2100\ldots0.4453$. We then select \textsc{redMaGiC} galaxies with true redshift in this range and determine their counts around the same \textsc{healpix} pixel centers and within the same aperture radii. \bblue{The \textsc{redMaGiC} mock catalogs have a complex mask similar to that of real DES data, which adds complication because the fraction of masked area in each aperture must be equal in order to meaningfully sort lines of sight by galaxy count. To this end,} we convert all counts to a masking fraction of 20 per-cent of area within the aperture radius using the stochastic method of \citet[][their section 2.1]{Gruen2017}. This leaves us with simulated 2D maps of $\delta_{m,T}$ and $N_T$ within a DES-Y1 shaped mask.

We can then measure the variances of these maps, $\mathrm{Var}(\delta_{m,T})$ and $\mathrm{Var}(N_T)$, as well as their covariance $\mathrm{Cov}(N_T, \delta_{m, T})$. These fulfill the relations
\begin{equation}
\mathrm{Var}(N_T) = \bar N + \bar N^2\ b^2 \mathrm{Var}(\delta_{m,T}) \ .
\end{equation}
and
\begin{equation}
\mathrm{Cov}(N_T, \delta_{m, T}) = \bar N\ b\ r\ \mathrm{Var}(\delta_{m,T})
\end{equation}
which then fixes $b$ and $r$. The values determined in this way are shown in table \ref{table:bg_r}.

\begin{figure*}
  \includegraphics[width=\textwidth]{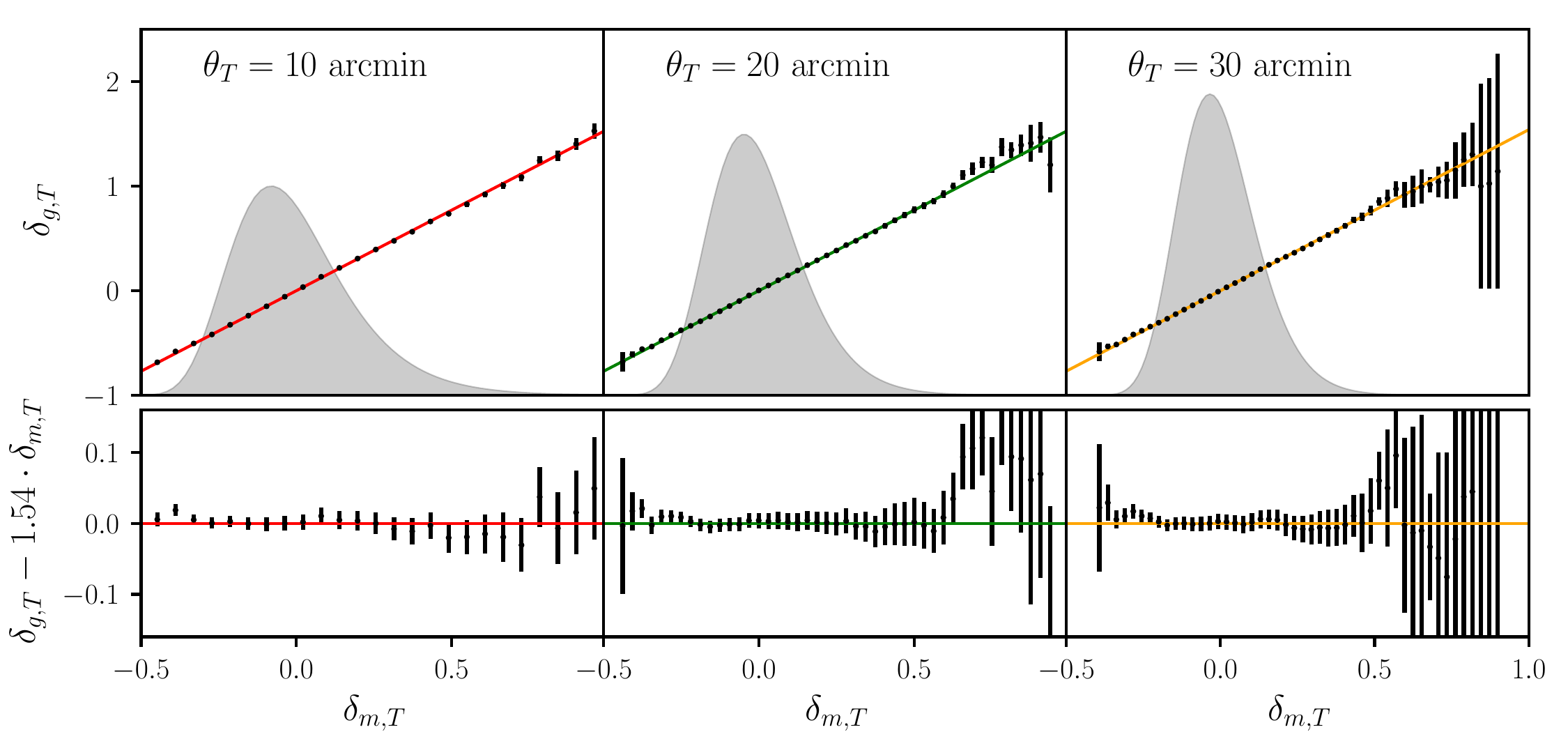}
   \caption{Average galaxy overdensity $\delta_{g,T}$ as a function of matter overdensity, $\delta_{m,T}$, for our simulated maps at different smoothing scales: 10 arcmin [left], 20 arcmin [middle] and 30 arcmin [right]. Solid lines show a linear bias model with the bias parameters obtained from maximizing the likelihood in equation \ref{eq:shot_noise_likelihood} and the residual between the two are shown in the bottom panels. Note that the coefficient of linearity found with \ref{eq:shot_noise_likelihood} ($\approx 1.54$) is almost identical to the value of the product $b\cdot r$ determined with equation \ref{eq:linear_approximation_of_joint_lognormal}. To indicate the range of $\delta_{m,T}$ that is relevant to our computation, we also show the density PDFs of figure \ref{fi:p_of_delta} as shaded regions. The errorbars were derived from a jackknife approach.}
  \label{fi:deltag_vs_deltam}
\end{figure*}

\begin{figure}
  \includegraphics[width=0.5\textwidth]{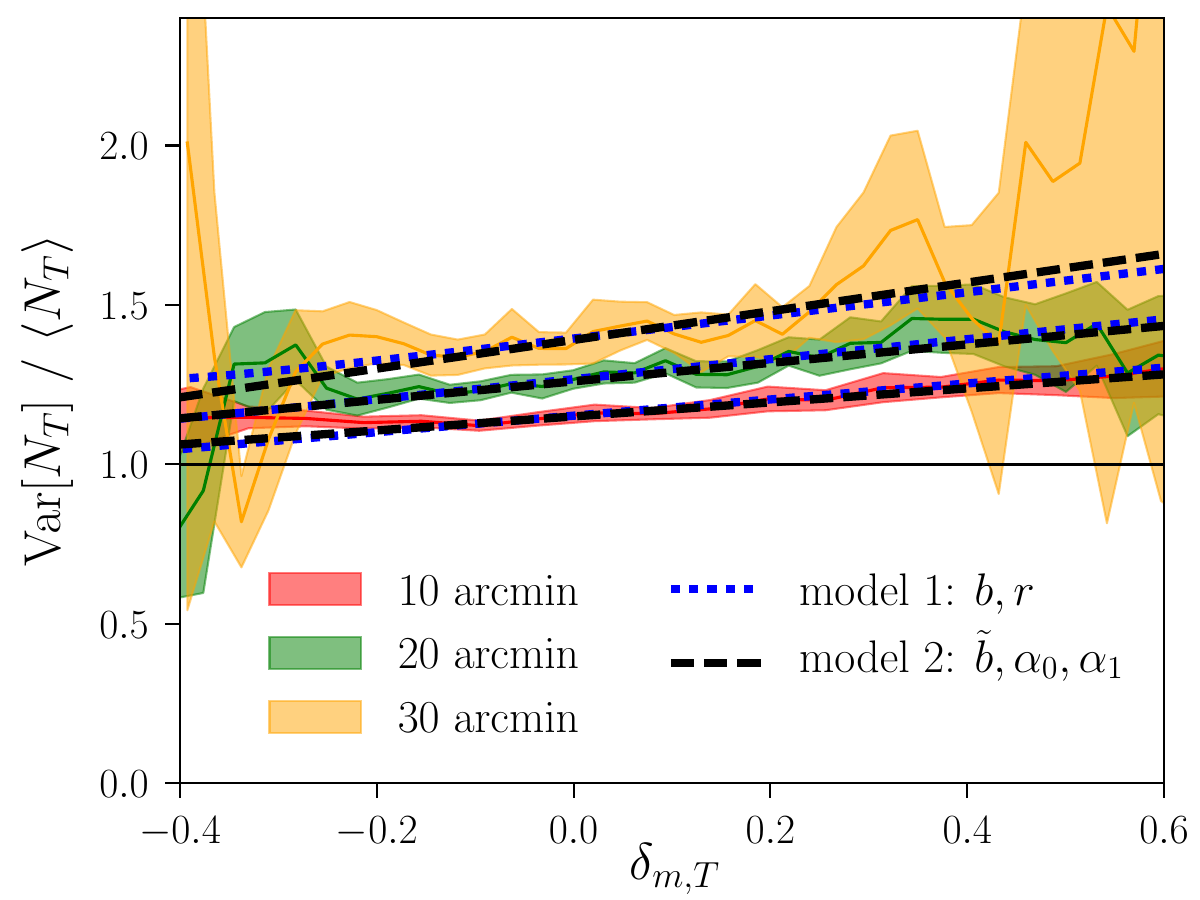}
   \caption{Ratio of the variance in the galaxy count distribution to average galaxy counts in our simulated maps as a function of matter overdensity. Differently coloured solid lines show the result for each smoothing scale. The dashed black lines show the predictions of the 2-parametric shot-noise model described in section \ref{sec:parametric_shot_noise}. The blue dotted lines show the corresponding prediction of the alternative, 1-parametric model described in section \ref{sec:stochasticity}. The horizontal solid line shows the expectation if shot noise was purely Poissonian. The coloured regions show the 95$\%$ confidence limits derived from jackknife resampling.
   %\jab{[Any idea for why the models work less well for underdense regions? Also, have you tried comparing to a $(b_1,b_2)$ model of galaxy biasing (or similar)?]}
   }
  \label{fi:subpoisson}
\end{figure}

\begin{figure}
  \includegraphics[width=0.5\textwidth]{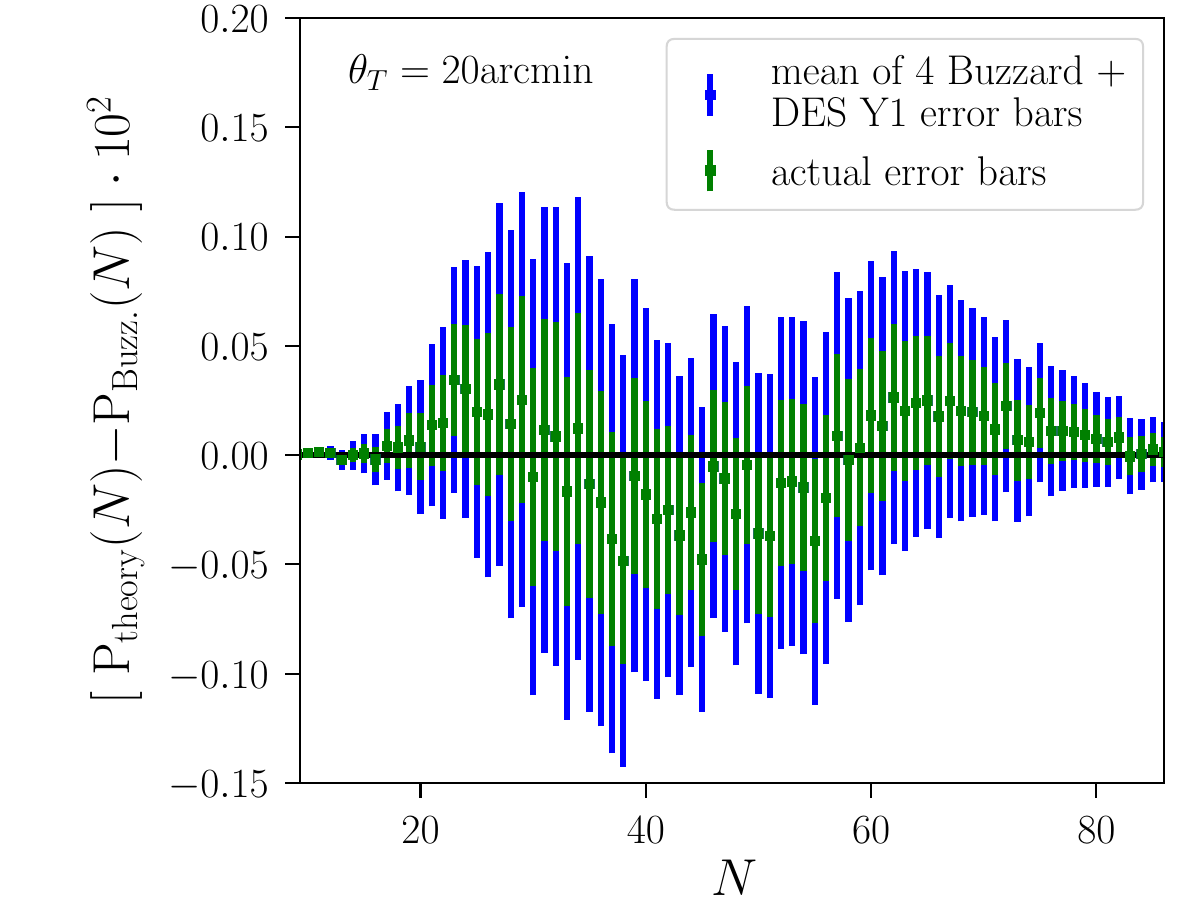}
   \caption{For the fiducial smoothing radius of our data analysis presented in \citet{Gruen2017}, $\theta_T=20$arcmin, we show the residuals between our baseline prediction of the counts-in-cells histogram and the average of measurements in 4 Buzzard realizations of DES Y1 data. Blue error bars represent the uncertainties we expect for DES Y1 while green error bars show the actual uncertainties of the mean measurement from our simulations.}
  \label{fi:pofn_residuals}
\end{figure}

We now need to check whether these value for $b$ and $r$ together with our assumption of a log-normal PDF for $\delta_{m,T}$ and $\delta_{g,T}$ describe the properties of our tracer galaxies well. Using our simulated maps of $\delta_{m,T}$ and $N_T$ we can measure the expectation value
\begin{equation}
\langle \delta_{g,T}|\delta_{m,T} \rangle = \frac{\langle N_T |\delta_{m,T} \rangle}{\bar N}-1
\end{equation}
as a function of $\delta_{m, T}$. Within the log-normal model (cf. appendix \ref{app:stochasticity} for the relevant formulae) this is very well approximated by
\begin{equation}
\label{eq:linear_approximation_of_joint_lognormal}
\langle \delta_{g,T}| \delta_{m,T} \rangle \approx r b\ \delta_{m,T}
\end{equation}
which becomes exact for Gaussian random variables. In Figure \ref{fi:deltag_vs_deltam} we compare measurements of $\langle \delta_{g,T}|\delta_{m,T} \rangle$ with the different smoothing radii $\theta_T=10,20,30$~arcmin to the prediction of the log-normal model. We find that in our simulations $\langle \delta_{g,T}|\delta_{m,T} \rangle$ is consistent with a linear relation in $\delta_{m,T}$. Interestingly, the scale dependence of $b$ and $r$ we find in table \ref{table:bg_r} almost perfectly cancels to give a scale independent proportionality coefficient
\begin{equation}
rb \approx 1.54\ .
\end{equation}
Next, we also measure the variance of galaxy counts $N_T$ as a function of $\delta_{m,T}$ in our simulated maps and compare to the prediction of the log-normal model. In Figure \ref{fi:subpoisson} we indeed find that
\begin{equation}
\frac{\mathrm{Var}\left[ N_T | \delta_{m,T} \right]}{\langle N_T | \delta_{m,T} \rangle} \neq 1
\end{equation}
and that the $\delta_{m,T}$-dependence is very well described by the log-normal model and the values of $b$ and $r$ we determined before. \blue{Finally, in figure \ref{fi:pofn_residuals} we show the residuals between our baseline prediction of the counts-in-cells histogram with $\theta_T=20$arcmin\footnote{This is the smoothing radius used in our data analysis \citep{Gruen2017}.} and the average of measurements in 4 Buzzard realizations of DES Y1 data (cf$.$ figure \ref{fi:quantile_illustration}). The residuals are well contained within DES Y1 errorbars.}

\begin{table}
\centering
\begin{tabular}{c|cc}
\hline\hline
Smoothing Scale & $b$ &  $r$ \\
$[$arcmin$]$ &  &  \\
\hline
10 & 1.644 $\pm$ 0.008 & 0.938 $\pm$ 0.001 \\
20 & 1.618 $\pm$ 0.008 & 0.956 $\pm$ 0.001 \\
30 & 1.603 $\pm$ 0.008 & 0.961 $\pm$ 0.001 \\
\hline\hline
\end{tabular}
\caption{Best-fit values galaxy bias and correlation coefficients of our simulated tracer galaxies within the model presented in section \ref{sec:stochasticity}. Error bars are estimated from a jackknife approach.}
\label{table:bg_r}
\end{table}

\begin{table}
\centering
\begin{tabular}{c|ccc}
\hline\hline
Smoothing Scale & $\tilde b$ &  $\alpha_0$ &  $\alpha_1$ \\
$[$arcmin$]$ &  &   &  \\
\hline
10 & 1.54 $\pm$ 0.001 & 1.15 $\pm$ 0.001 & 0.22 $\pm$ 0.003 \\
20 & 1.54 $\pm$ 0.002 & 1.26 $\pm$ 0.002 & 0.29 $\pm$ 0.010 \\
30 & 1.54 $\pm$ 0.002 & 1.39 $\pm$ 0.003 & 0.45 $\pm$ 0.020 \\
\hline\hline
\end{tabular}
\caption{Best-fit values galaxy bias and shot-noise parameters of our simulated tracer galaxies within the model presented in section \ref{sec:parametric_shot_noise}. Error bars are again estimated from a jackknife approach.}
\label{table:bg_a0_a1}
\end{table}

\subsubsection{Shot-noise model 2: Parametric model for super-Poissonianity}
\label{sec:parametric_shot_noise}

Our model of shot-noise based on galaxy bias $b$ and galaxy matter correlation coefficient $r$ describes our simulated tracer catalog well. But it contains the arbitrary assumption that both the variance and the skewness of $\delta_{m,T}$ and $\delta_{g,T}$ are related through the bias parameter $b$ (cf. equation \ref{eq:variance_and_skewness_relation}).
%\jab{[How arbitrary is this? I haven't thought too much about it, but is seems at least somewhat motivated.]}
To account for the possibility that real galaxies might behave in a more complicated way, we also consider a more flexible model of the conditional distribution $P(N_T| \delta_{m,T})$.

\citet{Gruen2016} assumed that there is no stochasticity in the relation of $\delta_{m,T}$ and $\delta_{g,T}$ and that galaxies trace the matter density with a linear bias and Poissonian shot noise. This means they set
\begin{eqnarray}
P(N_T = N | \delta_{m,T} = s) = \nonumber \\
\exp\left(N\ln[\bar N (1 + \tilde b s)]-[\bar N (1 + \tilde b s)]-\ln \Gamma(N+1)\right)\ ,\nonumber \\
\end{eqnarray}
where $\tilde b$ is the galaxy bias, and where we now use a generalizable definition of the Poisson distribution based on the Gamma function $\Gamma$, for reasons that will appear later. The galaxy bias $\tilde b$ defined this way is not identical to the definition in our fiducial model. We now rather have
\begin{equation}
\delta_{g,T} \equiv \tilde b\ \delta_{m,T}\ .
\end{equation}
%\jab{[This is more similar to the standard definition of $b$.]}
We test this in our simulated maps of $N_T$ and $\delta_{m,T}$ by fitting a linear relation to the mean smoothed galaxy contrast as a function of dark matter contrast that was shown in Figure \ref{fi:deltag_vs_deltam}. We indeed find that this linear biasing model describes the simulations very well and that $\tilde b \approx br$ as expected from our arguments in section \ref{sec:stochasticity}.

The model used by \citet{Gruen2016} however predicts that
\begin{equation}
\frac{\mathrm{Var}\left[ N_T | \delta_{m,T} \right]}{\langle N_T | \delta_{m,T} \rangle} \equiv 1
\end{equation}
which is not what we find in Figure \ref{fi:subpoisson}. To account for the deviations we observe from pure Poissonian shot-noise, we hence model the distribution of $N_T$ given $\delta_{m,T}$ as
\begin{eqnarray}
\label{eqn:pgeneral}
P(N_T = N | \delta_{m,T} = s) = \mathcal{N} \times \ \ \ \ \ \ \nonumber \\
\exp\left\lbrace \frac{N}{\alpha} \ln\left[\frac{\bar N_T}{\alpha} (1 + b s)\right] - \ln\Gamma\left(\frac{N}{\alpha}+1\right) - \frac{\bar N_T}{\alpha} (1 + b s) \right\rbrace\ ,\nonumber \rule[-1.5em]{-30pt}{0pt} \\ 
\end{eqnarray}
where the parameter $\alpha>0$ generalizes the distribution to one where groups of $\alpha$ galaxies appear with Poissonian noise and where the normalization coefficient $\mathcal{N}$ is needed to ensure that $\int P(N_T=N) \, \mathrm{d}N = 1$. We find $\mathcal{N}$ to be very close to $\alpha^{-1}$ and identical to $\alpha^{-1}$ in the case where $\alpha$ is an integer value.

To account for the observed increase of super-Poissonianity with density, we allow $\alpha$ to depend on $\delta_{m,T}$, 
\begin{equation}
\alpha(\delta_{m,T}) = \alpha_0 + \alpha_1 \times \delta_{m,T} \; .
\end{equation}
This indeed leads to a $\delta_{m,T}$-dependence of the variance of galaxy counts that is close to the relation mentioned in eq. \ref{eq:Var_N_ratio}. \blue{In our analysis we treat $\alpha_0$ and $\alpha_1$ as free parameters within the ranges
\begin{eqnarray}
\alpha_0 &\in& [0.1,3.0]\nonumber\\
\alpha_1 &\in& [-1.0,4.0]\ .
\end{eqnarray}
In principle we could allow any value $\alpha_0 > 0$ but we choose the boundary $0.1 < \alpha_0$ because it is numerically difficult (and slow) to predict the CiC histogram for values close to $\alpha_0 = 0.0$. The other boundaries roughly enclose the 2-$\sigma$ region of the posterior distribution of $\alpha_0$ and $\alpha_1$ we infer with DES Y1 like errors around the mean signal measured in Buzzard (after marginalizing over our other model parameters, cf$.$ appendix \ref{app:validate_alternative_model}). Also, the contraints on $\alpha_0$ and $\alpha_1$ we derive in \cite{Gruen2017} on DES data are well contained within our prior distributions.}

\blue{Nevertheless, these priors must be considered mildly informative. We expect that even stonger priors can be motivated theoretically. \cite{Baldauf2013} find that for their most massive halos shot-noise is reduced wrt$.$ Poisson expectation by a factor of $\approx 2$, indicating that $\alpha_0 \gtrsim 0.5$, while for halo masses comparable to redmaGiC halo masses (cf$.$ \citet{Clampitt2017}) they find shot-noise to be close to Poissonian. Also, there is evidence that the fraction of red galaxies that are satellites (resp. the fraction of satellite galaxies that are red) increases with environment density (see e.g$.$ \cite{Mandelbaum2006, Peng2012}). According to \cite{Baldauf2013} this will cause an increase of galaxy stochasticity with environment density, corresponding to $\alpha_1 > 0.0$. We intend to investigate implications of models for halo occupation distributions (HOD) on our shot-noise parametrizations in future studies (see e.g$.$ the work by \cite{Cacciato2012, Dvornik2018} on connecting HOD models and parametric models of galaxy bias and stochasticity).}

To compare this parametric shot-noise model to our simulations, we are nevertheless interested in the particular value of $\alpha_0$ and $\alpha_1$ that describe these simulations. From the tuples of $(N_T, \delta_{m,T})_i$ measured in our simulated maps, we can constrain bias and the $\alpha_{0/1}$ parameters with a likelihood $L$ that is simply the product of the probabilities of the individual tuples from \autoref{eqn:pgeneral},
\begin{equation}
\label{eq:shot_noise_likelihood}
\begin{split}
\ln L = \sum_i & \left[N_{i}/\alpha(\delta_{m,T}^{i})\right] \ln \left[ (\bar{N}_{i}/\alpha(\delta_{m,T}^{i}))(1+b\times\delta_{m,T}^{i}) \right] \\ - & \left[\bar{N}_{i}/\alpha(\delta_{m,T}^{i}) \right](1+b \times \delta_{m,T}^{i}) \\ - & \ln \Gamma \left[ N_{i}/\alpha(\delta_{m,T}^{i})+1\right] - \ln\alpha(\delta_{m,T}^{i}) \; .
\end{split}
\end{equation}
Because the tuples have correlated counts and densities, this is not an exact expression for the likelihood of our measurements, but it should be sufficient to obtain reasonable best fit values for $b$, $\alpha_0$ and $\alpha_1$. We determine the uncertainties of these best-fit values by finding the maximum of~\autoref{eq:shot_noise_likelihood} on jackknife resamplings of the simulations. The resulting parameter values are shown in~\autoref{table:bg_a0_a1} and displayed in \autoref{fi:deltag_vs_deltam} and \autoref{fi:subpoisson}. We find that our simulated \textsc{redMaGiC} galaxies are indeed well described as linearly biased tracers of the density field with a small, but significant, scale and density dependent super-Poissonian shot-noise.

\subsection{Summary of fiducial model and approximations therein}
\label{sec:model_summary}

For each ingredient (i) to (iii) of the framework described in~\autoref{sec:cosmology} we have introduced at least two different modeling ansatzes. We want to once more describe our baseline model \bblue{built} from these ansatzes (cf. also section \ref{sec:modeling_DT}). This is the model we consider in~\autoref{sec:log-normal} and which we use in the data analysis presented in \citet{Gruen2017}.
\begin{trivlist}
\item[(i)] \underline{$p(\delta_{m,T})$:} We find that the log-normal model (section \ref{sec:lognormal_for_PDF}) and our model based in cylindrical collapse (section \ref{sec:Tree_level_for_PDF}) describe the PDF of projected density contrast equally well. The computations based on the log-normal model are however significantly faster. Hence in our fiducial analysis we employ the log-normal model.
\\
\item[(ii)] \underline{$\langle\kappa_{<\theta} | \delta_{m,T}\rangle$:} We also introduced a log-normal model (section \ref{sec:lognormal_for_kappa}) and a model based on cylindrical collapse (section \ref{sec:cyl_collapse_for_kappa}) for the convergence profile around lines-of-sight with fixed density contrast $\delta_{m,T}$. Both models lead to almost identical predictions for the density split lensing signal. Hence we again choose the log-normal model for our fiducial analysis, because of the shorter computation time.
\\
\item[(iii)] \underline{$P(N_T | \delta_{m,T})$:} We introduced two models for the distribution of tracer counts $N_T$ in lines-of-sight of matter density $\delta_{m,T}$. The first was based on linear galaxy bias $b$ and galaxy-matter-correlation coefficient $r$ (section \ref{sec:stochasticity}). The second was based on an alternative definition of galaxy bias and on two parameters $\alpha_0$ and $\alpha_1$ describing density dependent deviations from Poissonian shot-noise (section \ref{sec:parametric_shot_noise}). Both models describe the behaviour of our simulated tracer galaxies \bblue{in Buzzard-v1.1} similarly well. But anticipating that real galaxies might behave in a more complicated way, we will consider both ansatzes in our fiducial analysis.
\end{trivlist}
\noindent In the following list, we are summarizing the approximations that went into the derivation of our baseline model.
\begin{trivlist}
\item[1.)] We assumed, that for fixed value of $\delta_{m,T}$ the convergence within angular radius $\theta$ is not dependent on $N_T$ (cf. equation \ref{eq:trough_radius_approx}).
\item[2.)] All second order moments in our formalism are computed with a \emph{halofit} power spectrum \citep{Takahashi2012} using an analytic approximation for the transfer function \citep{Eisenstein1998}.
\item[3.)] Equations \ref{eq:Limber_projection_connected_moment}, \ref{eq:Limber_projection_generating_function}, \ref{eq:Limber_projection_generating_function_2_variables} and \ref{eq:Limber_projection_G} employ a small angle and a Limber-like approximation \citep[following][]{Bernardeau2000}.
\item[4.)] We compute the cumulant generating function of density contrast in long cylinders by means of the cylindrical collapse approximation (cf. section \ref{sec:Tree_level_for_PDF}).
\item[5.)] We assume that the tree-level result of the cumulant generating function can be corrected for the full non-linear evolution of the density field by means of equations \ref{eq:Bernardeau_rescaling} and \ref{eq:phi_rescaling}.
\item[6.)] We approximate the PDF of $\delta_{m,T}$ resulting from a cylindrical collapse approximation by a log-normal PDF (cf. section \ref{sec:lognormal_for_PDF}).
\item[7.)] We employed approximations similar to 4.), 5.) and 6.) for the joint distribution of $\delta_{m,T}$ and $\kappa_{<\theta}$ (cf. section \ref{sec:cyl_collapse_for_kappa}, equation \ref{eq:rescaling_mixed_moments} and section \ref{sec:lognormal_for_kappa}).
\item[8.)] We assume that galaxies are linearly biased tracers of the density field. We consider two different models for shot-noise (resp. stochasticity), assuming that the full distribution $P(N_T|\delta_{m,T})$ is well described bei either two parameters ($b,r$) or three parameters ($\tilde b, \alpha_0, \alpha_1$).
\end{trivlist}

\noindent Despite this long list of approximations, this baseline model describes our measurements in the Buzzard simulations well within DES Y1 errorbars (cf. figure \ref{fi:quantile_illustration}). As shown in the next section, the model is also accurate enough to recover the true cosmology of our simulation within DES Y1 uncertainties in a simulated likelihood analysis. In \citet{Gruen2017} (and using an extended set of simulations) we furthermore show that the values of $\chi^2$ found between our fiducial model and individual simulation measurements are consistent with the $\chi^2$-distribution expected from our number of data points, and that the coverage (i.e.~the fraction of times the true simulation cosmology is within the confidence interval) matches expectations.

\section{Recovering cosmology in N-body simulations}
\label{sec:log-normal}

In this section we want to test whether the modeling that was described in sections \ref{sec:cosmology} and \ref{sec:model} is sufficient to recover the cosmology underlying a density split data vector measured in N-body simulations. The simulations we use are described in section \ref{sec:buzzard_description}. They are the same simulations against which we tested the ingredients of our model in the previous section. A likelihood analysis based on a density split data vector measured in these simulations in presented in section \ref{sec:buzzard_likelihood}. We only run a cosmological analysis on the mean data vector measured on 4 DES-Y1 realisations. The goal of this is to show that any possible systematic deviations between our modeling of density split statistics and the behaviour of our N-body simulations is smaller than the statistical uncertainties of DES-Y1. A more extensive validation of our likelihood pipeline is presented in \citet{Gruen2017}.

\begin{figure}  
\begin{centering}
  \includegraphics[width=0.48\textwidth]{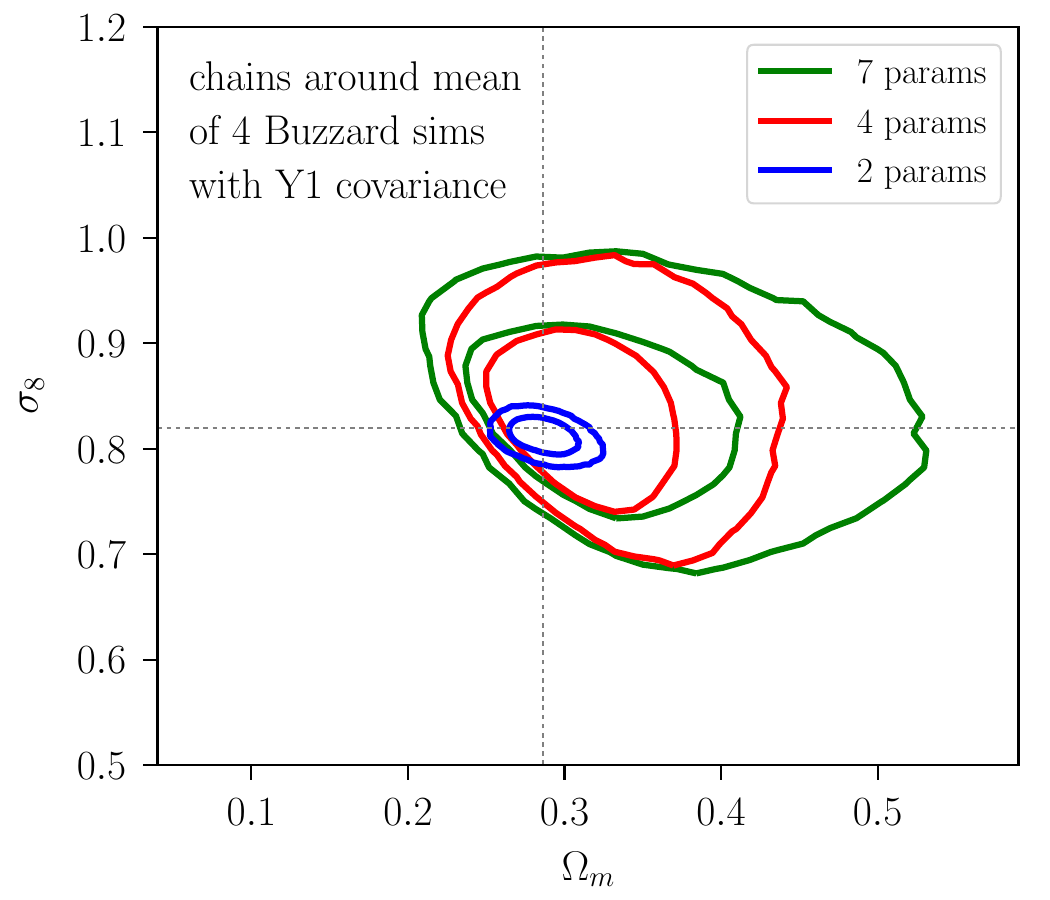}\\\includegraphics[width=0.48\textwidth]{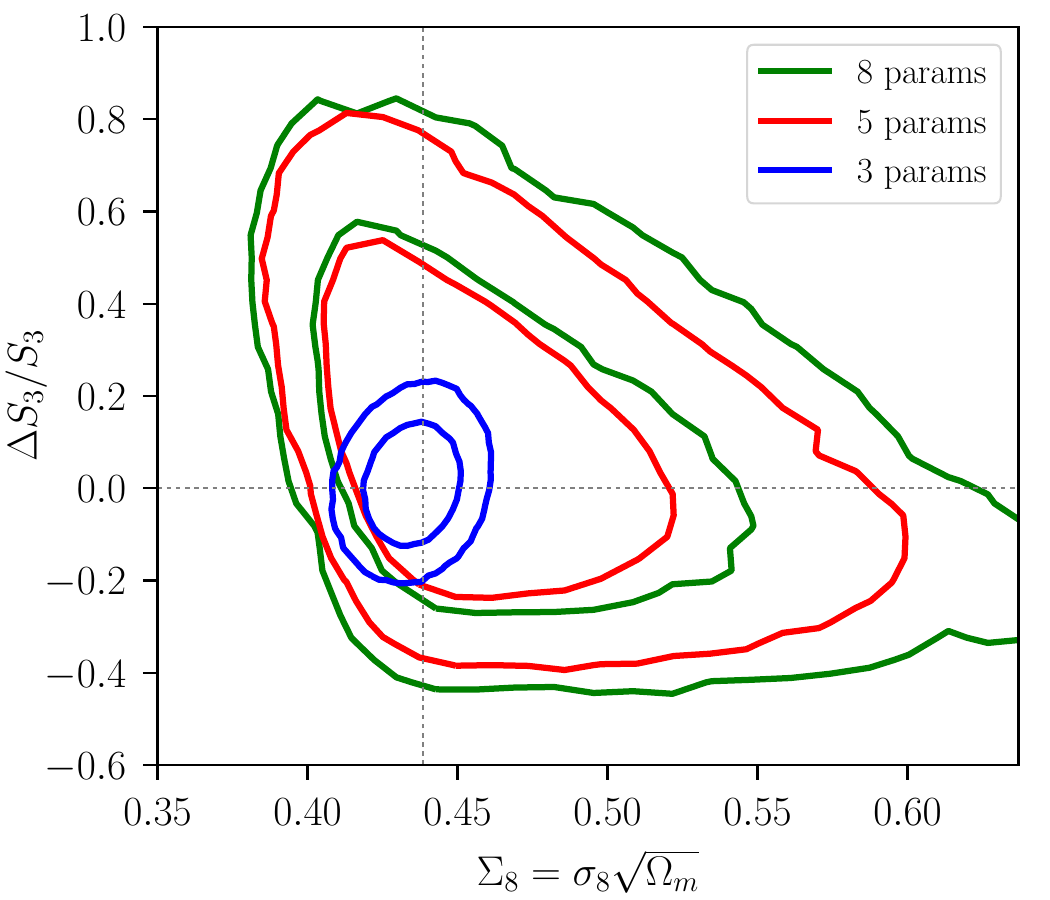}
  %\cprotect
  \caption{To test our model for possible systematic deviations from N-body simulations, we try to recover the Buzzard cosmology in a simulated likelihood analysis. \underline{Top panel:} $1\sigma$ and $2\sigma$ contours in the $\Omega_m$-$\sigma_8$ plane from a likelihood computed around the mean of 4 shape noise free realisations of DES Y1 (but assuming the full covariance matrix for a single DES Y1). The green contours are marginalized over $\Omega_b$, $n_{s}$, $h_{100}$, redMaGiC galaxy bias $b$ and galaxy-matter correlation coefficient $r$. For the parameters $\Omega_b$, $n_{s}$, $h_{100}$ we have assumed the same flat priors as used in the DES Y1 combined probes analysis presented in \citet{DES2017_short}. The red contours are marginalized only over $b$ and $r$ and the blue contours only vary $\Omega_m$ and $\sigma_8$. Even when going to this small parameter space, our model agrees with Buzzard within 1$\sigma$ errors of DES Y1. \underline{Bottom panel:} Same contours but in the $\Sigma_8$-$\Delta S_3 / S_3$ plane and varying one additional parameter, $\Delta S_3 / S_3$.}
  \label{fi:recover_cosmology}
  \end{centering}
\end{figure}

\subsection{Simulated likelihood analysis}
\label{sec:buzzard_likelihood}

We now measure the data vector that was described in section \ref{sec:using_DT} in 4 N-body realisations of DES-Y1. We always use the mean of these 4 data vectors. In order to further reduce the noise of this measurement, we turn off the shape noise in our simulated source catalogs, i.e.~we measure our signal directly from the gravitational shear acting each galaxy. We then run Monte-Carlo Markov Chains of our model around this data vector. For this we assume a Gaussian likelihood function with the full covariance (i.e. including shape noise) that was estimated by \citet{Gruen2017} for a DES-Y1 data set. The goal of this analysis is to test whether the fiducial cosmology of the Buzzard simulations is well contained within the 1$\sigma$ contraints derived from this likelihood analysis. A more extensive validation of our likelihood pipeline is presented in \citet{Gruen2017}.

In the top panel of Figure \ref{fi:recover_cosmology} we show the 1$\sigma$ and 2$\sigma$ contraints obtained from our simulated likelihood analysis in the $\Omega_m$-$\sigma_8$ plane after marginalizing over different sets of parameters. First, we only vary $\Omega_m$ and $\sigma_8$, \green{setting other cosmological parameters to the inputs of the Buzzard simulations and the parameters connecting galaxy count and matter density to the values we found from the Buzzard galaxy and density maps}. Note that those values would be inaccessible in a real measurement. The corresponding constraints are very tight, but the fiducial values of our parameters are still well contained in the 1$\sigma$ contour. Then, we also marginalize over the galaxy bias $b$ and the galaxy stochasticity $r$, demanding that $0 < r \leq 1$. The contours now widen, and the fiducial values of $\Omega_m$ and $\sigma_8$ are still located well within the corresponding 1$\sigma$ contour. Finally, we also marginalize over $\Omega_b$, $n_s$ and $h_{100}$, using the same informative priors that have been used in the DES-Y1 combined probes analysis \citep{DES2017_short}. The contours widen further, but our model and our simulations still agree well within 1$\sigma$ uncertainty.

In the bottom panel of Figure \ref{fi:recover_cosmology} we repeat this analysis, but now also vary the parameter $\Delta S_3 / S_3$ that was introduced in section \ref{sec:using_DT}. This parameter allows for deviations of the 3-point statistics of the density field from our fiducial model. Within our statistical uncertainties we find that the scaling between 3-point and 2-point statistics in our simulations is well described by our fiducial assumptions ($\Delta S_3 / S_3 = 0$).

\blue{We repeat this analysis with our alternative shot-noise model in appendix \ref{app:validate_alternative_model}.}

\section{Discussion \& conclusions}
\label{sec:discussion}

In this work we introduced density split statistics, a technique to separately measure contributions to weak lensing and counts-in-cells from regions of different foreground galaxy density. Based on the pioneering work of \citet{Bernardeau1994}, \citet{Bernardeau2000} and \citet{Valageas2002} (see also references therein) on modeling the cosmic density PDF we were able to model the density split lensing signal as well as the counts-in-cells histogram from basic principles. With the help of this model, we then showed that density split statistics has two features that make it a potentially powerful cosmological probe:
\begin{trivlist}
\item[$\bullet$] it is able to constrain the cosmological parameters $\Omega_m$ and $\sigma_8$ even if the relation of galaxy density and matter density is assumed to have 2 degrees of freedom: galaxy bias and galaxy-matter-correlation coefficient,
\item[$\bullet$] it is able to constrain the amplitude of 3-point statistics of the density field with almost no degeneracy to constraints on the amplitude of 2-point statistics.
\end{trivlist}
In our fiducial model we predict 3-point statistics from cosmological perturbation theory. Deviations from that fiducial prediction may hint to non-standard physics, that affect overdense and underdense parts of the matter field differently, or to any non-linear dynamics or small scale physics that break the scaling relations of $\Lambda$CDM perturbation theory. We showed that a DES-Y5 data set combined with data form the cosmic microwave background can measure the amplitude of 3-points statistics to a 1$\sigma$ accuracy of $\lesssim 5\%$. This is a conservative estimate since our projections neglect the fact that DES-Y5 will be a deeper data set than DES-Y1. Also, we so far neglected the possibility of a combined analysis including density split statistics and measurements of 2-point correlation functions.

Using measurements in high-resolution N-body simulations we showed that our model of the density split lensing signal and the counts-in-cells histogram is accurate to well within the statistical uncertainties of the DES-Y1 data set. Especially, in a mock likelihood analysis we were able to recover the input cosmology of our simulations to well within DES-Y1 parameter errors. Cosmological constraints from DES-Y1 data based on density split statistics are presented in \citet{Gruen2017}.

%%%%%%%%%%%%%%%%%%%%%%%%%%%%%%%%%%%%%%%%%%%%%%%%%%%%%%%
\section*{Acknowledgments}
%%%%%%%%%%%%%%%%%%%%%%%%%%%%%%%%%%%%%%%%%%%%%%%%%%%%%%%

OF was supported by SFB-Transregio 33 `The Dark Universe'  by  the  Deutsche  Forschungsgemeinschaft (DFG). Support for DG was provided by NASA through Einstein Postdoctoral Fellowship grant number PF5-160138 awarded by the  Chandra  X-ray Center, which is operated by the Smithsonian Astrophysical Observatory  for NASA under contract NAS8-03060. OF and SH acknowledge support by the DFG cluster of excellence \lq{}Origin and Structure of the Universe\rq{} (\href{http://www.universe-cluster.de}{\texttt{www.universe-cluster.de}}). Part of our computations have been carried out on the computing facilities of the Computational Center for Particle and Astrophysics (C2PAP).

This paper has gone through internal review by the DES collaboration. We want to thank all the members of the DES WL, LSS and Theory working groups that have contributed with helpful comments and discussions. We also want to thank the anonymous journal referee for very helpful comments.

Funding for the DES Projects has been provided by the U.S. Department of Energy, the U.S. National Science Foundation, the Ministry of Science and Education of Spain, 
the Science and Technology Facilities Council of the United Kingdom, the Higher Education Funding Council for England, the National Center for Supercomputing 
Applications at the University of Illinois at Urbana-Champaign, the Kavli Institute of Cosmological Physics at the University of Chicago, 
the Center for Cosmology and Astro-Particle Physics at the Ohio State University,
the Mitchell Institute for Fundamental Physics and Astronomy at Texas A\&M University, Financiadora de Estudos e Projetos, 
Funda{\c c}{\~a}o Carlos Chagas Filho de Amparo {\`a} Pesquisa do Estado do Rio de Janeiro, Conselho Nacional de Desenvolvimento Cient{\'i}fico e Tecnol{\'o}gico and 
the Minist{\'e}rio da Ci{\^e}ncia, Tecnologia e Inova{\c c}{\~a}o, the Deutsche Forschungsgemeinschaft and the Collaborating Institutions in the Dark Energy Survey. 
The DES data management system is supported by the National Science Foundation under Grant Number AST-1138766.

The Collaborating Institutions are Argonne National Laboratory, the University of California at Santa Cruz, the University of Cambridge, Centro de Investigaciones En{\'e}rgeticas, 
Medioambientales y Tecnol{\'o}gicas-Madrid, the University of Chicago, University College London, the DES-Brazil Consortium, the University of Edinburgh, 
the Eidgen{\"o}ssische Technische Hochschule (ETH) Z{\"u}rich, 
Fermi National Accelerator Laboratory, the University of Illinois at Urbana-Champaign, the Institut de Ci{\`e}ncies de l'Espai (IEEC/CSIC), 
the Institut de F{\'i}sica d'Altes Energies, Lawrence Berkeley National Laboratory, the Ludwig-Maximilians Universit{\"a}t M{\"u}nchen and the associated Excellence Cluster Universe, 
the University of Michigan, the National Optical Astronomy Observatory, the University of Nottingham, The Ohio State University, the University of Pennsylvania, the University of Portsmouth, 
SLAC National Accelerator Laboratory, Stanford University, the University of Sussex, and Texas A\&M University.

The DES participants from Spanish institutions are partially supported by MINECO under grants AYA2012-39559, ESP2013-48274, FPA2013-47986, and Centro de Excelencia Severo Ochoa SEV-2012-0234.
Research leading to these results has received funding from the European Research Council under the European Union’s Seventh Framework Programme (FP7/2007-2013) including ERC grant agreements 
 240672, 291329, and 306478.

\bibliographystyle{mnras}
\bibliography{main}

\appendix 

\section{Friedmann equations, linear growth, spherical collapse and cylindrical collapse}
\label{app:differential_equations}

Throughout this section we set $G = 1 = c$ and we assume a flat $\Lambda CDM$ universe. In proper co-moving time $t$ the Friedmann equations take the form
\begin{eqnarray}
H^2 &=& \frac{8\pi}{3}\left( \bar \rho_m + \bar \rho_\Lambda\right) \\
\frac{\mathrm d H}{\mathrm d t} + H^2 &=& -\frac{4\pi}{3}\left( \bar \rho_m - 2\bar \rho_\Lambda\right)\ ,
\end{eqnarray}
where $\smash{H = \frac{\mathrm d}{\mathrm d t} \ln a}$. In conformal time, defined by $\mathrm d t = a \mathrm d \tau$, this changes to
\begin{eqnarray}
\mathcal{H}^2 &=& \frac{8\pi}{3}a^2\left( \bar \rho_m + \bar \rho_\Lambda\right) \\
\frac{\mathrm d \mathcal H}{\mathrm d \tau} &=& -\frac{4\pi}{3}a^2\left( \bar \rho_m - 2\bar \rho_\Lambda\right)\ ,
\end{eqnarray}
where $\smash{\mathcal{H} = \frac{\mathrm d}{\mathrm d \tau} \ln a}$. We will from now put $\smash{\frac{\mathrm d}{\mathrm d \tau} \equiv \dot{}\ }$.

In the Newtonian approximation, i.e. on scales much smaller that the curvature horizon of the universe, the evolution of a spherical, cylindrical or planar perturbation $\delta$ is given by the equation
\begin{equation}
\frac{\mathrm d^2 \delta}{\mathrm d t^2} + 2 H \frac{\mathrm d \delta}{\mathrm d t} - \frac{N+1}{N} \frac{1}{1+\delta} \left(\frac{\mathrm d \delta}{\mathrm d t}\right)^2 = 4\pi \bar \rho_m \delta (1+\delta)\ , 
\end{equation}
where $N=3$ for a spherical perturbation, $N=2$ for a cylindrical perturlation and $N=1$ for a planar perturbation (see \citet{MukhanovBook} where this is demonstrated for $N=3$ and $N=1$).

In conformal time this equation reads
\begin{equation}
\ddot{\delta} + \mathcal{H} \dot{\delta} - \frac{N+1}{N} \frac{\dot{\delta}^2}{1+\delta} \ = 4\pi \bar \rho_m a^2 \delta (1+\delta)\ .
\end{equation}
To linear order in $\delta$ this becomes
\begin{equation}
\ddot{\delta} + \mathcal{H} \dot{\delta} \ = 4\pi \bar \rho_m a^2 \delta \ ,
\end{equation}
which is indeed independent of the particular shape of the perturbation.  

%\begin{strip}
\begin{widetext}

%\parbox{\textwidth}{
%\strip{\textwidth}{

\section{$\Lambda$CDM perturbation theory}
\label{app:PT_for_skewness}

In the following we are using the convention that the Fourier transform of a real space function $f(\mathbf{x})$ is given by 
\begin{equation}
\tilde f(\mathbf{k}) = \int \frac{\mathrm{d}^3x}{(2\pi)^3} f(\mathbf{x})\ e^{-i\mathbf{x}\mathbf{k}}\ .
\end{equation}
Consider the matter density contrast $\delta$ and the divergence of the velocity field $\theta = \boldsymbol{\nabla} \mathbf{v}$. In the Newtonian approximation the Fourier space equations of motion of $\delta$ and $\theta$ are \citep[cf.][]{Bernardeau}
\begin{eqnarray}
\label{eq:eom_for_PT}
\frac{\partial \tilde \delta(\mathbf{k}, \tau)}{\partial \tau} + \tilde\theta(\mathbf{k}, \tau) &=& -\int \mathrm{d}^3k_1\mathrm{d}^3k_2\ \delta_D(\mathbf{k} - \mathbf{k}_{12})\ \alpha(\mathbf{k}_1, \mathbf{k}_2)\ \tilde\delta(\mathbf{k}_1, \tau)\ \tilde\theta(\mathbf{k}_2, \tau)\nonumber \\
\frac{\partial \tilde \theta(\mathbf{k}, \tau)}{\partial \tau} + \mathcal{H}\ \tilde\theta(\mathbf{k}, \tau) + \frac{3\Omega_m^0H_0^2}{2a}\tilde\delta(\mathbf{k}, \tau) &=& -\int \mathrm{d}^3k_1\mathrm{d}^3k_2\ \delta_D(\mathbf{k} - \mathbf{k}_{12})\ \beta(\mathbf{k}_1, \mathbf{k}_2)\ \tilde\theta(\mathbf{k}_1, \tau)\ \tilde\theta(\mathbf{k}_2, \tau)\ ,
\end{eqnarray}
where $\mathbf{k}_{12} = \mathbf{k}_1 + \mathbf{k}_2$ and $\alpha$ and $\beta$ are given by
\begin{eqnarray}
\alpha(\mathbf{k}_1, \mathbf{k}_2) &=& 1 + \frac{1}{2} \frac{\mathbf{k}_1\cdot \mathbf{k}_2}{k_1k_2} \left(\frac{k_1}{k_2} + \frac{k_2}{k_1}\right) \nonumber \\
\beta(\mathbf{k}_1, \mathbf{k}_2) &=& \frac{1}{2} \frac{\mathbf{k}_1\cdot \mathbf{k}_2}{k_1k_2} \left(\frac{k_1}{k_2} + \frac{k_2}{k_1}\right) + \frac{(\mathbf{k}_1\cdot \mathbf{k}_2)^2}{k_1^2k_2^2} \ .
\end{eqnarray}
In the following we will abbreviate the integrals involving $\alpha$ and $\beta$ as $\alpha[\tilde\delta, \tilde\theta, \mathbf{k}]$ and $\beta[\tilde\theta, \tilde\theta, \mathbf{k}]$. In perturbation theory we write $\tilde \delta$ and $\tilde \theta$ as
\begin{equation}
\tilde \delta(\mathbf{k}, \tau) = \sum_{n=1}^{\infty} \delta_n(\mathbf{k}, \tau)\ \ \mathrm{and}\ \ \tilde \theta(\mathbf{k}, \tau) = -\frac{\partial \ln D_+(\tau)}{\partial \tau}\sum_{n=1}^{\infty} \theta_n(\mathbf{k}, \tau)\ ,
\end{equation}
where $\delta_n$ and $\theta_n$ are of order $n$ in the linearly approximated fields $\delta_1$ and $\theta_1$ and $D_+(\tau)$ is the linear growth factor. (We will ignore the decaying mode of linear growth here.) At linear order we have
\begin{equation}
\delta_1(\mathbf{k}, \tau) = \theta_1(\mathbf{k}, \tau) = \frac{D_+(\tau)}{D_+(\tau_0)}\delta_1(\mathbf{k}, \tau_0)\equiv D_+(\tau)\delta_{1,1}(\mathbf{k})\ ,
\end{equation}
where we have assumed that $D_+(\tau_0) = 1$ at present time $\tau_0$ and introduced the notation $\delta_{1,1}(\mathbf{k}) = \delta_1(\mathbf{k}, \tau_0)$ whose purpose will become clear at the end of this section. To get $\tilde\delta$ at second order we first note that
\begin{equation}
\frac{\partial \tilde \theta(\mathbf{k}, \tau)}{\partial \tau} + \mathcal{H}\ \tilde\theta(\mathbf{k}, \tau) = \frac{1}{a}\frac{\partial}{\partial \tau}\left(a\tilde \theta(\mathbf{k}, \tau)\right)\ .
\end{equation}
Hence, multiplying the first of equations \ref{eq:eom_for_PT} with $a$ and differentiating wrt. $\tau$ and then multiplying with $1/a$ we get
\begin{equation}
\frac{\partial^2 \tilde \delta(\mathbf{k}, \tau)}{\partial^2 \tau} + \mathcal{H}\frac{\partial \tilde \delta(\mathbf{k}, \tau)}{\partial \tau} + \frac{\partial \tilde \theta(\mathbf{k}, \tau)}{\partial \tau} + \mathcal{H}\ \tilde\theta(\mathbf{k}, \tau) = -\ \alpha\left[\frac{\partial\tilde\delta}{\partial \tau}, \tilde\theta, \mathbf{k}\right]-\alpha\left[\tilde\delta, \frac{\partial\tilde\theta}{\partial \tau}, \mathbf{k}\right]-\mathcal{H}\ \alpha\left[\tilde\delta, \tilde\theta, \mathbf{k}\right]\ .
\end{equation}
Now the second of equations \ref{eq:eom_for_PT} can be used to eliminate $\tilde\theta$ from the right-hand-side, giving
\begin{equation}
\frac{\partial^2 \tilde \delta(\mathbf{k}, \tau)}{\partial^2 \tau} + \mathcal{H}\frac{\partial \tilde \delta(\mathbf{k}, \tau)}{\partial \tau} - \frac{3\Omega_m^0 H_0^2}{2a}\tilde\delta(\mathbf{k}, \tau) = \beta[\tilde\theta, \tilde\theta, \mathbf{k}] -\ \alpha[\frac{\partial\tilde\delta}{\partial \tau}, \tilde\theta, \mathbf{k}]-\alpha[\tilde\delta, \frac{\partial\tilde\theta}{\partial \tau}, \mathbf{k}]-\mathcal{H}\ \alpha[\tilde\delta, \tilde\theta, \mathbf{k}]\ .
\end{equation}
At second order in perturbation theory this equation becomes
\begin{eqnarray}
\label{eq:genaral_second_order_equations}
\frac{\partial^2 \delta_2(\mathbf{k}, \tau)}{\partial^2 \tau} + \mathcal{H}\frac{\partial \delta_2(\mathbf{k}, \tau)}{\partial \tau} - \frac{3\Omega_m^0H_0^2}{2a}\delta_2(\mathbf{k}, \tau) &=& \left(\frac{\partial D_+}{\partial\tau}\right)^2\beta[\delta_{1,1}, \delta_{1,1}, \mathbf{k}] + \left(D\frac{\partial^2 D_+}{\partial\tau^2} + D\mathcal{H}\frac{\partial D_+}{\partial\tau} + \left(\frac{\partial D_+}{\partial\tau}\right)^2\right)\alpha[\delta_{1,1}, \delta_{1,1}, \mathbf{k}] \nonumber \\
&=& \left(\frac{\partial D_+}{\partial\tau}\right)^2\beta[\delta_{1,1}, \delta_{1,1}, \mathbf{k}] + \left(\frac{3\Omega_m^0H_0^2}{2a} D^2 + \left(\frac{\partial D_+}{\partial\tau}\right)^2\right)\alpha[\delta_{1,1}, \delta_{1,1}, \mathbf{k}] \nonumber \\
&=& \alpha[\delta_{1,1}, \delta_{1,1}, \mathbf{k}] \left(\frac{3\Omega_m^0H_0^2}{2a} D^2 + 2\left(\frac{\partial D_+}{\partial\tau}\right)^2\right)\nonumber \\
&& +\left(\beta[\delta_{1,1}, \delta_{1,1}, \mathbf{k}] - \alpha[\delta_{1,1}, \delta_{1,1}, \mathbf{k}] \right) \left(\frac{\partial D_+}{\partial\tau}\right)^2\ .
\end{eqnarray}
This is solved by
\begin{equation}
\delta_2(\mathbf{k}, \tau) = D_{2,1}(\tau) \delta_{2,1}(\mathbf{k}) + D_{2,2}(\tau) \delta_{2,2}(\mathbf{k}) 
\end{equation}
where
\begin{equation}
D_{2,1}(\tau) \equiv D_+^2(\tau)\ ,\ \delta_{2,1}(\mathbf{k}) = \alpha[\delta_{1,1}, \delta_{1,1}, \mathbf{k}]\ ,\ \delta_{2,2}(\mathbf{k}) = \beta[\delta_{1,1}, \delta_{1,1}, \mathbf{k}] - \alpha[\delta_{1,1}, \delta_{1,1}, \mathbf{k}]
\end{equation}
and $D_{2,2}$ is given by the differential equation
\begin{equation}
\frac{\partial^2 D_{2,2}(\tau) }{\partial^2 \tau} + \mathcal{H}\frac{\partial D_{2,2}(\tau) }{\partial \tau} - \frac{3\Omega_m^0H_0^2}{2a}D_{2,2}(\tau) = \left(\frac{\partial D_+}{\partial\tau}\right)^2\ .
\end{equation}

\subsection{Second order of $\delta$ in Einstein-de Sitter universe}

Let us define $1-\mu \equiv D_{2,2}/D_+^2$. Then the general solution to \ref{eq:genaral_second_order_equations} is given by
\begin{equation}
\delta_2(\mathbf{k}, \tau) = D_+^2 \left( [1-\mu] \beta[\delta_{1,1}, \delta_{1,1}, \mathbf{k}] + \mu\alpha[\delta_{1,1}, \delta_{1,1}, \mathbf{k}] \right)\ .
\end{equation}
In an Einstein-de Sitter Universe where $\Omega_m^0 = 1$ and $D \equiv a$ we have
\begin{equation}
D_{2,2} = \frac{2}{7} D_+^2\ ,\ \mu = \frac{5}{7}
\end{equation}
and $\delta_2$ is hence given by
\begin{eqnarray}
\delta_2(\mathbf{k}, \tau) &=& D_+^2 \left( \frac{2}{7} \beta[\delta_{1,1}, \delta_{1,1}, \mathbf{k}] + \frac{5}{7}\alpha[\delta_{1,1}, \delta_{1,1}, \mathbf{k}] \right)\nonumber \\
&=&\int \mathrm{d}^3k_1\mathrm{d}^3k_2\ \delta_D(\mathbf{k} - \mathbf{k}_{12})\ F_2(\mathbf{k}_1, \mathbf{k}_2)\ \delta_{1,1}(\mathbf{k}_1)\ \delta_{1,1}(\mathbf{k}_2)\nonumber \\
\end{eqnarray}
with
\begin{eqnarray}
F_2(\mathbf{k}_1, \mathbf{k}_2) &=& \frac{5}{7}\alpha(\mathbf{k}_1, \mathbf{k}_2) + \frac{2}{7}\beta(\mathbf{k}_1, \mathbf{k}_2) \nonumber \\
&=& \frac{5}{7} + \frac{1}{2} \frac{\mathbf{k}_1\cdot \mathbf{k}_2}{k_1k_2} \left(\frac{k_1}{k_2} + \frac{k_2}{k_1}\right) + \frac{2}{7}\frac{(\mathbf{k}_1\cdot \mathbf{k}_2)^2}{k_1^2k_2^2} \ .
\end{eqnarray}

\subsection{Second order of $\delta$ in $\Lambda$CDM universe}

In a general $\Lambda$CDM universe the function $F_2$ becomes time dependent. It is given by
\begin{equation}
F_2(\mathbf{k}_1, \mathbf{k}_2, \tau) =  \mu(\tau) + \frac{1}{2} \frac{\mathbf{k}_1\cdot \mathbf{k}_2}{k_1k_2} \left(\frac{k_1}{k_2} + \frac{k_2}{k_1}\right) + [1 - \mu(\tau)]\frac{(\mathbf{k}_1\cdot \mathbf{k}_2)^2}{k_1^2k_2^2} \ .
\end{equation}
A useful property or this kernel is that
\begin{equation}
\label{eq:kernel_becoming_zero}
F_2(\mathbf{k}, -\mathbf{k}, \tau) =  \mu(\tau) + \frac{1}{2} \frac{-k^2}{k^2} \left(1+1\right) + [1 - \mu(\tau)]\frac{k^4}{k^4} = \mu(\tau) - 1 + 1 - \mu(\tau) = 0\ .
\end{equation}
Denoting the angle between $\mathbf{k}_1$ and $\mathbf{k}_2$ with $\phi$ one can also arive at the following form of $F_2(\mathbf{k}_1, \mathbf{k}_2, \tau)$ which will be useful when computing the skewness of matter inside a long cylinder:
\begin{equation}
\label{eq:kernel_in_terms_of_phi}
F_2(\mathbf{k}_1, \mathbf{k}_2, \tau) =  \frac{1}{2}\left\lbrace \left(1 + \frac{k_1}{k_2}\cos \phi\right) + \left(1 + \frac{k_2}{k_1}\cos \phi\right) \right\rbrace+ [1 - \mu(\tau)](\cos^2\phi - 1) \ .
\end{equation}

\subsection{Bispectrum and 3-point function at leading order}

The bispectrum $B(k_1, k_2, k_3, \tau)$ of $\delta$ is defined by
\begin{equation}
\langle \tilde \delta(\mathbf{k}_1, \tau) \tilde \delta(\mathbf{k}_2, \tau) \tilde \delta(\mathbf{k}_3, \tau) \rangle = \delta_D(\mathbf{k}_1 + \mathbf{k}_2 + \mathbf{k}_3) \ B(k_1, k_2, k_3, \tau)\ .
\end{equation}
At leading order in perturbation theory this can be calculated as
\begin{eqnarray}
\langle \tilde \delta(\mathbf{k}_1, \tau) \tilde \delta(\mathbf{k}_2, \tau) \tilde \delta(\mathbf{k}_3, \tau) \rangle_{2\mathrm{nd.}} &=& D_+^2 \int \mathrm{d}^3q_1\mathrm{d}^3q_2\ \delta_D(\mathbf{k}_3 - \mathbf{q}_{12})\ F_2(\mathbf{q}_1, \mathbf{q}_2, \tau)\ \langle \delta_{1,1}(\mathbf{k}_1)\ \delta_{1,1}(\mathbf{k}_2) \delta_{1,1}(\mathbf{q}_1)\ \delta_{1,1}(\mathbf{q}_2) \rangle\nonumber \\
&& +\ \mathrm{cycl.}\ ,
\end{eqnarray}
where $'\mathrm{cycl.}'$ indicates that the integral on the right-hand-side should appear for all possible permutations of $\mathbf{k}_1$, $\mathbf{k}_3$ and $\mathbf{k}_3$. Since we assume the linear density field to be a Gaussian random field, the expectation value on the left-hand-side factorizes as
\begin{eqnarray}
\langle \delta_{1,1}(\mathbf{k}_{1})\delta_{1,1}(\mathbf{k}_{2})\delta_{1,1}(\mathbf{q}_{1})\delta_{1,1}(\mathbf{q}_{2}) &=& \langle \delta_{1,1}(\mathbf{k}_{1})\delta_{1,1}(\mathbf{k}_{2}) \rangle\langle \delta_{1,1}(\mathbf{q}_{1})\delta_{1,1}(\mathbf{q}_{2}) \rangle + \langle \delta_{1,1}(\mathbf{k}_{1})\delta_{1,1}(\mathbf{q}_{1}) \rangle\langle \delta_{1,1}(\mathbf{k}_{2})\delta_{1,1}(\mathbf{q}_{2}) \rangle\nonumber \\
&&+ \langle \delta_{1,1}(\mathbf{k}_{1})\delta_{1,1}(\mathbf{q}_{2}) \rangle\langle \delta_{1,1}(\mathbf{k}_{2})\delta_{1,1}(\mathbf{q}_{1}) \rangle \nonumber \\
&=& \delta_D(\mathbf{k}_{1} + \mathbf{k}_{2})\delta_D(\mathbf{q}_{1} + \mathbf{q}_{2}) P_{\mathrm{lin},0}(k_1)P_{\mathrm{lin},0}(q_1)
+ \delta_D(\mathbf{k}_{1} + \mathbf{q}_{1})\delta_D(\mathbf{k}_{2} + \mathbf{q}_{2}) P_{\mathrm{lin},0}(k_1)P_{\mathrm{lin},0}(q_2) \nonumber \\
&& + \delta_D(\mathbf{k}_{1} + \mathbf{q}_{2})\delta_D(\mathbf{k}_{2} + \mathbf{q}_{1}) P_{\mathrm{lin},0}(k_1)P_{\mathrm{lin},0}(q_1)
\end{eqnarray}
Because of equation \ref{eq:kernel_becoming_zero} the contribution of the first term to the bispectrum is zero. Using the symmetry $1\leftrightarrow 2$ between the second and third term we hence get
\begin{eqnarray}
\langle \tilde \delta(\mathbf{k}_1, \tau) \tilde \delta(\mathbf{k}_2, \tau) \tilde \delta(\mathbf{k}_3, \tau) \rangle_{2\mathrm{nd.}} &=& 2D_+^2\ \delta_D(\mathbf{k}_1 + \mathbf{k}_2 + \mathbf{k}_3) \ F_2(\mathbf{k}_1, \mathbf{k}_2, \tau) P_{\mathrm{lin},0}(k_1) P_{\mathrm{lin},0}(k_2)\nonumber \\
&& +\ \mathrm{cycl.}\ .
\end{eqnarray}

\subsection{Variance and skewness of long cylinder at leading order in perturbation theory}

Consider a cylinder with radius $R$ and length $L$. In Fourier space the tophat filter for this cylinder is given by
\begin{equation}
\label{eq:full_filter}
W_{R, L}(\mathbf{k}) = \frac{1}{(2\pi)^3} W_L(k_\|) W_R(\mathbf{k}_\perp)
\end{equation}
where we denote the component of $\mathbf{k}$ parallel to the cylinder axis with $k_\|$ and the components orthogonal to it are represented by the two-dimensional vector $\mathbf{k}_\perp$ and $W_L$ and $W_R$ given by
\begin{equation}
W_L(k_\|) = \frac{\sin(Lk_\|/2)}{L k_\|/2} \ ,\ W_R(\mathbf{k}_\perp) = \frac{2J_1(Rk_\perp)}{Rk_\perp}\ .
\end{equation}
Here $k_\perp = |\mathbf{k}_\perp|$ and $J_\nu$ are the cylindrical bessel functions. At leading order or tree level in perturbation theory the variance of matter contrast within the cylinder is then given by
\begin{eqnarray}
\langle \delta_{R, L}^2 \rangle_{\mathrm{tree}}(\tau) &=& D_+^2(\tau) \int \mathrm{d}k_{\|,1} \mathrm{d}k_{\|,2} \mathrm{d}^2k_{\perp,1} \mathrm{d}^2k_{\perp,2}\ W_L({k}_{\|,1})\ W_L({k}_{\|,2}) \ W_R(\mathbf{k}_{\perp,1})\ W_R(\mathbf{k}_{\perp,2}) \langle \delta_{1,1}(\mathbf{k}_{1})\delta_{1,1}(\mathbf{k}_{2}) \rangle \nonumber \\
&=& D_+^2(\tau) \int \mathrm{d}k_{\|} \mathrm{d}^2k_\perp\ W_L({k}_\|)^2\ W_R(\mathbf{k}_\perp)^2\ P_{\mathrm{lin},0}(\mathbf{k})\ ,
\end{eqnarray}
where $P_{\mathrm{lin},0}(\mathbf{k})$ is today's linear power spectrum and in our Fourier convention the factors of $2\pi$ in Eqn.~\ref{eq:full_filter} cancel with corresponding factors from the convolution theorem. For $L \gg R$ we can actually approximate $W_L$ by
\begin{equation}
W_L({k}_\|)^2 \approx \frac{2\pi}{L} \delta_D({k}_\|)
\end{equation}
such that in this limit we get
\begin{eqnarray}
\langle \delta_{R, L}^2 \rangle_{\mathrm{tree}}(\tau) &\approx & \frac{(2\pi)^2 D_+^2}{L} \int \mathrm{d}k\ k\ W_R(k)^2\ P_{\mathrm{lin},0}(k) \ .
\end{eqnarray}
The third moment at tree level is given by
\begin{eqnarray}
\langle \delta_{R, L}^3 \rangle_{\mathrm{tree}}(\tau) &=& 3D_+^2 \int \mathrm{d}k_{\|,1} \mathrm{d}k_{\|,2} \mathrm{d}q_\| \mathrm{d}^2k_{\perp,1} \mathrm{d}^2k_{\perp,2} \mathrm{d}^2q_\perp\ W_L({k}_{\|,1})\ W_L({k}_{\|,2})\ W_L({q}_{\|})\ W_R(\mathbf{k}_{\perp,1})\ W_R(\mathbf{k}_{\perp,2})\ W_R(\mathbf{q}_\perp)\nonumber \\
&&\langle \delta_{1,1}(\mathbf{k}_{1})\delta_{1,1}(\mathbf{k}_{2})\delta_{2}(\mathbf{q}, \tau) \rangle \nonumber \\
&=& 3D_+^4 \int \mathrm{d}k_{\|,1} \mathrm{d}k_{\|,2} \mathrm{d}q_\| \mathrm{d}^2k_{\perp,1} \mathrm{d}^2k_{\perp,2} \mathrm{d}^2q_\perp  \mathrm{d}^3q_1  \mathrm{d}^3q_2\ W_L({k}_{\|,1})\ W_L({k}_{\|,2})\ W_L({q}_{\|})\ W_R(\mathbf{k}_{\perp,1})\ W_R(\mathbf{k}_{\perp,2})\ W_R(\mathbf{q}_\perp)\nonumber \\
&&\delta_D(\mathbf{q}-\mathbf{q}_{1}-\mathbf{q}_{2}) F_2(\mathbf{q}_{1}, \mathbf{q}_{2}, \tau) \langle \delta_{1,1}(\mathbf{k}_{1})\delta_{1,1}(\mathbf{k}_{2})\delta_{1,1}(\mathbf{q}_{1})\delta_{1,1}(\mathbf{q}_{2}) \rangle  \nonumber \\
&=& 3D_+^4 \int \mathrm{d}k_{\|,1} \mathrm{d}k_{\|,2} \mathrm{d}^2k_{\perp,1} \mathrm{d}^2k_{\perp,2} \mathrm{d}^3q_1  \mathrm{d}^3q_2\ W_L({k}_{\|,1})\ W_L({k}_{\|,2})\ W_R(\mathbf{k}_{\perp,1})\ W_R(\mathbf{k}_{\perp,2})\nonumber \\
&& W_R(\mathbf{q}_{\perp, 1} + \mathbf{q}_{\perp, 2})\ W_L({q}_{\|,1} + \mathbf{q}_{\|,2})F_2(\mathbf{q}_{1}, \mathbf{q}_{2}, \tau) \langle \delta_{1,1}(\mathbf{k}_{1})\delta_{1,1}(\mathbf{k}_{2})\delta_{1,1}(\mathbf{q}_{1})\delta_{1,1}(\mathbf{q}_{2}) \rangle \ .
\end{eqnarray}
Since we assume the linear density field to be a Gaussian random field, the expectation value on the left-hand-side factorizes as
\begin{eqnarray}
\langle \delta_{1,1}(\mathbf{k}_{1})\delta_{1,1}(\mathbf{k}_{2})\delta_{1,1}(\mathbf{q}_{1})\delta_{1,1}(\mathbf{q}_{2}) &=& \langle \delta_{1,1}(\mathbf{k}_{1})\delta_{1,1}(\mathbf{k}_{2}) \rangle\langle \delta_{1,1}(\mathbf{q}_{1})\delta_{1,1}(\mathbf{q}_{2}) \rangle + \langle \delta_{1,1}(\mathbf{k}_{1})\delta_{1,1}(\mathbf{q}_{1}) \rangle\langle \delta_{1,1}(\mathbf{k}_{2})\delta_{1,1}(\mathbf{q}_{2}) \rangle\nonumber \\
&&+ \langle \delta_{1,1}(\mathbf{k}_{1})\delta_{1,1}(\mathbf{q}_{2}) \rangle\langle \delta_{1,1}(\mathbf{k}_{2})\delta_{1,1}(\mathbf{q}_{1}) \rangle \nonumber \\
&=& \delta_D(\mathbf{k}_{1} + \mathbf{k}_{2})\delta_D(\mathbf{q}_{1} + \mathbf{q}_{2}) P_{\mathrm{lin},0}(k_1)P_{\mathrm{lin},0}(q_1)
+ \delta_D(\mathbf{k}_{1} + \mathbf{q}_{1})\delta_D(\mathbf{k}_{2} + \mathbf{q}_{2}) P_{\mathrm{lin},0}(k_1)P_{\mathrm{lin},0}(q_2) \nonumber \\
&& + \delta_D(\mathbf{k}_{1} + \mathbf{q}_{2})\delta_D(\mathbf{k}_{2} + \mathbf{q}_{1}) P_{\mathrm{lin},0}(k_1)P_{\mathrm{lin},0}(q_1)
\end{eqnarray}
Because of equation \ref{eq:kernel_becoming_zero} the contribution of the first term to the skewness is zero. Using the symmetry $1\leftrightarrow 2$ between the second and third term we hence get
\begin{eqnarray}
\langle \delta_{R, L}^3 \rangle_{\mathrm{tree}}(\tau)\rangle
&=& 6D_+^4 \int \mathrm{d}q_{\|,1} \mathrm{d}q_{\|,2} \mathrm{d}^2q_{\perp,1} \mathrm{d}^2q_{\perp,2}\ W_L({q}_{\|,1})\ W_L({q}_{\|,2})\ W_L({q}_{\|,1}+{q}_{\|,2})\ W_R(\mathbf{q}_{1})\ W_R(\mathbf{q}_{2})\ W_R(\mathbf{q}_1+\mathbf{q}_2)\nonumber \\
&&P_{\mathrm{lin},0}(q_1) P_{\mathrm{lin},0}(q_2) F_2(\mathbf{q}_{1}, \mathbf{q}_{2}, \tau)\ . \nonumber \\
\end{eqnarray}
For $L \gg R$ we can use the approximation
\begin{equation}
W_L({q}_{\|,1})\ W_L({q}_{\|,2})\ W_L({q}_{\|,1}+{q}_{\|,2}) \approx \frac{(2\pi)^2}{L^2} \delta_D^2({q}_{\|,1},{q}_{\|,2})\ .
\end{equation}
This gives
\begin{equation}
\langle \delta_{R, L}^3 \rangle_{\mathrm{tree}}(\tau)\rangle
= 6D_+^4\frac{(2\pi)^2}{L^2} \int \mathrm{d}^2q_{1} \mathrm{d}^2q_{2}\ W_R(\mathbf{q}_{1})\ W_R(\mathbf{q}_{2})\ W_R(\mathbf{q}_1+\mathbf{q}_2)\ P_{\mathrm{lin},0}(q_1) P_{\mathrm{lin},0}(q_2) F_2(\mathbf{q}_{1}, \mathbf{q}_{2}, \tau)\ ,
\end{equation}
where we will consider all vectors to be 2-dimensional from now on. Using equation \ref{eq:kernel_in_terms_of_phi} to express $F_2$ interms of $q_1$, $q_2$ and $\phi$ we can simplify this to
\begin{equation}
\langle \delta_{R, L}^3 \rangle_{\mathrm{tree}}(\tau)\rangle
= \frac{(2\pi)^3 6 D_+^4}{L^2} \int \mathrm{d}q_{1} \mathrm{d}q_{2}\ q_1 W_R({q}_{1})\ q_2 W_R({q}_{2})\ P_{\mathrm{lin},0}(q_1) P_{\mathrm{lin},0}(q_2) \int \mathrm d\phi\ W_R\left[\sqrt{q_1^2+q_2^2 + 2q_1q_2\cos\phi}\right] F_2(q_{1}, q_{2}, \phi, \tau)\ .
\end{equation}
Using relations given in Bernardeau (1995) or Buchalter et al. (2000) one can simplify the integral over $\phi$ as
\begin{eqnarray}
&&\int \mathrm{d}\phi\ W_R\left[\sqrt{q_1^2+q_2^2 + 2q_1q_2\cos\phi}\right] F_2(q_{1}, q_{2}, \phi, \tau) \nonumber \\
&=& \frac{1}{2} \int \mathrm{d}\phi\ W_R\left[\sqrt{q_1^2+q_2^2 + 2q_1q_2\cos\phi}\right] \left\lbrace \left(1 + \frac{k_1}{k_2}\cos \phi\right) + \left(1 + \frac{k_2}{k_1}\cos \phi\right) \right\rbrace \nonumber \\
&& + [1 - \mu(\tau)]\int \mathrm{d}\phi\ W_R\left[\sqrt{q_1^2+q_2^2 + 2q_1q_2\cos\phi}\right]  (\cos^2\phi - 1)  \nonumber \\
&=& 2\pi W_R(q_1) W_R(q_2) + \frac{\pi}{2} \left\lbrace W_R(q_1)\ R q_2 \left.\frac{\partial W_R(x)}{\partial x}\right|_{x=Rq_2} + W_R(q_2)\ R q_1 \left.\frac{\partial W_R(x)}{\partial x}\right|_{x=Rq_1} \right\rbrace  - \pi [1 - \mu(\tau)] W_R(q_1) W_R(q_2) \nonumber \\
&=& \pi [1 + \mu(\tau)] W_R(q_1) W_R(q_2) + \frac{\pi}{2} \frac{\partial}{\partial \ln R} \left\lbrace W_R(q_1) W_R(q_2) \right\rbrace\ .
\end{eqnarray}
For the third moment of $\delta_{R, L}$ this gives
\begin{eqnarray}
\langle \delta_{R, L}^3 \rangle_{\mathrm{tree}}(\tau) &=& [1 + \mu(\tau)]3 D_+^4(\tau)\frac{(2\pi)^4}{L^2} \int \mathrm{d}q_1 \mathrm{d}q_2\ q_1 q_2\ W_R(q_{1})^2\ W_R(q_{2})^2\  P_{\mathrm{lin},0}(q_1) P_{\mathrm{lin},0}(q_2)  \nonumber \\
&& + 3 D_+^4(\tau)\frac{(2\pi)^4}{2L^2} \int \mathrm{d}q_1 \mathrm{d}q_2\ q_1 q_2\ W_R(q_{1})\ W_R(q_{2})\ \frac{\partial}{\partial \ln R} \left\lbrace W_R(q_1) W_R(q_2) \right\rbrace  P_{\mathrm{lin},0}(q_1) P_{\mathrm{lin},0}(q_2) \nonumber \\
&=& 3[1 + \mu(\tau)]\left(\frac{(2\pi)^2 D_+^2(\tau)}{L} \int \mathrm{d}q_1\ q\ W_R(q)^2\  P_{\mathrm{lin},0}(q)\right)^2 \nonumber \\
&& +\frac{3 D_+^4(\tau)}{4} \frac{(2\pi)^4}{L^2} \frac{\partial}{\partial \ln R}\int \mathrm{d}q_1 \mathrm{d}q_2\ q_1 q_2 W_R(q_{1})^2\ W_R(q_{2})^2\  P_{\mathrm{lin},0}(q_1) P_{\mathrm{lin},0}(q_2) \nonumber \\
&=& 3[1 + \mu(\tau)]\left(\frac{(2\pi)^2 D_+^2(\tau)}{L} \int \mathrm{d}q_1\ q\ W_R(q)^2\  P_{\mathrm{lin},0}(q)\right)^2 \nonumber \\
&& + \frac{3}{4} \frac{\partial}{\partial \ln R}\left(\frac{(2\pi)^2 D_+^2(\tau)}{L} \int \mathrm{d}q_1\ q\ W_R(q)^2\  P_{\mathrm{lin},0}(q)\right)^2 \nonumber \\
&=& 3[1 + \mu(\tau)] \left( \langle \delta_{R, L}^2 \rangle_{\mathrm{tree}}(\tau) \right)^2 + \frac{3}{4} \frac{\partial}{\partial \ln R} \left( \langle \delta_{R, L}^2 \rangle_{\mathrm{tree}}(\tau) \right)^2\ .
\end{eqnarray}
Especially we have
\begin{equation}
S_3 \equiv \frac{\langle \delta_{R, L}^3 \rangle_{\mathrm{tree}}(\tau)}{\langle \delta_{R, L}^2 \rangle_{\mathrm{tree}}(\tau)^2} = 3[1 + \mu(\tau)] + \frac{3}{2} \frac{\partial \ln \langle \delta_{R, L}^2 \rangle_{\mathrm{tree}}(\tau)}{\partial \ln R} \ .
\end{equation}
For an Einstein-de Sitter universe and a power law power spectrum $P(k) \sim k^n$ this gives $S_3 = 36/7 - 3/2\ (n+2)$. The leading order prediction for $S_3$ is surprisingly good, even in the mildly non-linear regime \citep[see][and references therein]{Bernardeau}. Hence in order to predict the non-linear skewness, we simply employ the approximation
\begin{equation}
\label{eq:2nd_to_last_eq_of_app_B}
\langle \delta_{R, L}^3 \rangle_{\mathrm{non}-\mathrm{lin.}}(\tau)\ \approx\ S_3\ \langle \delta_{R, L}^2 \rangle_{\mathrm{non}-\mathrm{lin.}}(\tau)^2\ ,
\end{equation}
where we compute the non-linear variance with the use of halofit as detailed in \citet{Takahashi2014} which is a revised version of \citet{Smith2003}.

\subsection{The moment $\langle \delta_{R_A , L}^2\ \delta_{R_B , L}\rangle_{\mathrm{tree}} $}
\label{app:third_moment_rescaling}

For predicting the density split lensing signal we are also interested in the moment $\langle \delta_{R_A , L}^2\ \delta_{R_B , L}\rangle_{\mathrm{tree}} $, where $R_A$ and $R_B$ are two different Radii. The above derivations can be generalized to give
\begin{comment}
\begin{eqnarray}
\label{eq:last_eq_of_app_B}
\langle \delta_{R_A , L}^2\ \delta_{R_B , L}\rangle_{\mathrm{tree}}(\tau) &=& [1 + \mu(\tau)] \left[\ 2\ \mathrm{Var}(R_A)\ \mathrm{Cov}(R_A, R_B) + \mathrm{Cov}(R_A, R_B)^2\  \right]  \nonumber \\
&& + \mathrm{Var}(R_A)\mathrm{Cov}(R_A, R_B) \left[ \frac{1}{2}\frac{\partial \ln \mathrm{Var}(R_A)}{\partial \ln R_A}  + \frac{\partial \ln \mathrm{Cov}(R_A, R_B)}{\partial \ln R_A}\right]\ , \nonumber \\
&& + \mathrm{Cov}(R_A, R_B)^2\ \frac{\partial \ln \mathrm{Cov}(R_A, R_B)}{\partial \ln R_B}
\end{eqnarray}
\end{comment}
\begin{eqnarray}
\label{eq:last_eq_of_app_B}
\langle \delta_{R_A , L}^2\ \delta_{R_B , L}\rangle_{\mathrm{tree}}(\tau) &=& \mathrm{Var}(R_A)\ \mathrm{Cov}(R_A, R_B)\left\lbrace 2[1 + \mu(\tau)] +  \frac{1}{2}\frac{\partial \ln \mathrm{Var}(R_A)}{\partial \ln R_A}  + \frac{\partial \ln \mathrm{Cov}(R_A, R_B)}{\partial \ln R_A} \right\rbrace  \nonumber \\
&& +\ \mathrm{Cov}(R_A, R_B)^2 \left\lbrace [1 + \mu(\tau)] + \frac{\partial \ln \mathrm{Cov}(R_A, R_B)}{\partial \ln R_B}\right\rbrace
\end{eqnarray}
where we defined
\begin{equation}
\mathrm{Var}(R_A) = \langle \delta_{R_A , L}^2 \rangle_{\mathrm{tree}}\ ,\ \mathrm{Cov}(R_A, R_B) = \langle \delta_{R_A , L}\ \delta_{R_B , L}\rangle_{\mathrm{tree}}\ .
\end{equation}
To correct this expression for the non-linear evolution of the power spectrum, we compute $\mathrm{Var}(R_A)$ and $\mathrm{Cov}(R_A, R_B)$ with our halofit power spectrum whenever they appear outside of the logarithmic derivatives. This is a generalization of the rescaling of $\mathrm{Var}(R_A)$ by means of $S_3$.

For $R_B \gg R_A$ this rescaling is dominated by first term on the right hand side of equation \ref{eq:last_eq_of_app_B}. For $R_B \approx R_A$ it reduces to equation \ref{eq:2nd_to_last_eq_of_app_B}. As a consequence, using the procedure described around \ref{eq:rescaling_mixed_moments} to correct for the non-linear power spectrum evolution yields a predcition for the density split lensing signal that is almost identical to the procedure described here. Also, it can be considered accurate to the extend that equation\ref{eq:2nd_to_last_eq_of_app_B} is accurate. We nevertheless rescale the 3rd order moments in the more elaborate way described here.

\end{widetext}

\subsection{The moment $\langle \delta_{R_A , L}^n\ \delta_{R_B , L}\rangle_{\mathrm{tree}} $}
\label{app:nth_moment_rescaling}

Using a diagrammatic representation of perturbation theory (see e.g.$\ $\cite{Bernardeau}) one can see that the tree-level result for the moment $\langle \delta_{R_A , L}^n\ \delta_{R_B , L}\rangle_{c}$ will consist of terms that scale as
\begin{equation}
\sim \mathrm{Cov}(R_A, R_B)^{k}\ \mathrm{Var}(R_A)^{n-k}\ ,\ 1\leq k \leq n\ .
\end{equation}
For $R_B \approx R_A$ each of these scalings reduces to $\sim \mathrm{Var}(R_A)^{n}$ (cf.$\ $\ref{eq:2nd_to_last_eq_of_app_B} and the definition of $S_{n+1}$ in \ref{eq:Bernardeau_rescaling}). On the other hand, for $R_B \gg R_A$ the terms scaling as $\sim\mathrm{Cov}(R_A, R_B)\ \mathrm{Var}(R_A)^{n-1}$ are the dominant contributions (cf.$\ $ the last section for the case $n=2$). This is why we use \ref{eq:rescaling_mixed_moments} when rescaling moments with $n > 2$ in $G_{\mathrm{cyl.}, \theta_T w, \theta w, L}(q_l(w) L y, w)$ (see also \ref{eq:Limber_projection_G}).

%}

%\end{strip}

\section{Comparison with Millennium simulation}
\label{app:appendix_C}

\begin{figure}
\center{
  \includegraphics[width=8.0cm]{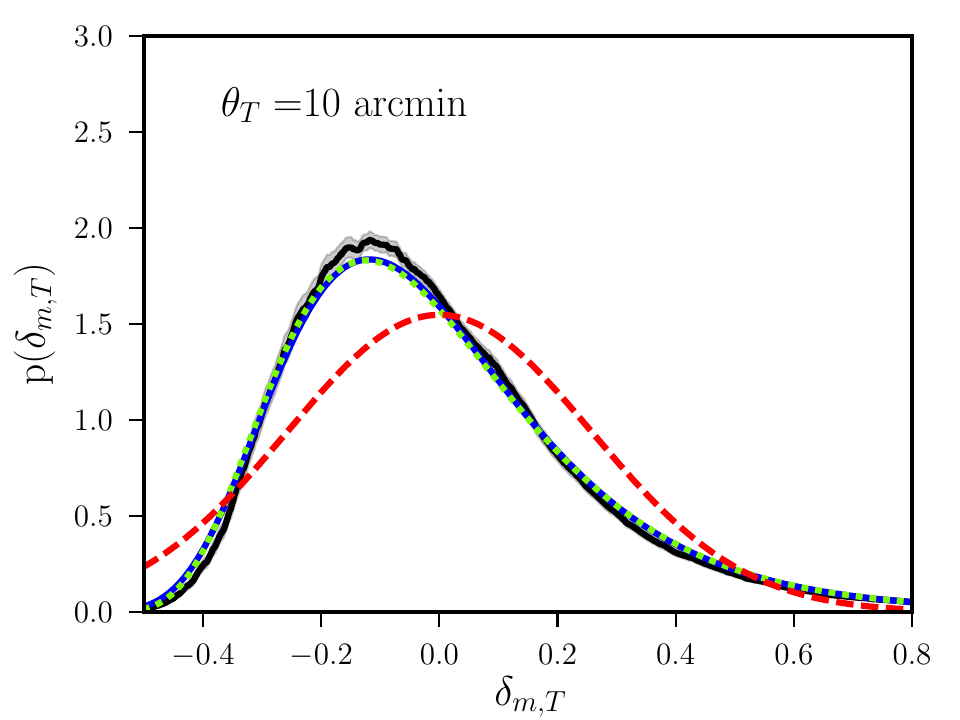}
  \includegraphics[width=8.0cm]{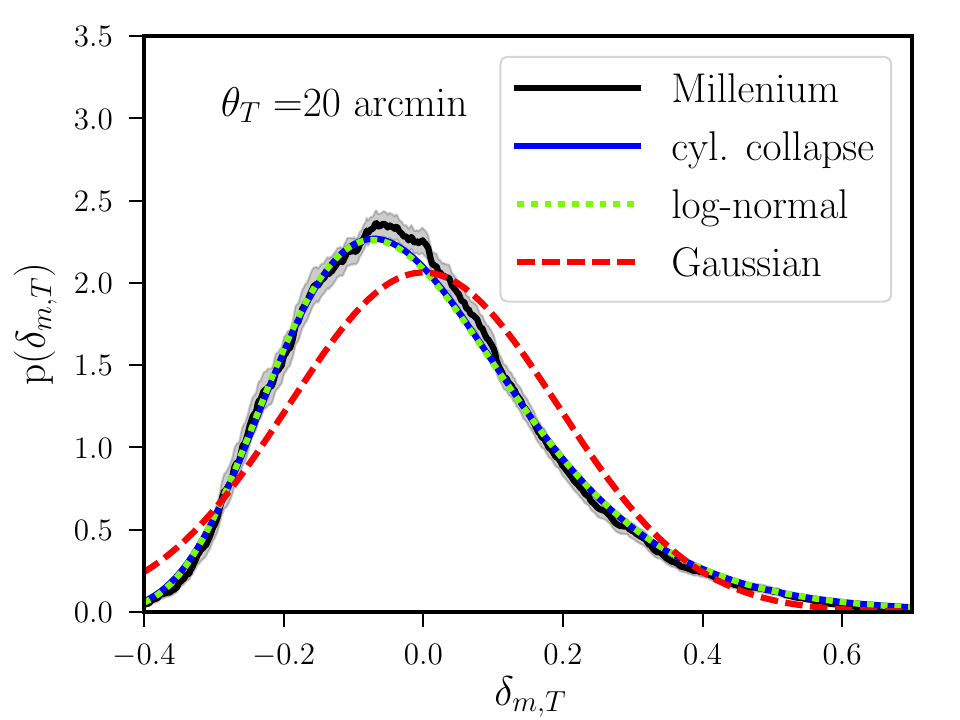}
  \includegraphics[width=8.0cm]{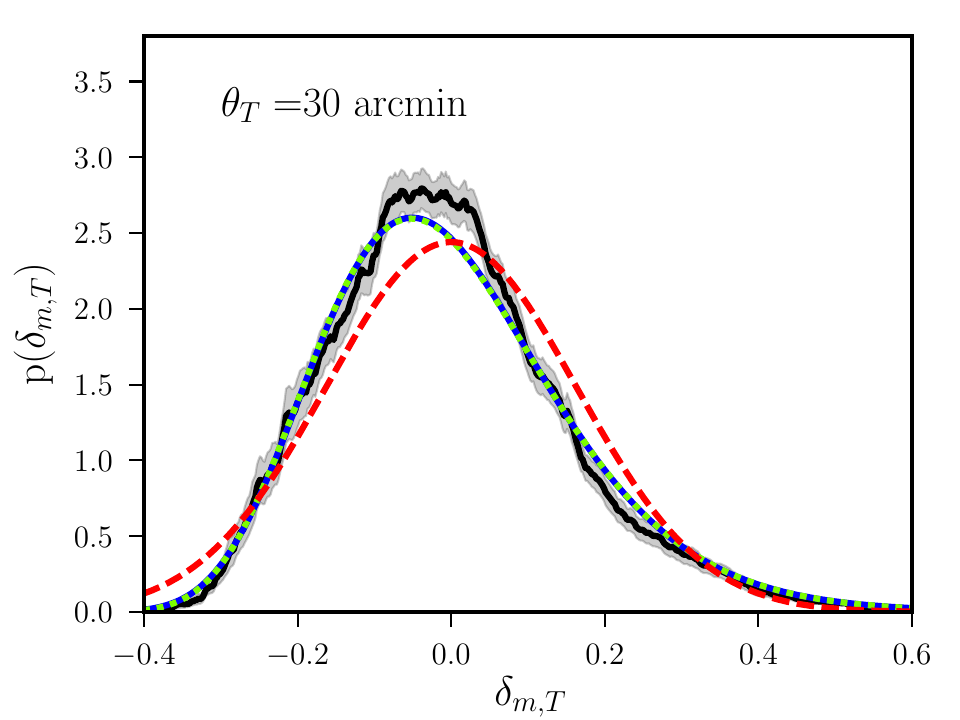}
  }
   \caption{The PDF of projected density contrast $\delta_{{m,T}}$ in the Millennium Run (MR) compared to our model. In each plot, the black line shows a histogram of $\delta_{{m,T}}$ measured from 64 patches of $4\times4\,\text{deg}^2$ made from the MR by projecting the 3D density contrast with a constant selection function $q_{{l}}$ between $0.19 \lesssim  z \lesssim 0.43$, i.e. with a constant co-moving density between those redshifts. The blue lines display the PDF predicted by our PT-motivated log-normal model, and the red lines show a Gaussian PDF with the same variance. The grey band is using the subsample covariance to estimate the error on the mean of all patches \citep{Friedrich2016}.
}
  \label{fi:p_of_delta_MR}
\end{figure}

\bblue{In Figure \ref{fi:p_of_delta_MR}, we compare our model for the PDF of projected density contrast to another set of N-body simulations, the Millennium Run \citep[MR][]{2005Natur.435..629S}. The MR has a smaller simulation volume of only $(500h^{-1}\,\text{Mpc})^3$ co-moving, but a force resolution of $5h^{-1}\,\text{kpc}$ that is 4-10 times higher than that of the Buzzard simulations. The fiducial model and the log-normal model describe the distribution of $\delta_{{m,T}}$ measured from the MR well considering the large statistical uncertainty on $p(\delta_{{m,T}})$ due to the limited simulated sky area.
}

\begin{figure}  
\begin{centering}
  \includegraphics[width=0.48\textwidth]{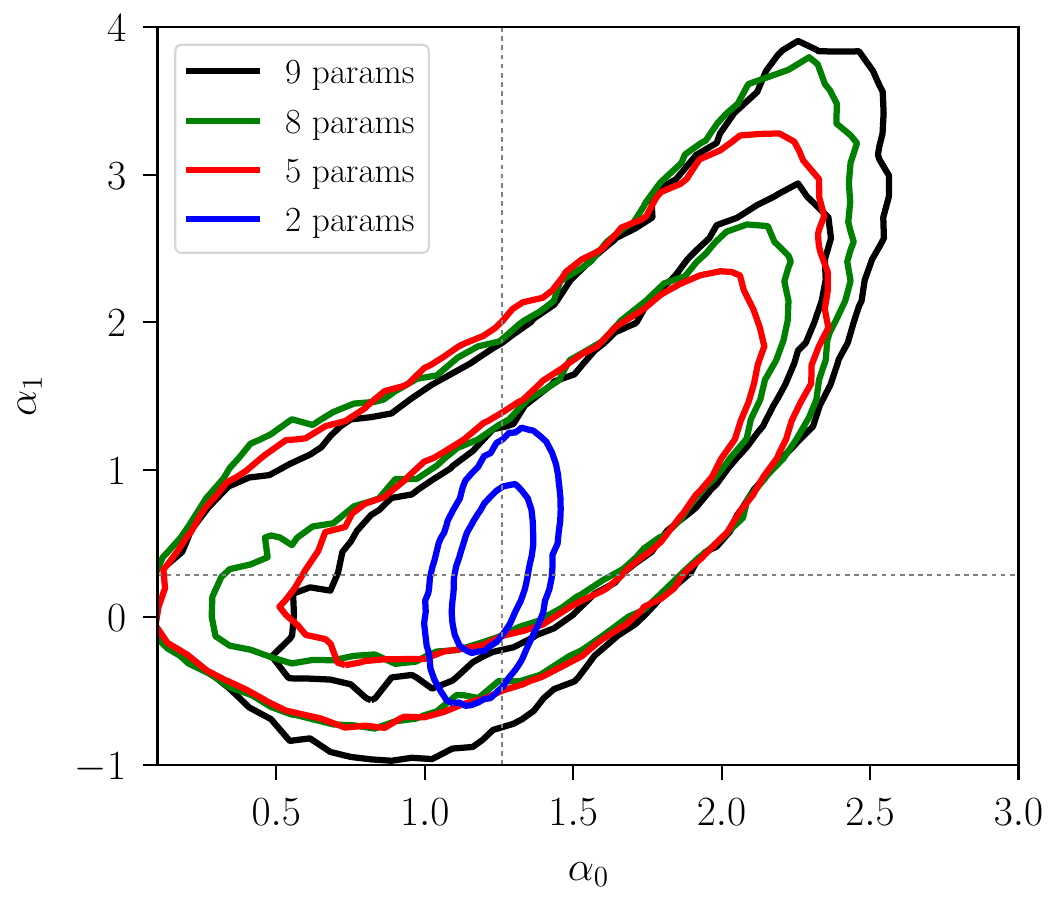}
  %\cprotect
  \caption{$1\sigma$ and $2\sigma$ contours in the $\alpha_0$-$\alpha_1$ plane from a likelihood computed around the mean of 4 shape noise free realisations of DES Y1 (but assuming the full covariance matrix for a single DES Y1). The blue contour only varies $\alpha_0$ and $\alpha_1$. The red contour marginalizes over $\Omega_m$, $\sigma_8$ and galaxy bias $b$. The green contour additionally marginalizes over  $\Omega_b$, $n_{s}$, $h_{100}$, assuming the priors used by \cite{DES2017_short}. And the black contour also allows variation of the parameter $\Delta S_3 / S_3$. Dotted lines show the values of $\alpha_0$ and $\alpha_1$ that were found to describe our mock data best in section \ref{sec:parametric_shot_noise}.}
  \label{fi:a0_a1_contour}
  \end{centering}
\end{figure}

\begin{figure}  
\begin{centering}
  \includegraphics[width=0.48\textwidth]{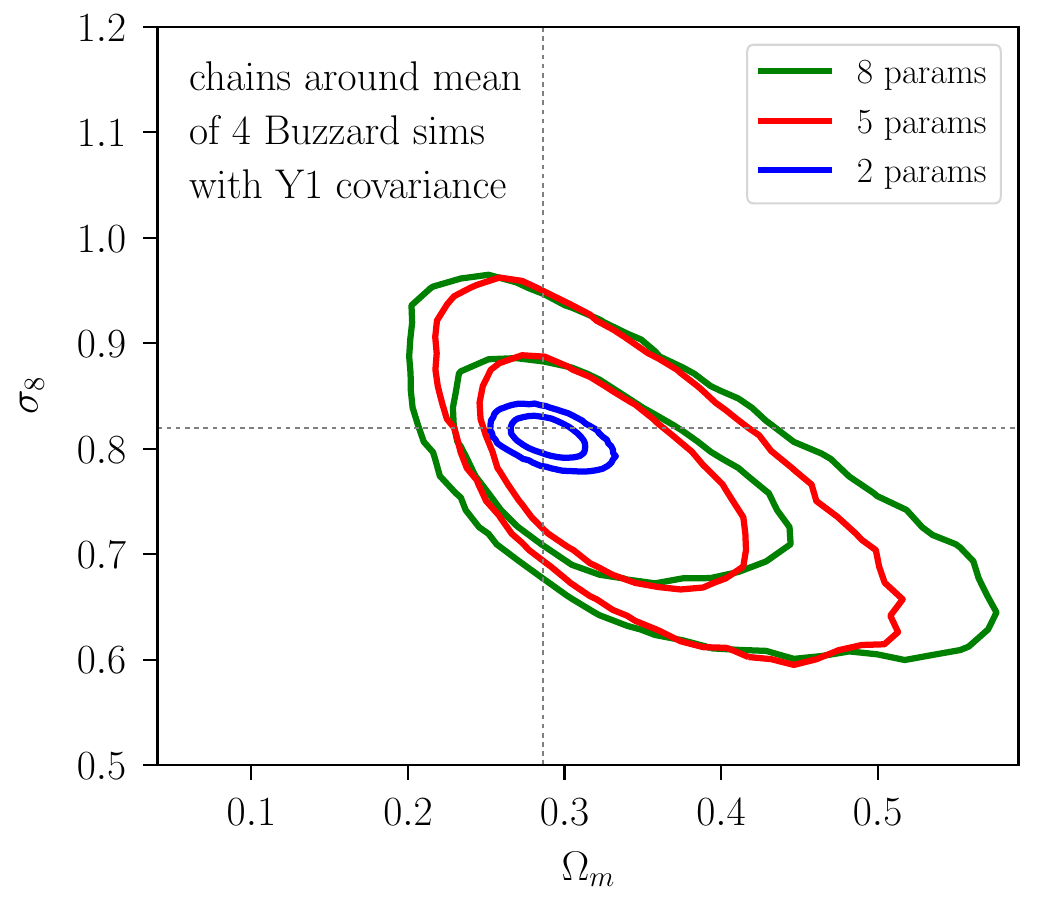}\\\includegraphics[width=0.48\textwidth]{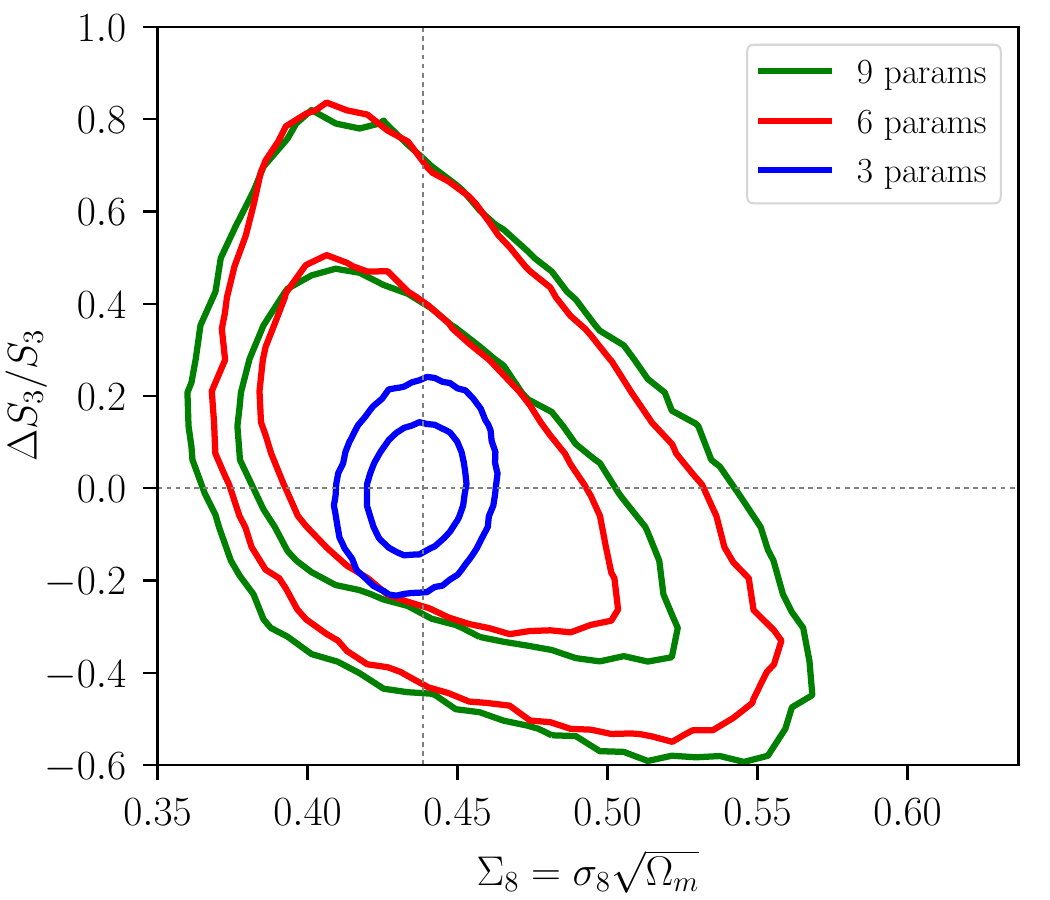}
  %\cprotect
  \caption{In analogy to figure \ref{fi:recover_cosmology}, we test whether our alternative shot-noise parametrization can recover the Buzzard cosmology in a simulated likelihood analysis. \underline{Top panel:} $1\sigma$ and $2\sigma$ contours in the $\Omega_m$-$\sigma_8$ plane from a likelihood computed around the mean of 4 shape noise free realisations of DES Y1 (but assuming the full covariance matrix for a single DES Y1). The green contours are marginalized over $\Omega_b$, $n_{s}$, $h_{100}$, redMaGiC galaxy bias $b$ as well as the shot-noise parameters $\alpha_0$ and $\alpha_1$. For the parameters $\Omega_b$, $n_{s}$, $h_{100}$ we have assumed the same flat priors as used in the DES Y1 combined probes analysis presented in Abbott et al. (in prep.). The red contours are marginalized only over bias and shot-noise parameters and the blue contours only vary $\Omega_m$ and $\sigma_8$. Even when going to this small parameter space, our model agrees with Buzzard within 1$\sigma$ errors of DES Y1. \underline{Bottom panel:} Same contours but in the $\Sigma_8$-$\Delta S_3 / S_3$ plane and varying one additional parameter, $\Delta S_3 / S_3$. Dotted lines show the true Buzzard cosmology and our fiducial value of $\Delta S_3 / S_3 = 0$.}
  \label{fi:recover_cosmology_alternative}
  \end{centering}
\end{figure}

\section{Galaxy stochasticity}
\label{app:stochasticity}

Consider the field of galaxy density contrast $\delta_{g,T}$ and the field of matter density contrast $\delta_{m,T}$, where both fields are assumed to be smoothed over a fix circular aperture. The number of galaxies found inside such an aperture is assumed to be a Poissonian random variable with first and second moments for a given value of $\delta_{g,T}$ are given by
\begin{equation}
\langle \hat N | \delta_{g,T} \rangle = \bar N (1+\delta_{g,T})
\end{equation}
and
\begin{equation}
\langle \hat N^2 | \delta_{g,T} \rangle = \bar N (1+\delta_{g,T}) + \bar N^2 (1+\delta_{g,T})^2\ .
\end{equation}
Let $\mathrm{Var}_m$ be the variance of $\delta_{m,T}$ and $\mathrm{Var}_g = b^2 \mathrm{Var}_m$ the variance of $\delta_{g,T}$, where $b$ is the galaxy bias. Then the galaxy stochasticity $r$ is defined by $\mathrm{Cov}_{mg} = rb \mathrm{Var}_m$, i.e. it is the correlation coefficient of $\delta_{g,T}$ and $\delta_{m,T}$.

We will now assume both $\delta_{g,T}$ and $\delta_{m,T}$ to be joint log-normal random variables, i.e.
\begin{eqnarray}
\delta_{m,T} &=& \left[e^{n_m} - 1\right]\delta_{m,0} \nonumber \\
\delta_{g,T} &=& \left[e^{n_g} - 1\right]\delta_{g,0} \ ,
\end{eqnarray}
where $n_m$ and $n_g$ have a joint Gaussian distribution and $\delta_{g,0} = b \delta_{m,0}$. The variances of $n_m$ and $n_g$ are given by
\begin{eqnarray}
\sigma_m^2 &=& \ln\left\lbrace 1 + \frac{\mathrm{Var}_m}{\delta_{m,0}^2} \right\rbrace\nonumber \\
\sigma_g^2 &=& \ln\left\lbrace 1 + \frac{\mathrm{Var}_g}{\delta_{g,0}^2} \right\rbrace\nonumber \\
&=& \sigma_m^2
\end{eqnarray}
and their covariance is given by
\begin{eqnarray}
\xi_{mg} &=& \ln\left\lbrace 1 + \frac{\mathrm{Cov}_{mg}}{\delta_{m,0}\delta_{g,0}} \right\rbrace\nonumber \\
&=& \ln\left\lbrace 1 + r\frac{\mathrm{Var}_{m}}{\delta_{m,0}^2} \right\rbrace\ .
\end{eqnarray}
Let us denote the correlation coefficient of the Gaussian field by
\begin{eqnarray}
\rho &=& \frac{\xi_{mg}}{\sigma_m^2} \nonumber \\ 
 &=& \frac{\ln\left\lbrace 1 + r\frac{\mathrm{Var}_{m}}{\delta_{m,0}^2} \right\rbrace }{\ln\left\lbrace 1 + \frac{\mathrm{Var}_m}{\delta_{m,0}^2} \right\rbrace } \ .
\end{eqnarray}
Note that $\rho$ will depend on scale even of $b$ and $r$ do not.

Now we want to compute the conditional moments $\langle \delta_{g,T} | \delta_{m,T} \rangle$ and $\langle \delta_{g,T}^2 | \delta_{m,T} \rangle$. First,
\begin{eqnarray}
 \langle e^{n_g} | n_m \rangle &=& e^{\langle n_g | n_m\rangle + \sigma_g^2(1-\rho^2)/2} \nonumber \\
 &=& e^{\rho(n_m + \sigma_m^2/2) - \sigma_g^2\rho^2/2} \nonumber \\
 &=& e^{\sigma_g^2 (\rho - \rho^2)/2} e^{\rho n_m} \ .
\end{eqnarray}
Second,
\begin{eqnarray}
 \mathrm{Var}\left( e^{n_g} | n_m \right) &=& \left( e^{\sigma_g^2(1-\rho^2)} -1 \right) e^{2\langle n_g | n_m\rangle + \sigma_g^2(1-\rho^2)} \nonumber \\
 &=& \left( e^{\sigma_g^2(1-\rho^2)} -1 \right) e^{\sigma_g^2 (\rho - \rho^2)} e^{2\rho n_m} \ .
\end{eqnarray}
Now what is $\mathrm{Var}(\hat N | \delta_{m,T})$?
\begin{eqnarray}
\langle \hat N^2 | \delta_{g,T} \rangle &=& \langle \hat N^2 | n_m \rangle \nonumber \\
&=& \int \mathrm{d}\delta_{g,T}\ p(\delta_{g,T} | n_m)\times \nonumber \\
&& \times\left( \bar N [1+\delta_{g,T}] + \bar N^2 [1+\delta_{g,T}]^2 \right) \nonumber\\
&=&\bar N + \bar N^2 + \int \mathrm{d}\delta_{g,T}\ p(\delta_{g,T} | n_m)\times \nonumber \\
&& \times\left( \delta_{g,T} [\bar N\ + 2\bar N^2] + \bar N^2\delta_{g,T}^2 \right) \nonumber\\
&=&\bar N + \bar N^2 + [\bar N\ + 2\bar N^2]\langle \delta_{g,T} | n_m \rangle + \nonumber\\
&& + \bar N^2 \left( \mathrm{Var}\left( \delta_{g,T} | n_m \right) + \langle \delta_{g,T} | n_m \rangle^2 \right) \nonumber
\end{eqnarray}
\begin{eqnarray}
\Rightarrow \mathrm{Var}(\hat N | \delta_{m,T}) &=& \bar N ( 1 + \langle \delta_{g,T} | n_m \rangle) + \bar N^2 \mathrm{Var}\left( \delta_{g,T} | n_m \right)  \nonumber\\
&=& \bar N ( 1 + \langle \delta_{g,T} | n_m \rangle) + \bar N^2 \delta_{g,0}^2 \mathrm{Var}\left( e^{n_g} | n_m \right)  \nonumber\\
\end{eqnarray}
The probability $P(N_A | \delta_{m,T})$ can be computed in a similar way, by numerically evaluating
\begin{equation}
P(N_A | \delta_{m,T}) = \int \mathrm{d}\delta_{g,T}\ p(\delta_{g,T}|\delta_{m,T})\ P(N_A | \delta_{g,T})\ ,
\end{equation}
where $p(\delta_{g,T}|\delta_{m,T})$ can be computed from basic relations for joint log-normal random variables.

\section{Validation of alternative shot-noise model}
\label{app:validate_alternative_model}

In our data analysis \citep{Gruen2017} we investigate both shot-noise parametrizations introduced in section \ref{sec:CiC_computation}. We hence check whether our alternative shot-noise parametrization, i.e$.$ the one that uses three parameters to describe the relation between matter and galaxies ($b$, $\alpha_0$ and $\alpha_1$, cf$.$ section \ref{sec:parametric_shot_noise}), recovers the true cosmology and shot-noise parameters of our mock data.

In figure \ref{fi:a0_a1_contour} we show the posterior constraints derived for the two shot-noise parameters $\alpha_0$ and $\alpha_1$, when marginalizing over different sets of model parameters. Our priors $0.1 < \alpha_0 < 3.0$ and $-1.0 < \alpha_1 < 4.0$ are mildly informative. We however expect that even stronger priors can be motivated (cf$.$ our discussion in section \ref{sec:parametric_shot_noise}) and will investigate this in future work. Figure \ref{fi:recover_cosmology_alternative} shows that our alternative shot-noise parametrization also recovers the correct Buzzard cosmology (cf$.$ figure \ref{fi:recover_cosmology}, which presents the same test for our baseline model).

\end{document}